\documentclass{aa}

\usepackage{graphicx}
\usepackage{natbib}
\usepackage{ulem}
\usepackage{lscape} 
\usepackage[varg]{txfonts}
\usepackage{verbatim}

\bibpunct{(}{)}{;}{a}{}{,} 

\usepackage{epsfig}

\usepackage{amssymb}

\usepackage{amsmath}
 
\usepackage{multirow}
\usepackage{multicol}
\usepackage{xcolor} 

\usepackage{xspace} 

\newcommand{\ind}[1]{_{\mathrm{#1}}}

\newcommand{\kep}{\textit{Kepler}\xspace}

\newcommand\dnurotcore{\delta\nu\ind{rot,core}}

\defcitealias{Balona_2015}{Bal15}
\defcitealias{Ceillier_2017}{Cei17}
\defcitealias{Costa_2015}{Cos15}
\defcitealias{Davenport_2016}{Dav16}
\defcitealias{Deacon_2016}{Dea16}
\defcitealias{Godoy-Rivera_2018}{GR18}
\defcitealias{Howell_2016}{Hwl16}
\defcitealias{Honda_2016}{Hnd16}
\defcitealias{Katsova_2018}{Kat18} 
\defcitealias{Kumar_2018}{Kum18} 
\defcitealias{Pugh_2016}{Pug16}
\defcitealias{Schwamb_2012}{Sch12}
\defcitealias{Tang_2012}{Tan12}
\defcitealias{Tayar_2015}{Tay15}

\begin{document}

\title{KIC 7955301: a hierarchical triple system with eclipse timing variations and an oscillating red giant}


\author{Patrick Gaulme\inst{\ref{inst1},\ref{inst2}} 
\and Tam\'as Borkovits\inst{\ref{inst3a},\ref{inst3b},\ref{inst3c},\ref{inst3d},\ref{inst3e}} 
\and Thierry Appourchaux\inst{\ref{inst_app}}
\and Kre\v{s}imir Pavlovski\inst{\ref{inst_pav}} 
\and Federico Spada \inst{\ref{inst1}} 
\and Charlotte Gehan\inst{\ref{inst1},\ref{inst8}}
\and Joel Ong\inst{\ref{inst_yale}}
\and Andrea Miglio\inst{\ref{inst_bo1},\ref{inst_bo2},\ref{inst_bhm1}} 
\and Andrew Tkachenko\inst{\ref{inst66}}
\and Beno\^{\i}t Mosser\inst{\ref{inst6}}
\and Mathieu Vrard\inst{\ref{inst7}} 
\and Mansour Benbakoura\inst{\ref{inst5}}
\and S. Drew Chojnowski\inst{\ref{inst9}} 
\and Jean Perkins\inst{\ref{inst_monterey}} 
\and Anne Hedlund\inst{\ref{inst2}} 
\and Jason Jackiewicz\inst{\ref{inst2}} 
}

\institute{Max-Planck-Institut f\"{u}r Sonnensystemforschung, Justus-von-Liebig-Weg 3, 37077, G\"{o}ttingen, Germany \email{gaulme@mps.mpg.de}\label{inst1}
\and 
Department of Astronomy, New Mexico State University, P.O. Box 30001, MSC 4500, Las Cruces, NM 88003-8001, USA \label{inst2}
\and
Baja Astronomical Observatory of University of Szeged, H-6500 Baja, Szegedi \'ut, Kt. 766, Hungary \label{inst3a}
\and
ELKH-SZTE Stellar Astrophysics Research Group, H-6500 Baja, Szegedi \'ut, Kt. 766, Hungary \label{inst3b}
\and
Konkoly Observatory, Research Centre for Astronomy and Earth Sciences,  H-1121 Budapest, Konkoly Thege Mikl\'os \'ut 15-17, Hungary \label{inst3c}
\and
ELTE Gothard Astrophysical Observatory, H-9700 Szombathely, Szent Imre h. u. 112, Hungary \label{inst3d}
\and
MTA-ELTE Exoplanet Research Group, H-9700 Szombathely, Szent Imre h. u. 112, Hungary \label{inst3e}
\and
Université Paris-Saclay, Institut d’Astrophysique Spatiale, UMR 8617, CNRS, Bâtiment 121, 91405 Orsay Cedex, France \label{inst_app}
\and
Department of Physics, Faculty of Science, University of Zagreb, Bijeni\v{c}ka
cesta 32, 10000 Zagreb, Croatia\label{inst_pav}
\and
Instituto de Astrof\'isica e Ci\^encias do Espa\c{c}o, Universidade do Porto, CAUP, Rua das Estrelas, 4150-762, Porto, Portugal\label{inst8}
\and
Department of Astronomy, Yale University, P.O. Box 208101, New Haven, CT 06520-8101, USA \label{inst_yale}
\and
Dipartimento di Fisica e Astronomia Augusto Righi, Università degli Studi di Bologna, Via Gobetti 93/2, I-40129 Bologna, Italy \label{inst_bo1}
\and
INAF-Osservatorio di Astrofisica e Scienza dello Spazio di Bologna, via Gobetti 93/3, I-40129 Bologna, Italy
 \label{inst_bo2}
\and
School of Physics and Astronomy, University of Birmingham, Birmingham B15 2TT, UK \label{inst_bhm1}
\and
Institute of Astronomy, KU Leuven, Celestijnenlaan 200D, B-3001 Leuven, Belgium
\label{inst66}
\and
LESIA, Observatoire de Paris, Universit\'e PSL, CNRS, Sorbonne Universit\'e, Universit\'e de Paris, 92195 Meudon, France\label{inst6}
\and
Department of Astronomy, The Ohio State University, 140 West 18th Avenue, Columbus OH 43210, USA \label{inst7}
\and
Laboratoire AIM, DRF/IRFU/SAp, CEA Saclay, 91191 Gif-sur-Yvette Cedex, France\label{inst5}
\and
 Department of Physics, Montana State University, P.O. Box 173840, Bozeman, MT 59717-3840, USA\label{inst9}
 \and
Monterey Institute for Research in Astronomy, 200 8th Street, Marina, CA 93933, USA \label{inst_monterey}
}

\titlerunning{KIC 7955301: a hierarchical triple system with oscillating red giant}
\authorrunning {Gaulme et al.}
\abstract{KIC 7955301 is a hierarchical triple system with clear eclipse timing and depth variations that was discovered by the \kep satellite during its original mission.  It is composed of a non-eclipsing primary star at the bottom of the red giant branch on a 209-day orbit with a K/G-type main-sequence inner eclipsing binary, orbiting in 15.3 days. This system was noted for the large amplitude of its eclipse timing variations (over 4 hours), and the detection of clear solar-like oscillations of the red-giant component, including p-modes of degree up to $l=3$ and mixed $l=1$ modes. The system is a single-lined spectroscopic triple, meaning that only spectral lines from the red giant are trackable along the orbit. We perform a dynamical model by combining the 4-year-long \kep photometric data, eclipse timing variations and radial-velocity data obtained with the high resolution spectrometers ARCES of the 3.5-m ARC telescope at Apache Point observatory and SOPHIE of the 1.93-m telescope at Haute Provence Observatory. The ``dynamical'' mass of the red-giant component is determined with a 2\,\% precision at $1.30^{+0.03}_{-0.02} M_\odot$. We perform asteroseismic modeling based on the global seismic parameters and on the individual frequencies. Both methods provide an estimate of the mass of the red giant that matches the dynamical mass within the uncertainties. Asteroseismology also reveals the rotation rate of the core ($\approx 15$ days), the envelope ($\sim 150$ days), and the inclination ($\sim75^\circ$) of the red giant. Three different approaches lead to estimating the age to range between 3.3 and 5.8 Gyr, which highlights the difficulty of determining stellar ages despite the exceptional wealth of information available for this system. On short timescales, the inner binary exhibits eclipses with varying depths during a $7.3$-year long interval, and no eclipses during the consecutive $11.9$\,years. This is why \textit{Kepler} could detect its eclipses, TESS will not, and the future ESA PLATO mission should. Over the long term, the system appears to be stable and owes its evolution to the evolution of its individual components. This triple system could end its current smooth evolution by merging by the end of the red giant branch of the primary star because the periastron distance is $\approx 142 R_\odot$, which is close to the expected radius of the red giant at the tip of the RG branch. 
}

\maketitle


\section{Introduction}
\label{sect_intro}
Among the 5000 stars that are visible with the naked eye, about 2000 are known to be multiple-star systems. Naked-eye stars are a small fraction of the Milky Way, but are reasonably representative of the incidence of binarity, which is estimated to be between 50 and almost 100\,\% \citep[e.g.,][]{Eggleton_2006}. Some systems are close enough to be in contact, others are far apart enough to evolve almost independently. Binary systems are known with orbital periods as short as 0.2 days or as long as thousands of years. Studying multiple-star systems is thus important for understanding the evolution of our Galaxy. 

Stars in multiple systems are also precious benchmarks for calibrating asteroseismology when it is possible to measure the mass of the oscillating star independently from its oscillation properties. This is the case with double-lined spectroscopic binary (SB2) where the components eclipse each other, or optically resolved binary systems. Hitherto, all solar-like oscillators belonging to eclipsing binary stars (EBs) are red giants (RGs) detected by the \textit{Kepler} mission \citep[][]{Hekker_2010, Gaulme_2013, Gaulme_2014, Beck_2014,Beck_2015, Kuszlewicz_2019, Benbakoura_2021}. So far, fourteen wide SB2/EBs including an oscillating RG have been fully characterized with the help of ground-based radial-velocity support \citep[][]{Frandsen_2013,Rawls_2016, Gaulme_2016a, Brogaard_2018,Themessl_2018, Benbakoura_2021}. Three more bona fide oscillating RGs in EBs were reported by \citet{Gaulme_Guzik_2019}, and are currently under study. Beyond EBs, it is also possible to determine the masses of stars belonging to visual multiple-systems, where individual components are spatially resolved, which allows retrieval of their projected orbits, provided radial velocities are available too. So far, very few of such systems include solar-like pulsators \citep[][]{Marcadon_2018, Metcalfe_2020}. The drawback of such systems is that the orbits are long and getting accurate orbital parameters can take decades and rely on heterogeneous datasets. Finally, hierarchical triple systems are promising types of benchmarks; they are composed of a close binary with a relatively distant companion \citep[e.g.,][]{Tokovinin_1997,Ford_2000,Borkovits_2003}. Depending on their orbital configurations, the presence of eclipses and significant eclipse timing variations (ETVs), it can be possible to determine the masses of their components \citep{Derekas_2011, Carter_2011, Borkovits_2016}.

KIC 7955301 is a hierachical triple system composed of a young RG and a pair of small main-sequence stars that was observed for nearly four consecutive years by the original NASA \kep mission \citep{Borucki_2011} at long (29.4244-minute cadence). It was pointed out by \citet{Gaulme_2013} and \citet{Rappaport_2013} who noticed its outstanding eclipse timing, depth and duration variations. The pair of main-sequence stars shows partial eclipses, but the RG does not eclipse the inner binary. The eclipse timings show a periodicity of about 208.6 days, while the pair of MS stars orbit in 15.3 days. The precession of the orbital plane of the MS stars is quite visible as well, and has a period longer than the time series. This system was part of the cohort of systems with ETVs studied by \citet{Borkovits_2016}. From eclipse timing only, they estimated the mass of the RG to be $1.5\pm 0.5 M_\odot$ and the total mass of the MS pair to be $2.2\pm0.8 M_\odot$. From asteroseismic scaling relations and a quick look at the mixed dipole ($l=1$) modes, \citet{Gaulme_2013} inferred the RG mass to be $1.2\pm0.2 M_\odot$, the radius $5.9\pm0.2 R_\odot$, and the core rotation of about 30 days. By considering that asteroseismic scaling relations tend to slightly overestimate stellar masses for RGs \citep{Gaulme_2016a}, the RG is likely a descendent of an F-type star that had a convective envelope during the MS and that is expected to rotate slowly.

With this paper, we firstly aim at testing our ability to measure accurate masses in hierarchical triple systems. We know it is possible to determine the mass of a star that belongs to an eclipsing binary down to 1\,\% \citep[e.g.,][]{Maxted_2020}. However, no study has specifically been developed for measuring the mass of an oscillating star in a hierarchical triple that is a single-lined spectroscopic triple system (ST1). We note that the famous RG in the hierarchical triple system HD 181068 \citep{Derekas_2011}, observed by \kep, is triply eclipsing and does not show any oscillations, likely because of the mode suppression observed in short period multiple systems \citep{Gaulme_2020}. 
Given the clear ETVs and the clear oscillations of the RG, KIC 7955301 appears to be an ideal case for testing our methods on hierarchical triple systems with an oscillating component. Subsequently, the approaches described here should be extended to all known triple with ETVs that include an oscillating component \citep{Gaulme_2013}.

The second objective is a rather unique opportunity to study an RG star in greater detail, thanks to the multiple approaches used simultaneously to give a clear picture of the system. Our study makes use of the \kep photometric data and high-resolution optical spectra obtained with the 3.5-m telescope at Apache Point observatory, and with the échelle spectrograph SOPHIE of the 1.93-m telescope at the Haute-Provence Observatory (Sect. \ref{sect_observations}). The spectra are used to both determine the radial velocities and the atmospheric parameters (Sect. \ref{sect_spec_ana}). The \kep light curves are used to both analyze the eclipses for the asteroseismic analysis (Sect. \ref{sect_astero}) and the dynamical modeling (Sect. \ref{sect_dyn_mod}). Finally, the deduced parameters -- oscillation frequencies, orbital parameters, etc. -- are used to optimize a stellar evolution model (Sect. \ref{sect_stellar_model}).

\begin{table}
\centering
\caption{Archival properties of the KIC 7955301 triple System.}
\begin{tabular}{lc}
\hline
\hline
Parameter & Value   \\
\hline
RA (J2000) & $19:20:45$  \\  
Dec (J2000) &  $+43:43:26$  \\  
$K_\mathrm{p}$$^a$ & $12.672$ \\
$G$$^b$& $12.6283 \pm 0.0002$  \\
$G_{\rm BP}^b$ & $13.2017 \pm 0.0012$  \\
$G_{\rm RP}$$^b$ & $11.9340 \pm 0.0009$  \\
B$^c$ & $14.082 \pm 0.043$ \\
V$^d$ & $12.901 \pm 0.008$ \\
g$'^d$& $13.419 \pm 0.045$ \\
r$'^d$& $12.553 \pm 0.043$ \\
i$'^d$& $12.323 \pm 0.026$ \\
J$^e$ & $10.992 \pm 0.023$ \\
H$^e$ & $10.487 \pm 0.018$ \\
K$^e$ & $10.385 \pm 0.011$ \\
W1$^f$ & $10.323 \pm 0.023$ \\
W2$^f$ & $10.409 \pm 0.020$ \\
W3$^f$ & $10.310 \pm 0.056$ \\
W4$^f$ & $ 9.118$  \\
$T_{\rm eff}$ (K)$^c$ & $4805 \pm 75$  \\
Distance (pc)$^g$ & $1375\pm35$  \\ 
${\rm [M/H]}^c$ & $0.1163 \pm 0.0072$ \\ 
$E(B-V)^c$ & $0.044$ \\
$\mu_\alpha$ (mas ~${\rm yr}^{-1}$)$^h$ & $-0.86 \pm 0.02$   \\ 
$\mu_\delta$ (mas ~${\rm yr}^{-1}$)$^h$ &  $-8.59 \pm 0.02$   \\ 
\hline
\label{tbl:mags}  
\end{tabular}
\tablefoot{RA and Dec are the system coordinates in the J2000 frame, $K_\mathrm{p}$ is the Kepler magnitude, $G$, $G_\mathrm{BP}$ and $G_\mathrm{RP}$ are the visible, blue and red passband Gaia magnitudes from DR2, parameters from B to $W4$ are magnitudes whose origin is detailed in the notes underneath the Table, $T\ind{eff}$ is the effective temperature (Kelvin), [M/H] the metallicity (dex), $E(B-V)$ the reddening, and $\mu_\alpha$ and $\mu_\delta$ the parallaxes along the right ascension and declination respectively. \tablefoottext{a}{\kep Input Catalog \citep{Brown_2011}.} \tablefoottext{b}{Gaia DR2 \citep{GaiaDR2}.} \tablefoottext{c}{TESS Input Catalog v8.1 \citep{Stassun_2018}} \tablefoottext{d}{AAVSO Photometric All Sky Survey (APASS) DR9, \citep{Henden_2015}, \url{http://vizier.u-strasbg.fr/viz-bin/VizieR?-source=II/336/apass9}.} \tablefoottext{e}{2MASS catalog \citep{Skrutskie_2006}.} \tablefoottext{f}{WISE point source catalog \citep{Cutri_2013}.} \tablefoottext{g}{\citet{Bailer-Jones_2021}. (h) Gaia EDR3 \citep{GaiaEDR3}.} Note also, that for the SED analysis in Sect.~\ref{subsect:PARSECSEDetc} the uncertainties of the passband magnitudes were set to $\sigma_\mathrm{mag}=\mathrm{max}(\sigma_\mathrm{catalog},0.030)$ to avoid the strong overdominance of the extremely accurate Gaia magnitudes over the other measurements.}
\end{table} 

\section{Observations}
\label{sect_observations}
\begin{figure}[t!]
\includegraphics[width=8.5cm]{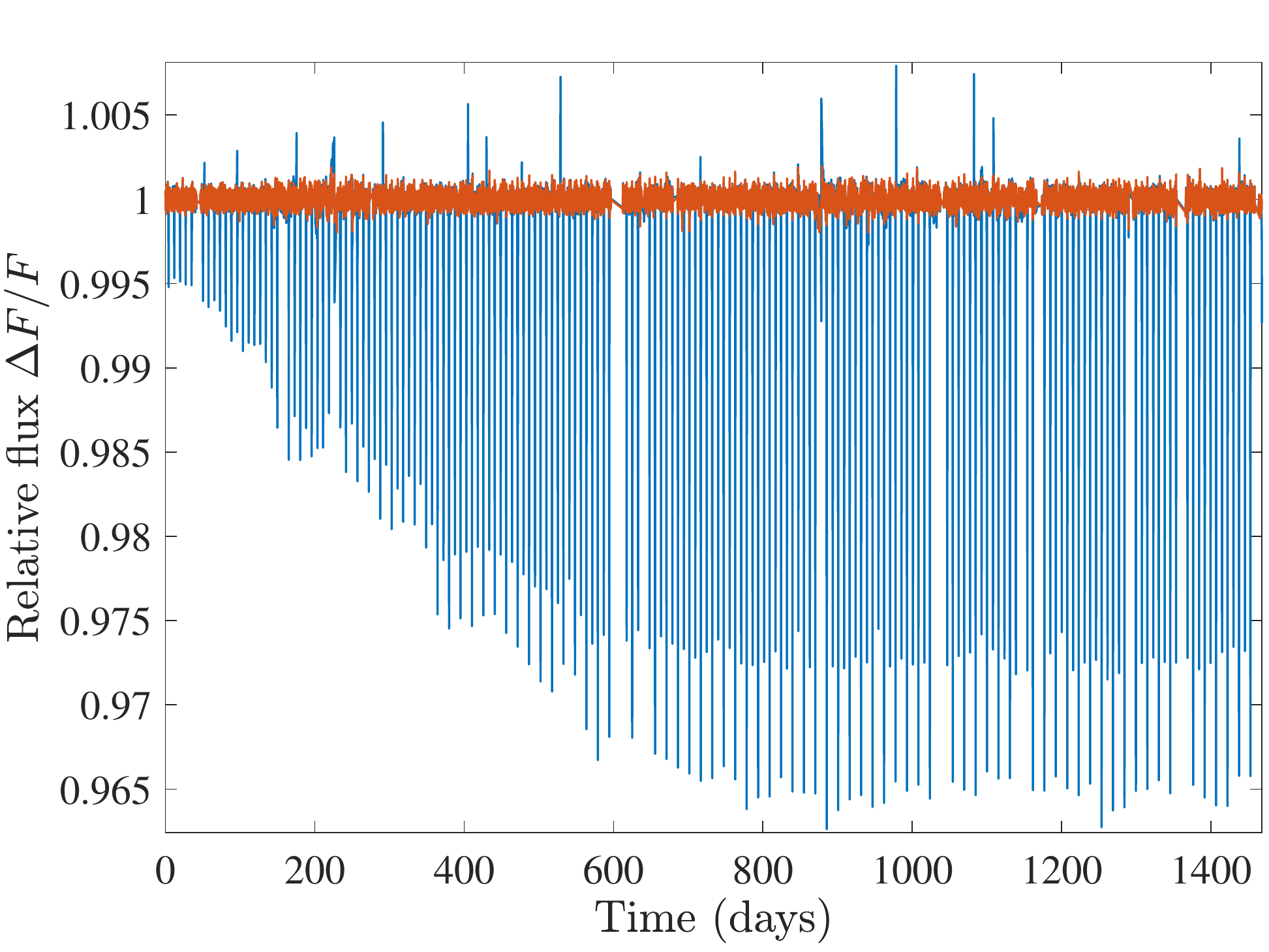} 
\includegraphics[width=8.5cm]{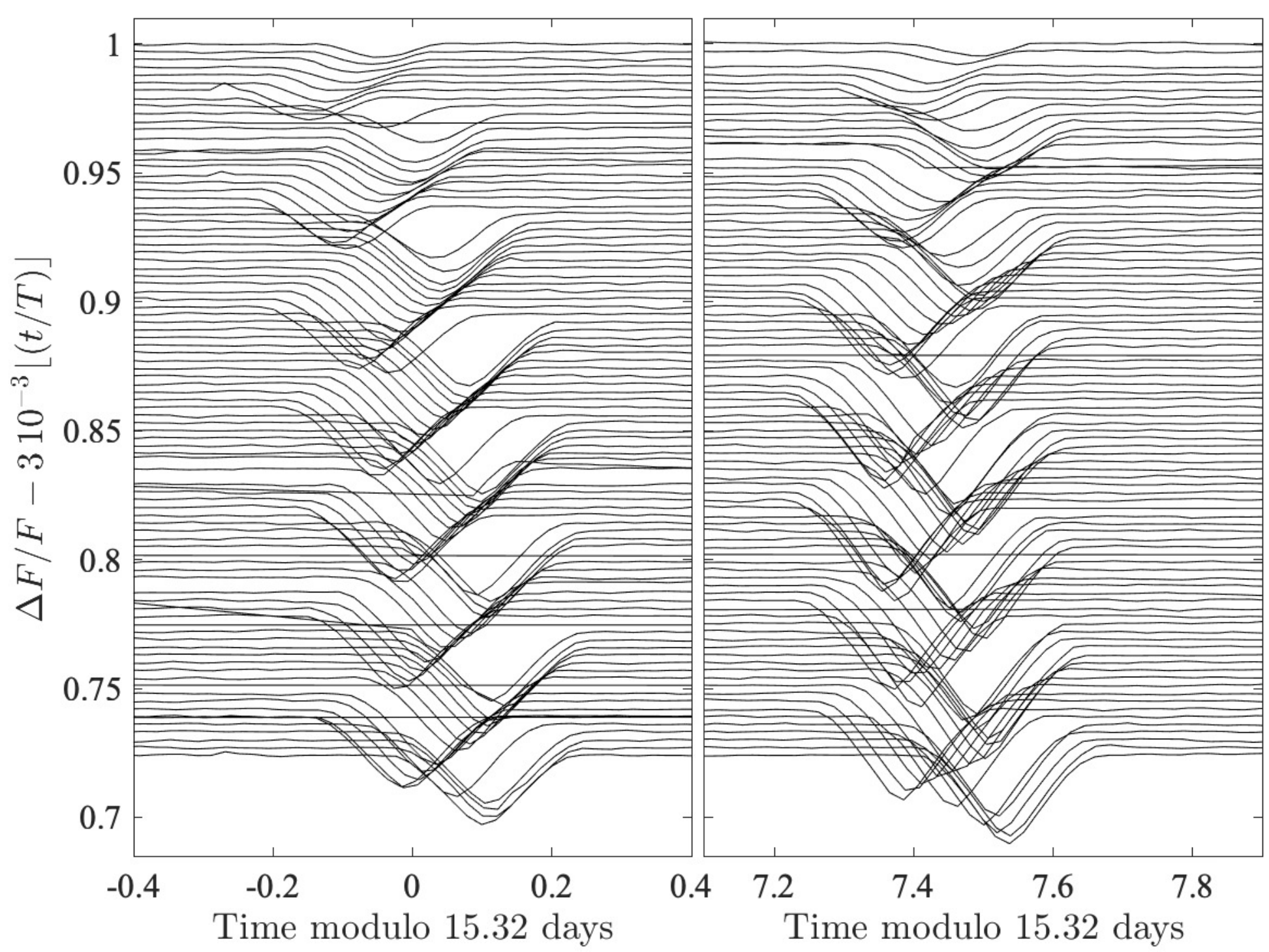} 
\caption{\kep light curve of KIC 7955301. Top panel: relative flux as a function of time, where $t=0$ is the first day of data. The blue curve is the stitched light curve, whereas the orange curve has the eclipses removed and is used for asteroseismic analysis. Bottom panel: stitched lightcurve folded over the inner binary orbital period (15.32 days), where consecutive eclipses are vertically offset by 0.003 to ease the visibility of the transit depth and timing variations.}
\label{fig_LC_clean}
\end{figure}

\subsection{The \kep light curve}
We worked with the \kep public light curves that are available on the Mikulski Archive for Space Telescopes (MAST)\footnote{\url{http://archive.stsci.edu/kepler/}}. Two types of time series are available: the Simple Aperture Photometry (SAP) and the Pre-search Data Conditioning Simple Aperture Photometry (PDC-SAP) light curves. The latter consists of time series that were corrected for discontinuities, systematic errors and excess flux due to aperture crowding \citep{Twicken_2010}. They do not meet our requirements for monitoring possible rotational modulation or eclipse depth, which are often altered during the process \citep[e.g.,][]{Garcia_2014,Gaulme_2014}. We thus made use of the SAP data to preserve any possible long-term signal. This choice entails our own detrending and stitching operation on the light curves while ensuring that the rotational modulation is preserved after each interruption of the time series. The methods employed to clean the time series are detailed in \citet{Gaulme_2016a}.

For asteroseismology, light curves with eclipses are an issue because their signal contaminates the Fourier domain, in which we perform the asteroseismic analysis. We tried two different approaches to remove the eclipses from the time series before computing its power density spectrum. The first approach is better in principle: it consists of modeling each eclipse with a typical eclipse function, for example that from \citet{Mandel_Agol_2002} for fitting exoplanetary transits, and then subtracting the model from the light curve. However, this approach, as noted in \citet{Gaulme_2016a}, is less efficient than simply  clipping out the eclipses, followed by gap filling with a second-order polynomial. Indeed, an eclipse model is always a little imperfect and when subtracted about 100 times (the number of eclipses during the \kep run) they still damage the Fourier transform, unlike eclipse removal and filling. Figure\,\ref{fig_LC_clean} (top panel) shows the original time series along with the time series with no eclipses that is used for analyzing the RG oscillations. The bottom panel highlights the eclipses variations in the form of a folded light curve where a shift was introduced between consecutive orbits. Eclipse depth varies from 0.5 to almost 4\,\%, timing by about 4 hours, and duration from approximately 4.3 to 6.7 hours.

\subsection{High-resolution optical spectra}
\label{sect_opt_spec}

From 2012 to 2019, we were granted time on the ARCES \'echelle spectrograph of the 3.5-m telescope of the Astrophysical Research Consortium (ARC) at Apache Point observatory (APO), which covers the whole visible domain at an average resolution of $31,000$ \citep{Wang_2003}. This allowed us to monitor KIC 7955301 23 times, among observations of other RGs in multiple systems. Even though the ARC \'echelle spectrograph was not designed for precise RV measurements, it has successfully been used for this purpose in earlier work \citep[e.g.,][]{Rawls_2016,Gaulme_2016a, Benbakoura_2021}. The measurement error reported in these papers is about 0.5 km s$^{-1}$ for an RG spectrum with a signal-to-noise ratio (S/N) between 10 and 20. In practise our spectra have a S/N ranging from 10 to 25. 
The ARCES optical spectra were processed and analyzed in the same way as in \citet[][]{Gaulme_2016a} and \citet{Benbakoura_2021} and we refer to these papers for details. 

In addition to the APO spectrograph, we were granted observing time on the SOPHIE \'echelle spectrograph at the 1.93-m telescope at the Haute-Provence Observatory (OHP). We obtained spectra of KIC 7955301 on June 8, 9 and October 9, 2018. The SOPHIE data processing pipeline directly provides the radial velocities from the spectra. We refer the reader to \citet{Santerne_2011a,Santerne_2011b} for details on the data reduction and radial-velocity measurement based on SOPHIE spectra. 

\section{Spectroscopic analysis}
\label{sect_spec_ana}
\begin{figure}
\includegraphics[width=8.5cm]{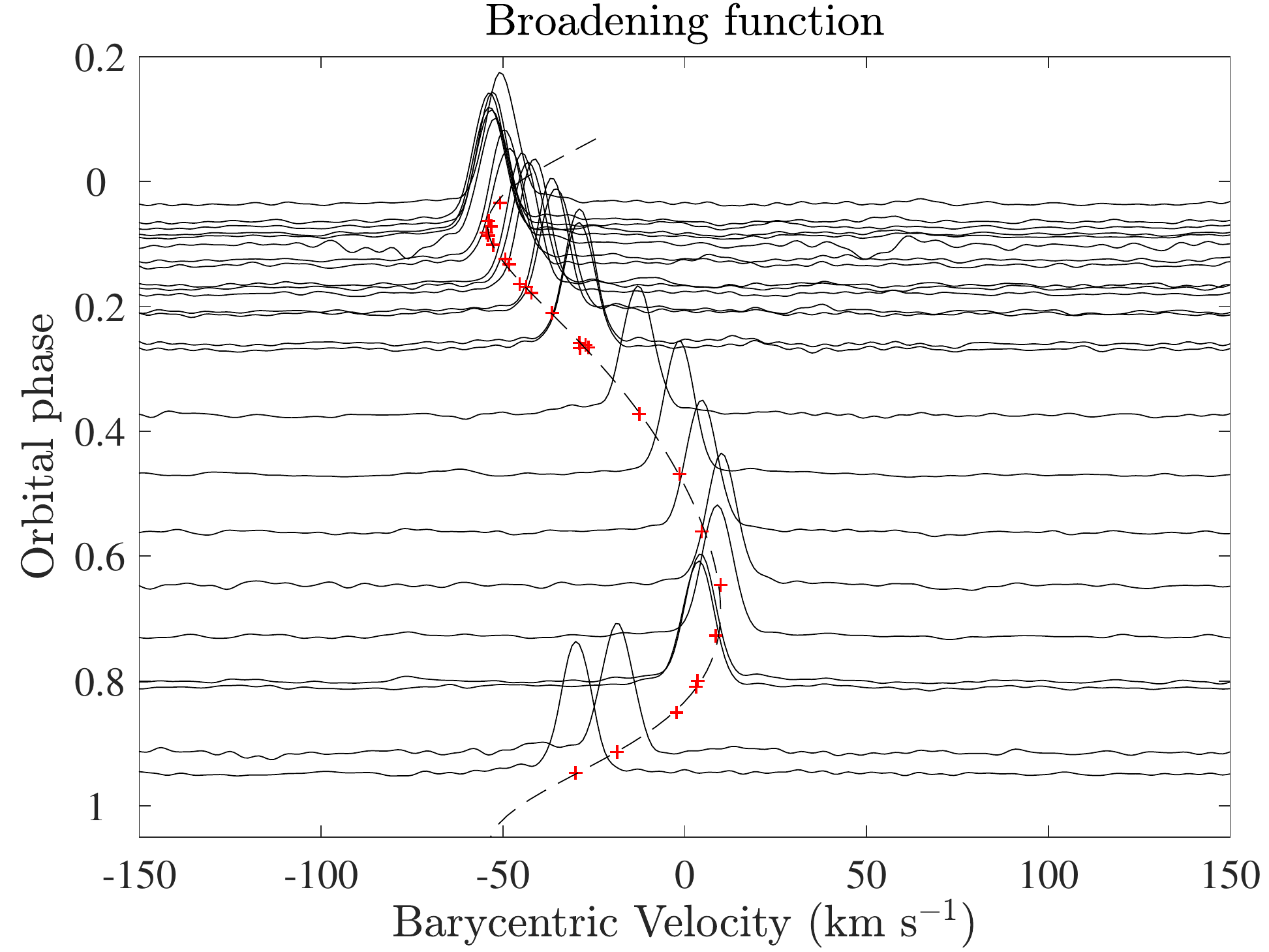} 
\caption{Broadening function of the 23 spectra taken with the \'echelle spectrometer of the 3.5-m ARC telescope at Apache Point observatory. Spectra were sorted by increasing orbital phase. The position along the $y$ axis corresponds with the orbital phase. Barycentric velocity corrections are included in the BF computation. The radial velocity data are indicated with the red plus symbols and their best-fit Keplerian-orbit model by a dashed line.}
\label{fig_BF}
\end{figure}
\subsection{Radial-velocities}
\label{sect_rad_vel}
From the reduced one-dimensional spectra obtained at APO, we computed the radial velocities with the Broadening-Function technique \citep[BF,][]{Rucinski_2002}. The fundamental hypothesis of this method is that the observed spectrum is the theoretical spectrum convolved with a broadening function that accounts for stellar rotation and instrumental effects. The BF technique deconvolves the observed spectrum by the theoretical one to extract the BF. For SB2 systems, the BF shows two peaks, one per component.

We employed the theoretical spectra generated by the PHOENIX BT-Settl code \citep{Allard_2003}, which were computed with the solar abundances derived by \citet{Asplund_2009}. As in \citet{Gaulme_2016a} and \citet{Benbakoura_2021}, we used templates of MS stars ($T_{\rm eff} = 6200$\,K) to maximize the chance of detecting the signal from the companion star. Such a choice actually also helps the deconvolution of the RG spectrum because the large number of absorption lines in RG spectra tends to increase the noise in the BF profiles when the original spectra do not have an excellent S/N, which is our case. We computed the BF by making use of the wavelength range 4500 and 5800\,\AA. Our final RV data were obtained after correcting the BF profiles from the barycentric corrections, which account for the rotation and revolution of the Earth with respect to the target. We computed the barycentric corrections with the \texttt{PyAstronomy}\footnote{\url{https://github.com/sczesla/PyAstronomy}} routine \texttt{helcorr}, which is based on the \citet{Piskunov_2002} algorithm. 

All the radial velocities we produced in this work are compiled in Table \ref{tab:radial_velocities} and displayed in Fig. \ref{fig_BF} together with the series of BF profiles. For illustration purposes we overplot data points with a fit performed with a simple Keplerian orbit. This allows us to estimate the standard deviation of the measurements to be about 0.70\,km\,s$^{-1}$. We note that employing a Keplerian orbit is a very simplistic approach as KIC 7955301 is a triple system. In Sect. \ref{sect_dyn_mod} we show that the argument of periastron of the outer orbit changed by about 20$^\circ$ during the four years of \kep observations, meaning that a simple Keplerian orbit is not sufficient to accurately model the RG orbit.

\subsection{Disentangling the spectra}

A proper estimate of the atmospheric parameters of the RG and possibly of the inner binary components requires disentangling the spectra. 
For a triple system with such a large flux contrast between the evolved primary component and the inner dwarf binary, it is not an easy task. The spectral lines of the three components are diluted and it is challenging to disentangle the individual contributions to the continuum without knowing the light ratio. It is straightforward to get the ratio from a light curve when all components eclipse each other. Unfortunately, in our case, the RG is not eclipsing with the inner pair. The remaining method for measuring the light ratio and distentangling the spectra consists of scaling the absorption lines from synthetic spectra to the series of observed spectra \citep[][and references therein]{Pavlovski_2009, Pavlovski_2018}.  

According to the dynamical model presented in the following sections, the contribution of the main-sequence components to the total luminosity is close to 10\,\%, which implies that we cannot visualize their signatures neither in the BF profile, nor in the spectra. Nevertheless, very faint components have been revealed by spectral disentangling even at a level of only 1-2\,\% of the total light, such as the M dwarf in the EB V530 Ori \citep{Torres_2014}, 
the Roche-lobe filling giant in the inner EB of the Algol triple 
system \citep{Kolbas_2015}, the main sequence companions of the RGs 
in EBs \citep[e.g.,][]{Gaulme_2016a,Helminiak_2017,Brogaard_2018,Themessl_2018, Benbakoura_2021}.

In spectral disentangling as originally formulated by 
\citet{Simon_Sturm_1994}, the spectra of the individual components are simultaneously reconstructed with an optimization of the orbital elements of the multiple-star system.  The problem consists of solving the matrix equation \mbox{\boldmath{$A \cdot x = b$}}, where the vector {\boldmath {$x$}} represents the unknown individual spectra of the components -- which we aim at extracting -- and {\boldmath {$b$}} all of the observed spectra. The design matrix {\boldmath {$A$}} is constructed from Doppler shifts for a given set of the orbital elements and exposure dates. {\boldmath {$A$}}  could also contain light dilution factors, if known. Since the set of linear equations are overdetermined, the solution may be calculated with the linear-algebra technique known as singular value decomposition. In our particular case, the RVs for the RG component are measured by the BF (Sect.~\ref{sect_rad_vel}, and Table \ref{tab:radial_velocities}). From the dynamical model of the system described in Sect.~\ref{sect_dyn_mod},  we computed the expected values of the RVs of the components of the inner binary. 

The spectral disentangling is performed with the code {\sc cres} \citep{Ilijic_2004}, which is based on the \citet{Simon_Sturm_1994} method for spectral disentangling in the visible domain. In {\sc cres}, the RVs, light dilution factors of each component, and the observed spectra are inputs. The observed spectra should be given in units of wavelength. The observed spectra do not need to be in an uniform scale, and the spectra from different spectrographs do not need to be resampled first. Any portion of the observed spectra could be masked out to exclude undesired wavelength regions. 

\begin{figure}
\includegraphics[width=8.5cm]{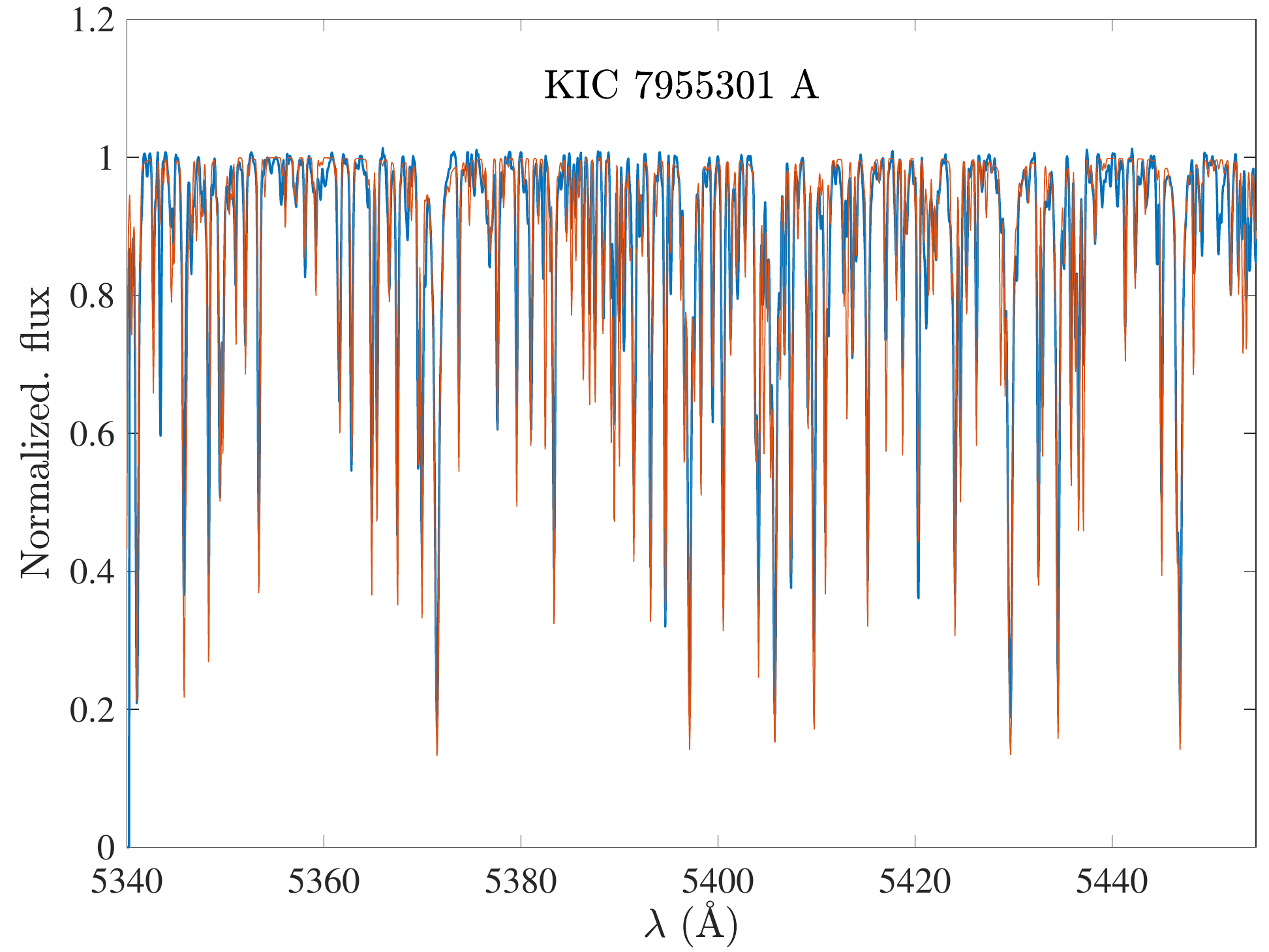} 
\includegraphics[width=8.5cm]{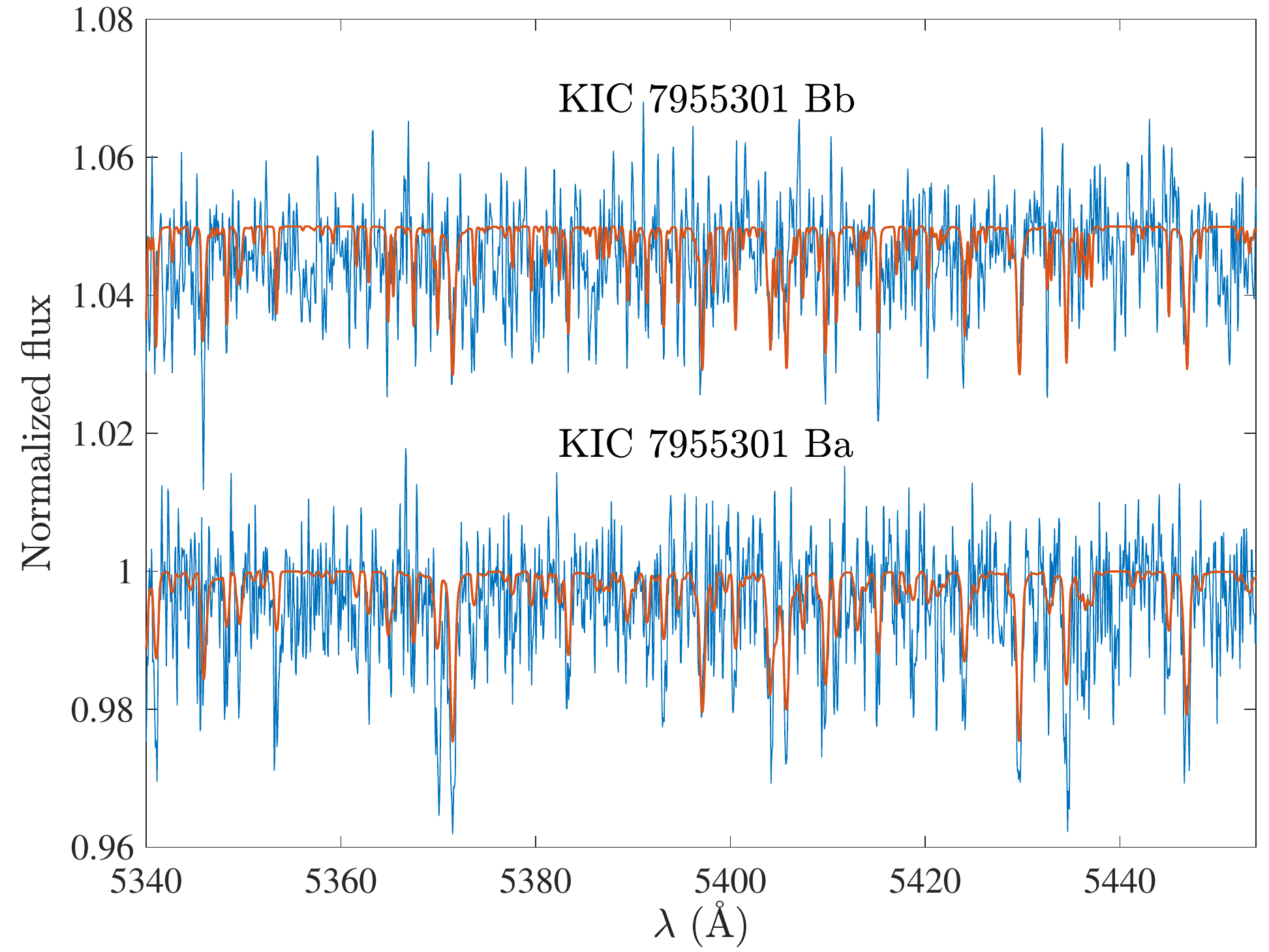} 
\caption{A sample region of the mean spectra of the individual components of the system obtained by disentangling the APO spectra. Top panel: spectrum of the RG component (blue line) and its best fit obtained with the {\sc gssp} code (red) from 5340 to 5450\AA. 
Bottom panel: spectra of the inner binary system Ba and Bb (blue lines) together with their best-fit synthetic spectra (red lines). The optimal values representing the best fits are listed in Table~\ref{tab:atm_param}.
}
\label{fig_disen_spec}
\end{figure}

Altogether, we use the 23 high-resolution \'echelle spectra obtained at APO (Sect.~\ref{sect_opt_spec}). Spectra have a S/N from $\sim5$ to 60, with an average S/N $\simeq 27$, as measured in several short line-free windows near 5500 - 5600\,{\AA}. The S/N was then used to assign weights to the observed spectra. We note that the code actually allows an assignment of the weights to individual pixels, another feature advantageous in the
wavelength-domain disentangling, but we did not use this option,
assuming it is sufficient to take the weight per spectrum, and not per pixel. The selected spectral segments had various lengths, from about  40 to 90\,{\AA},  always taking into account overlapping regions which serve to double-check 
the quality of our disentangling.

It is known that without any substantial change in the fractional light
of the components in the course of an orbital cycle, there is no unique
solution in reconstruction of the individual spectra of the components
\citep{Pavlovski_Hensberge_2005}. One possibility is to optimize the light
dilution factors in the calculations of the components' spectra, but with
dominant fractional light contribution of the RG component, which 
is not eclipsing with the inner pair of the MS stars, this would be
meaningless. The best option is to perform 
spectral disentangling in a pure separation mode, and then determine
fractional light contribution for each component from its spectral
characteristics. 

The reconstruction of the individual spectra of all three components was performed in the wavelength range $[5100,5650]$\,{\AA}. In addition, the spectral range of the disentangled spectrum for the RG component was extended to the range $[4700,6825]$ \AA. In the latter case, our intention was to cover the Balmer lines H$\alpha$, and H$\beta$, as well as the sodium doublet \ion{Na}{i} at $\lambda\lambda$ 5890, and 5896 {\AA}. Some portions of the reconstructed spectra are shown in Fig.~\ref{fig_disen_spec}. It is evident from a bottom panel of Fig.~\ref{fig_disen_spec} showing reconstructed spectra for the stars in the inner (eclipsing) pair that the signatures of the stellar spectra are successfully revealed, albeit the noise level is large. We can estimate the gain in the S/N for
the spectra of the disentangled components by  simple calculations, since spectral disentangling, in principle, is working as co-addition of the observed spectra. By assuming random noise with an average S/N $\simeq 27$, the 23 spectra should lead to total S/N of about 130. Of this, the RG spectrum benefits most since it contributes about 91\,\% of the total light, as is determined later in this section. The other two components share the remaining $\sim 9$\,\% -- given their contribution is similar (Sect. \ref{sect_dyn_mod}, Table \ref{tab: syntheticfit_KIC7955301}) --, meaning that the S/N of their reconstructed spectra is about 5-6. Thus, their spectral signatures are revealed only thanks to
the gain of S/N caused by the spectral disentangling. The successful isolation of the spectra for the stars in the inner system, which is based on the RVs calculated from the predicted orbit, is an encouraging confirmation of the correctness of the dynamical model.

The deepest absorption lines in the disentangled spectrum of the RG component are due to \ion{Mg}{i}$b$ triplet lines at  5167, 5172, and 5183\,{\AA}, and \ion{Na}{i} D doublet at 5890, and 5896\,{\AA}. The depths of these sets of spectral lines constrain the fractional light contribution of the RG to the total light of the system. A limit defined by the physical solution is at 89\%, which is corroborated with the optimal fitting (Table~\ref{tab:atm_param}).

It is well established by observational evidence that lithium is depleted in the majority of giants \citep{Brown_1989}. There are rare exceptions known as Li-rich giants with abundances $A({\rm Li}) > 1.5$ (a scale where the number of hydrogen atoms $\log N({\rm H}) = 12.0$). Examination of the disentangled spectrum of the RG component of KIC\,7955301 shows almost no trace of the \ion{Li}{i} resonance doublet at 6708 {\AA}. An estimate of an upper limit of the lithium abundance was made with a comparison to the synthetic line profiles. The calculations were performed assuming local thermodynamic equilibrium with model atmospheres calculated with Atlas9 \citep{Castelli_Kurucz_2003}, while line profiles for \ion{Li}{i} 6708 {\AA} are calculated with spectrum synthesis code {\sc uclsyn} \citep{Smith_1992}. The atomic data are from \citet{Yan_1998}. We were able to establish only an upper limit in lithium abundance $A({\rm Li}) < -1.0$ dex. This is a common lithium abundance found in modern massive spectroscopic surveys \citep{Luck_Heiter_2007, Buder_2018,  Charbonnel_2020}.

\subsection{Stellar atmospheric parameters}
\begin{table}
\caption{Best-fit parameters deduced from the analysis of the atmospheric parameters.}
\label{tab:atm_param}
\centering
\begin{tabular}{l l l l}
\hline
\multicolumn{4}{c}{Red giant (star A)}\\
\hline
\multirow{6}{1cm}{Run A} & $T\ind{eff}$ & [K]  & $4720 \pm 105$   \\
& $\log g$ &[dex] & 3.0 (fixed)\\
& $v\ind{micro}$ & [km s$^{-1}$] & $1.44 \pm 0.35$ \\
& $v\sin i$ & [km s$^{-1}$] & $6.4 \pm 1.2$\\
& dilution factor& & $0.91 \pm 0.08$ \\
& $[\mathrm{M}/\mathrm{H}]$& [dex] & $-0.01 \pm 0.12$\\
\hline
\multirow{6}{1cm}{Run B} & $T\ind{eff}$ & [K] & $4760 \pm 110$ \\
& $\log g$ & [dex]  & 3.1  (fixed)\\
& $v\ind{micro}$ & [km s$^{-1}$]  & $1.36 \pm 0.35$ \\
& $v\sin i$ & [km s$^{-1}$] & $6.3 \pm 1.2$ \\
& dilution factor & & $0.92 \pm 0.07$\\
& $[\mathrm{M}/\mathrm{H}]$ & [dex]  & $0.03 \pm 0.12$ \\
\hline
\multirow{6}{1cm}{Run C} & $T\ind{eff}$ & [K] & $4700 \pm 155$ \\
& $\log g$ & [dex] & $2.95 \pm 0.40$\\
& $v\ind{micro}$ & [km s$^{-1}$] & $1.31 \pm 0.35$\\
& $v\sin i$ & [km s$^{-1}$]  & $6.6 \pm 1.3$\\
& dilution factor  & & $0.91 \pm 0.08$\\
& $[\mathrm{M}/\mathrm{H}]$ & [dex]  & $-0.01 \pm 0.17$\\
\hline
\multicolumn{4}{c}{Inner binary (stars Ba and Bb)}\\
\hline
\multirow{6}{1cm}{Ba} & $T\ind{eff}$ & [K] & $5620 \pm 580$\\
& $\log g$ & [dex] & 4.5 (fixed)\\
& $v\ind{micro}$ & [km s$^{-1}$] & 2.0 (fixed) \\
& $v\sin i$ & [km s$^{-1}$] & $19.0 \pm 5.5$ \\
& dilution factor & &  $0.043 \pm 0.013$ \\
& $[\mathrm{M}/\mathrm{H}]$ & [dex]  & 0.0 (fixed) \\
\hline
\multirow{6}{1cm}{Bb} & $T\ind{eff}$ & [K] & $5330 \pm 550$\\
& $\log g$ & [dex] & 4.5 (fixed)\\
& $v\ind{micro}$ & [km s$^{-1}$] & 2.0 (fixed)\\
& $v\sin i$ & [km s$^{-1}$] & $8.9 \pm 7.5$\\
& dilution factor & &  $0.025 \pm 0.009$\\
& $[\mathrm{M}/\mathrm{H}]$ & [dex]  &  0.0 (fixed)\\
\hline
\end{tabular}
\tablefoot{Run A: $\log g$ is fixed to 3.0 dex (atmosphere model grid point closest to $\log g =  3.034$ dex reported in Table \ref{tab: syntheticfit_KIC7955301}, column "with SED+PARSEC”); wavelength range 4700 - 5700 {\AA}. Run B: $\log g$ is fixed to 3.1 dex (atmosphere model grid point closest to $\log g =  3.115$ dex reported in Table \ref{tab: syntheticfit_KIC7955301}, column "without SED+PARSEC”); wavelength range 4700 - 5700 {\AA}. Run C: $\log g$ is treated as a free parameter; wavelength range 4700 - 5700 {\AA}. Inner binary, both components; values of $\log g$ are fixed to 4.5 dex (Table \ref{tab: syntheticfit_KIC7955301}); fixing vmicro and [M/H] as well because spectra are really noisy; wavelength range 5100 - 5630 {\AA}.}
\end{table}
For the analysis of the disentangled spectra of all three components of the KIC 7955301 system, we employed the Grid Search in Stellar Parameters\footnote{\url{https://fys.kuleuven.be/ster/meetings/binary-2015/gssp-software-package}} \citep[{\sc gssp}][]{Tkachenko2015} software package, specifically its {\sc gssp\_single} module. It is a grid search-based spectrum analysis algorithm that (on the fly) generates synthetic spectra in an arbitrary wavelength range based on a pre-computed grid of model atmospheres, and performs a comparison between the observed spectrum and each synthetic spectrum from the grid in the $\chi^2$ statistical framework. The reported 1-$\sigma$ uncertainties are computed from the $\chi^2$-statistics and taking into account possible correlations between the free parameters. The {\sc gssp} algorithm allows for the simultaneous optimization of the effective temperature, surface gravity, micro- and macro-turbulent velocities, projected rotational velocity, and metallicity of the star. Optionally, one can also optimize for the degree of the light dilution in the spectrum due to the star being a member of a binary and/or higher order multiple system. 

In this study, we employ the grid of {\sc LLmodels} model atmospheres \citep{Shulyak2004} and opt for inclusion of the light dilution factor into the optimization to account for a priori unknown dilution of the disentangled spectrum of each of the three stellar components under the analysis. The light dilution factor is assumed to be wavelength independent, which is a fair assumption for the RG component given its by far most dominant ($>$ 90\%) contribution to the composite spectrum of the system, as well as for the two main-sequence components given the limited wavelength interval of some 500 \AA\ that could be used for the disentangling. We additionally note that the macro-turbulent velocity parameter was ignored in the analysis of all three stellar components of the system, and we additionally fixed $\log\,g$ for both main-sequence components to the values inferred from the light curve solution, micro-turbulent velocity to 2 km\,s$^{-1}$, and assumed solar chemical composition ([M/H] = 0.0 dex) for both of them. The choice to fix so many parameters in the spectrum analysis of the main-sequence components is dictated by their small cumulative contribution ($<$ 8-10\%) to the total light of the system, and hence their very noisy disentangled spectra. The consequence of fixing the macro-turbulent velocity parameter for the RG component is that the $v\,\sin i$ value reported  is in fact representative of the combined spectral line broadening due to the effects of rotation and macroturbulence. Finally we note that we explored three different options in the analysis of the RG spectrum, namely fixing the surface gravity to the two values inferred from the light curve solution and reported in Table~\ref{tab:atm_param}, and treating $\log\,g$ as a free parameter. 


\section{Asteroseismic analysis}
\label{sect_astero}
\begin{figure}[t!]
\includegraphics[width=8.5cm]{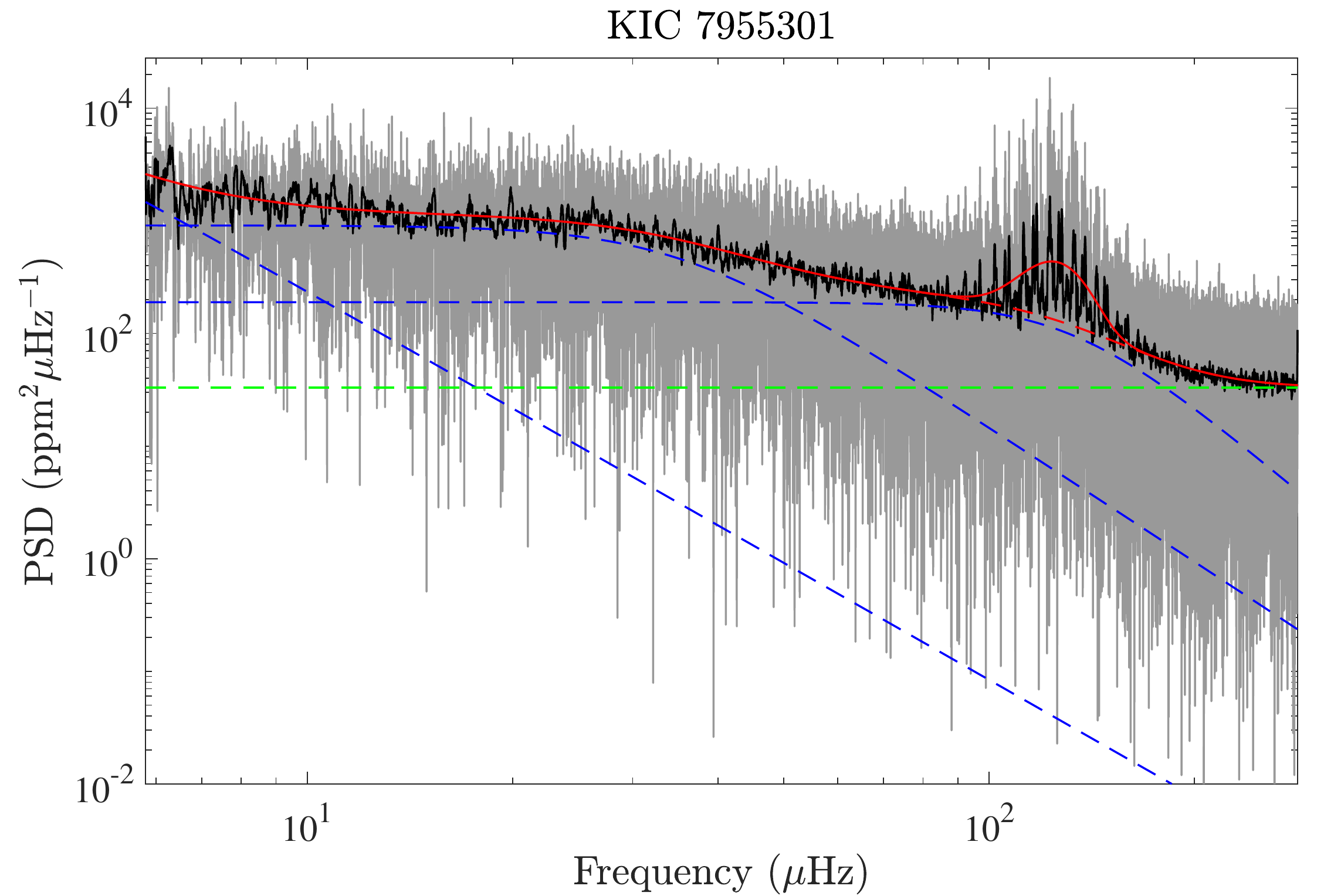} 
\includegraphics[width=8.5cm]{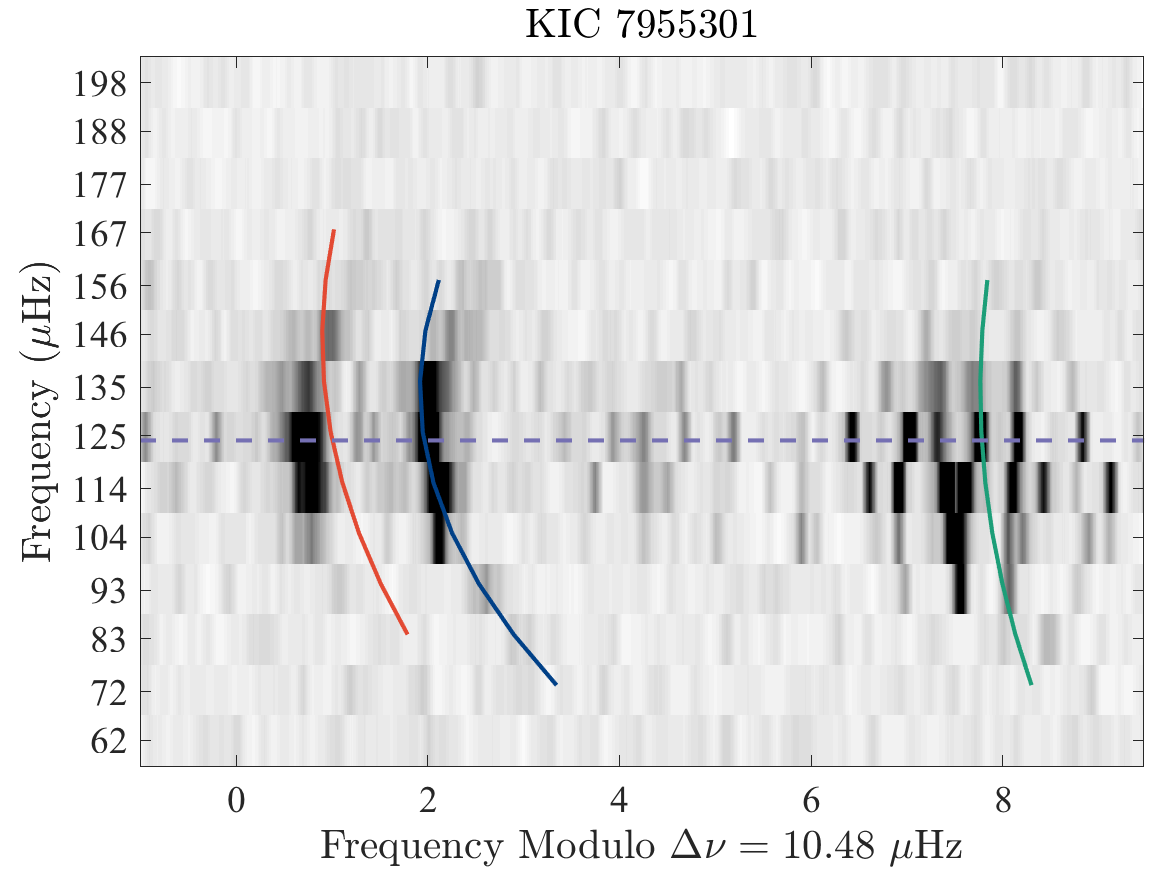} 
\caption{Oscillation spectrum of KIC 7955301. Top panel: power spectral density of the time series after eclipse removal and filling. The red dashed line represents the stellar background noise: it is the sum of the blue dashed lines (correlated noise Harvey profiles) and the green dashed line (white noise). The plain red line represents the fit of the power spectrum including the stellar background and the Gaussian envelope of the oscillations. Bottom panel: \'echelle diagram associated with the frequency spacing $\Delta\nu = 10.48\,\mu$Hz. Colored lines indicate the oscillation universal pattern, where blue is $l=0$, green $l=1$, and red $l=2$. Modes of degree $l=3$ are visible half way in between the $l=0$ and $l=1$ ridges. The horizontal dashed line indicates the location of $\nu_\mathrm{max}$.}
\label{fig_PSD_back}
\end{figure}
\begin{figure}[t!]
\includegraphics[width=8.5cm]{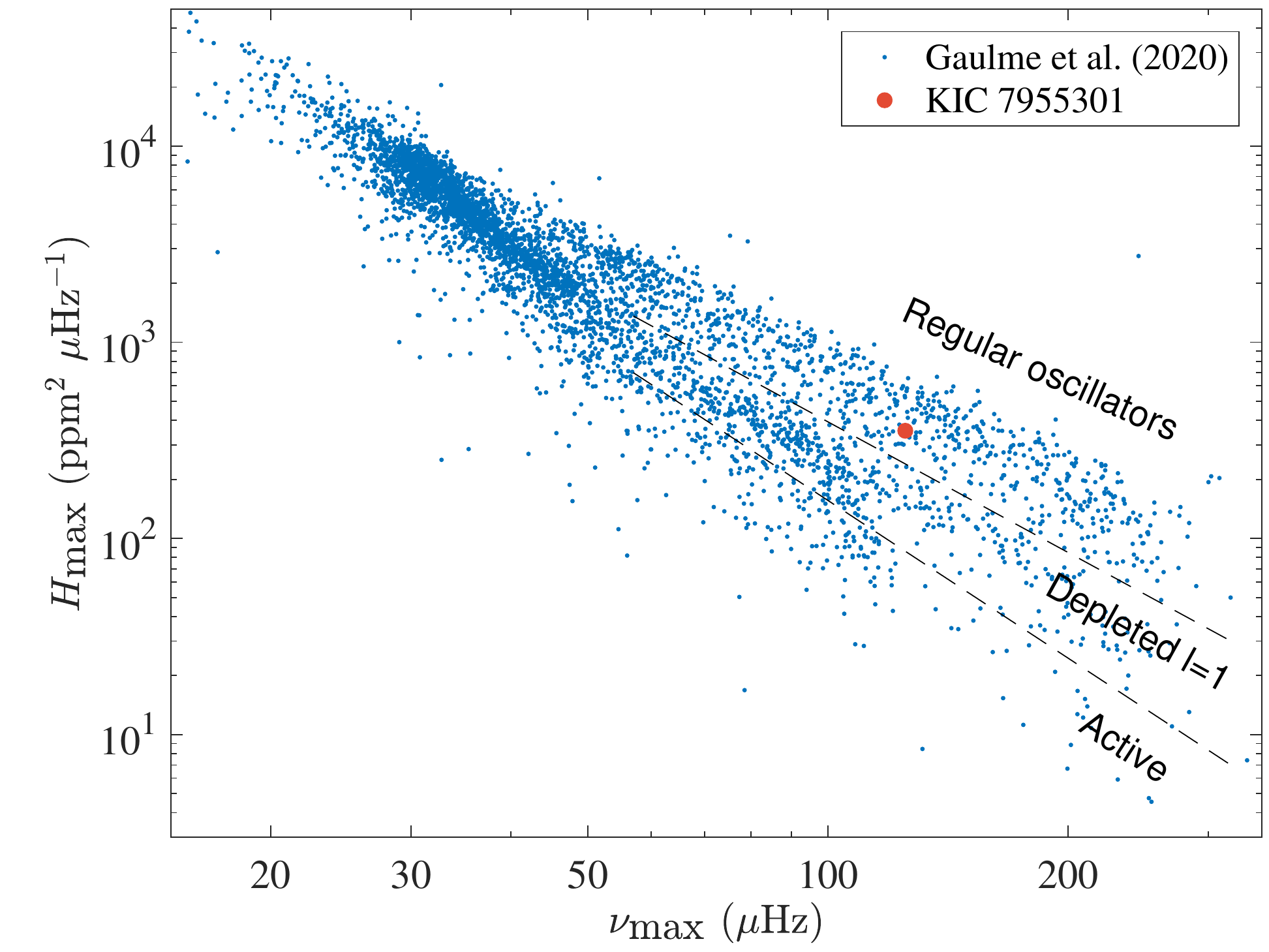} 
\includegraphics[width=8.5cm]{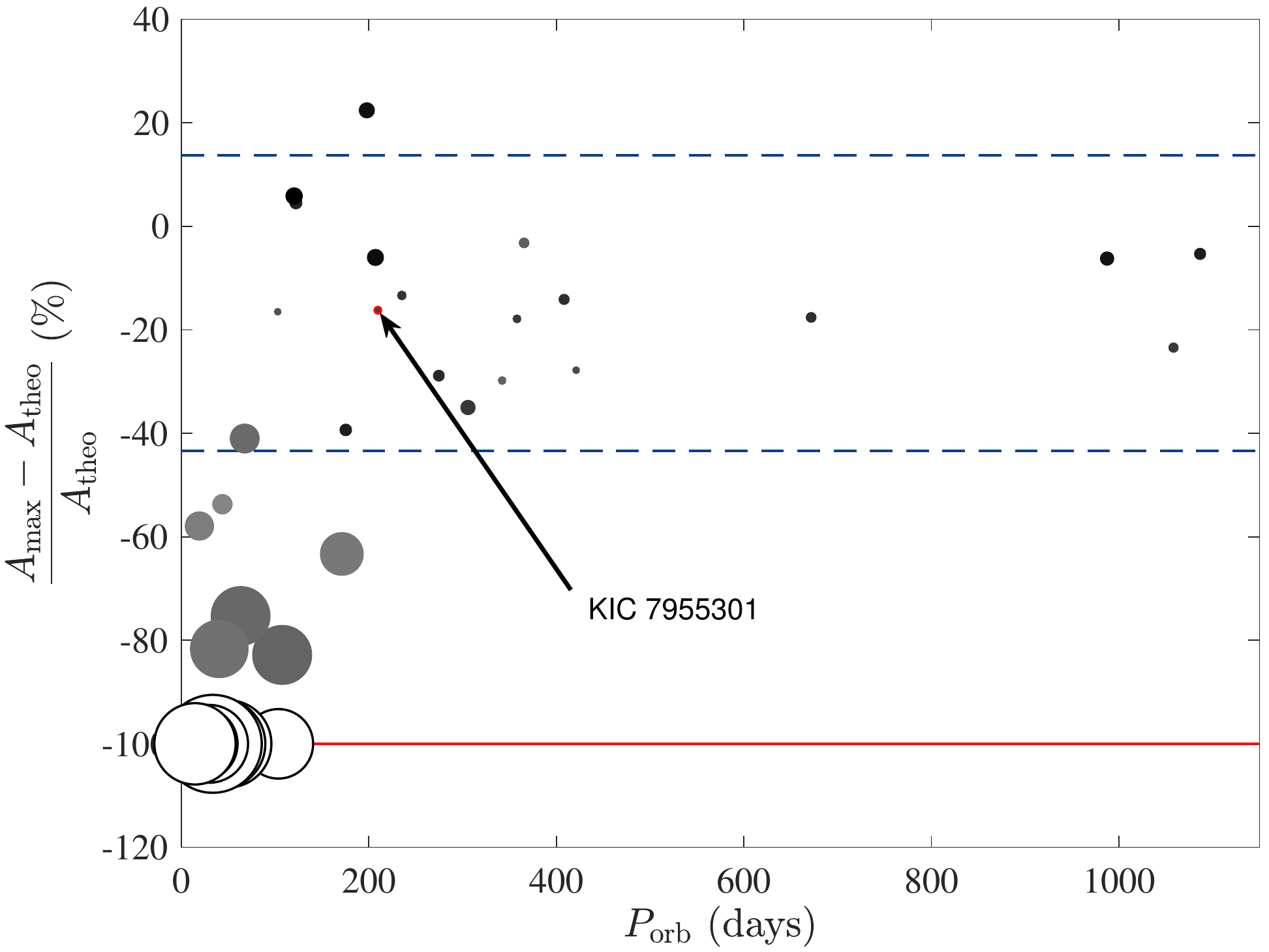} 
\caption{Oscillation amplitude. Top panel: height of the Gaussian envelope used to model the contribution of the oscillations to the stellar background. The red dot highlights the position of KIC 7955301 with respect to a sample of 4500 RGs analyzed by \citet{Gaulme_2020} indicated in blue. The sample splits into three categories: the oscillators with regular amplitude, those with depleted $l=1$ modes, and the active RGs that show a global mode suppression ($l=0,1,2,3$). KIC 7955301 falls in the regular oscillators. Bottom panel: amplitude of the largest $l=0$ mode of KIC 7955301 compared with those of the 35 RGs in EBs that were studied by \citet{Benbakoura_2021}. The figure displays the relative difference between expected and measured oscillation amplitudes (\%) as a function of orbital period $P\ind{orb}$ (days). The red line indicates 0, i.e., stars whose oscillations are not detected. The two dashed blue lines represent the region in which relative mode amplitude lies for systems with orbital periods longer than 180 days within two sigma. The size of each symbol represents the amplitude of stellar variability (large means variable; small not variable), and the gray scale indicates the pulsation mode amplitude (white-no modes; black-large amplitude).}
\label{fig_mode_amp}
\end{figure}

\subsection{Global oscillation properties}

The stellar granulation and accurate values of $\nu\ind{max}$ and mode amplitude $H\ind{max}$ are estimated by fitting the power spectrum (Fig. \ref{fig_PSD_back}) as commonly performed in asteroseismology \citep{Kallinger_2014}, and already used in \citet{Gaulme_2016a}. Following \citet{Kallinger_2014}, the power density spectrum is fitted by 
\begin{equation}
    S(\nu) = N(\nu) + \eta(\nu) \left[B(\nu) + G(\nu)\right],
\end{equation}
where the $N$ is the function describing the noise, $\eta$ is a damping factor originating from the data sampling, $B$ is the sum of three ``Harvey'' functions (super Lorentzian functions centered on 0), and G is the Gaussian function that accounts for the oscillation excess power
\begin{equation}
    G(\nu) =H\ind{max} \exp{\left[-\frac{(\nu-\nu\ind{max})^2}{2\sigma^2}\right]}
    .
\end{equation}
The terms $\nu\ind{max}$ and $H\ind{max}$ are the central frequency and height of the Gaussian function. The best-fit values are: $\nu\ind{max} = 124.89\pm0.33\ \mu$Hz, $H = 354.2\pm10.5$ ppm $\mu$Hz$^{-1}$, and $\sigma=12.5\pm0.4\,\mu$Hz.

A first estimate of $\Delta\nu$ is performed with the envelope of the autocorrelation function (EACF) developed by \citet[][]{Mosser_Appourchaux_2009} from the whitened power spectral density (power spectrum divided by background function). From the EACF, $\Delta\nu = 10.46 \pm 0.05$. 
Then, we used the universal pattern of RGs introduced by \citet{Mosser_2011} to correct it. The principle of this method is to compare the measured oscillation frequencies to a theoretical law, the so-called universal pattern, predicting the variations of these frequencies as a function of $\Delta\nu$ and the radial order. That way, the value of $\Delta\nu$ is revised to  $10.49\pm0.02\ \mu$Hz. 

We then compute a proxy of the stellar masses and radii using the asteroseismic scaling relations that were originally proposed by \citet{Kjeldsen_Bedding_1995} for SL MS oscillators, and then successfully applied to RGs \citep[e.g.,][]{Mosser_2013}. 
We employ the asteroseismic scaling relations as proposed by \citet{Mosser_2013} for RGs, i.e., where $\nu\ind{max,\odot} = 3104\ \mu$Hz, $\Delta\nu_\odot = 138.8\ \mu$Hz, $T\ind{eff,\odot} = 5777$ K, and where the observed $\Delta\nu$ is converted into an asymptotic one, such as $\Delta\nu\ind{as} = 1.038\,\Delta\nu$. By considering the temperature $T\ind{eff,A,sce.A} = 4720\pm105$ K that was found from the disentangled spectrum by fixing the surface gravity at the value obtained from the most complete model $\log g = 3.0$ (scenario A, Table \ref{tab:atm_param}), the stellar parameters are $R\ind{A} = 5.91 \pm 0.07\ R_\odot$, and  $M\ind{A} = 1.27 \pm 0.05\ M_\odot$. 


Finally, we check the amplitude of the oscillations to see whether tidal interactions in the system have altered their properties, as was observed in close systems by \citet{Gaulme_2014,Gaulme_2020}, and \citet{Benbakoura_2021}.
From the background fitting, the height of the Gaussian function is $354\pm11$ ppm $\mu$Hz$^{-1}$, which is a typical value for an RG with $\nu\ind{max} = 125\ \mu$Hz, according to sample of 4500 RGs analyzed by \citet[][Fig. 7]{Gaulme_2020} with the exact same codes. We note that the height of the oscillation envelope is a little underestimated because the RG only contributes approximately 91\,\% of the photometric flux, whereas most of the 4500 stars displayed in \citet[][Fig. 7]{Gaulme_2020} are not in multiple systems and do not suffer from this dilution factor. 

An alternative metric on the amplitude of the oscillation modes consists of measuring the amplitude of the largest $l=0$ mode, as was performed by \citet{Benbakoura_2021}. The fitting of the $l=0$ peaks with Lorentzian functions leads to a maximum $l=0$ amplitude of $7.06\pm0.08$ ppm, which is in agreement with the known RGs in EBs that do not show any alteration of their oscillation properties (Fig. \ref{fig_mode_amp}). We conclude that the oscillations of KIC 7955301 have regular amplitudes and do not show any global suppression, as was observed for RGs in short period binary systems. This result is consistent with the long-period binary systems observed so far \citep{Benbakoura_2021}.

\begin{figure}[t!]
\center
\includegraphics[width=8.5cm]{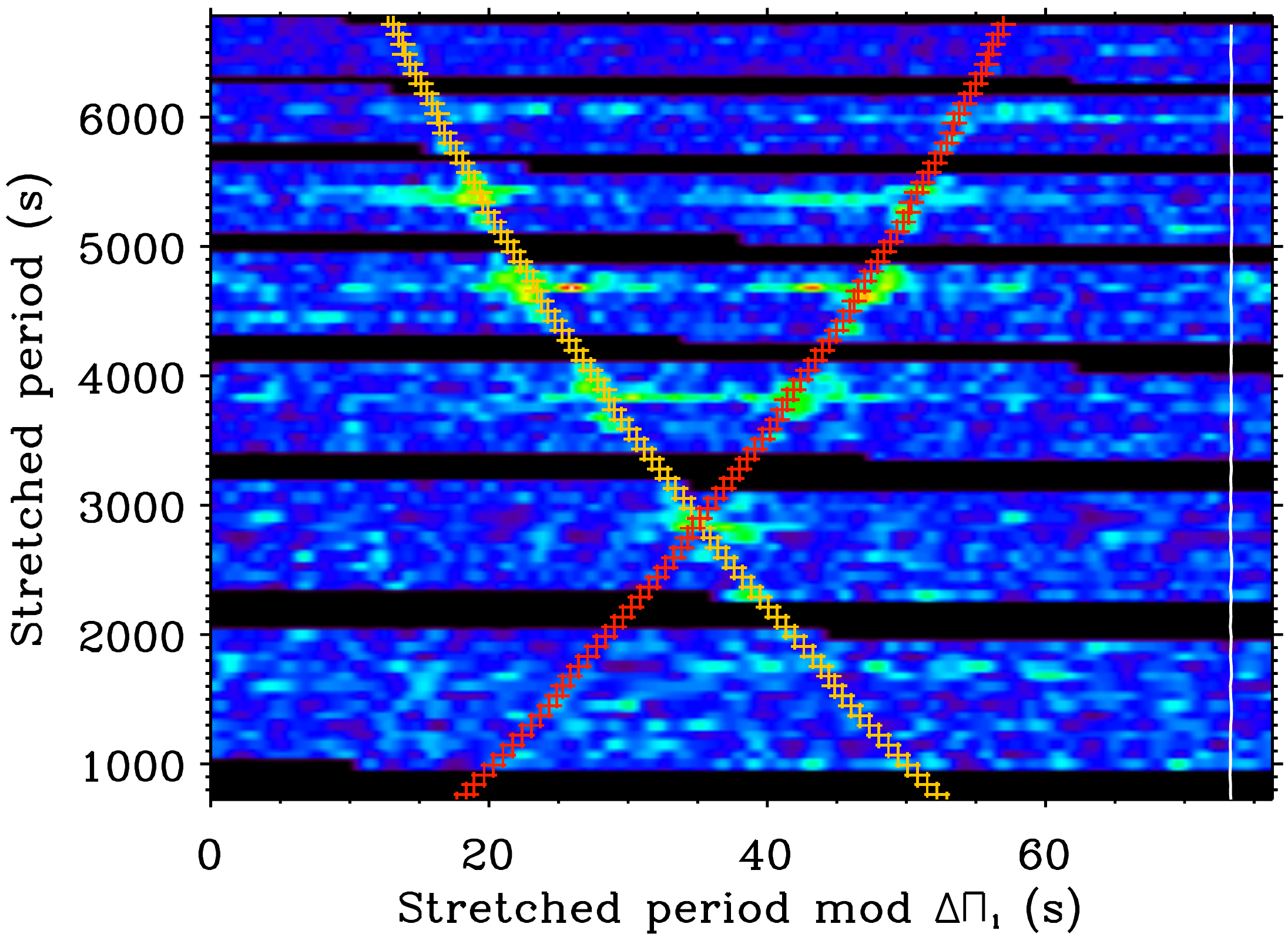} 
\includegraphics[width=8.5cm]{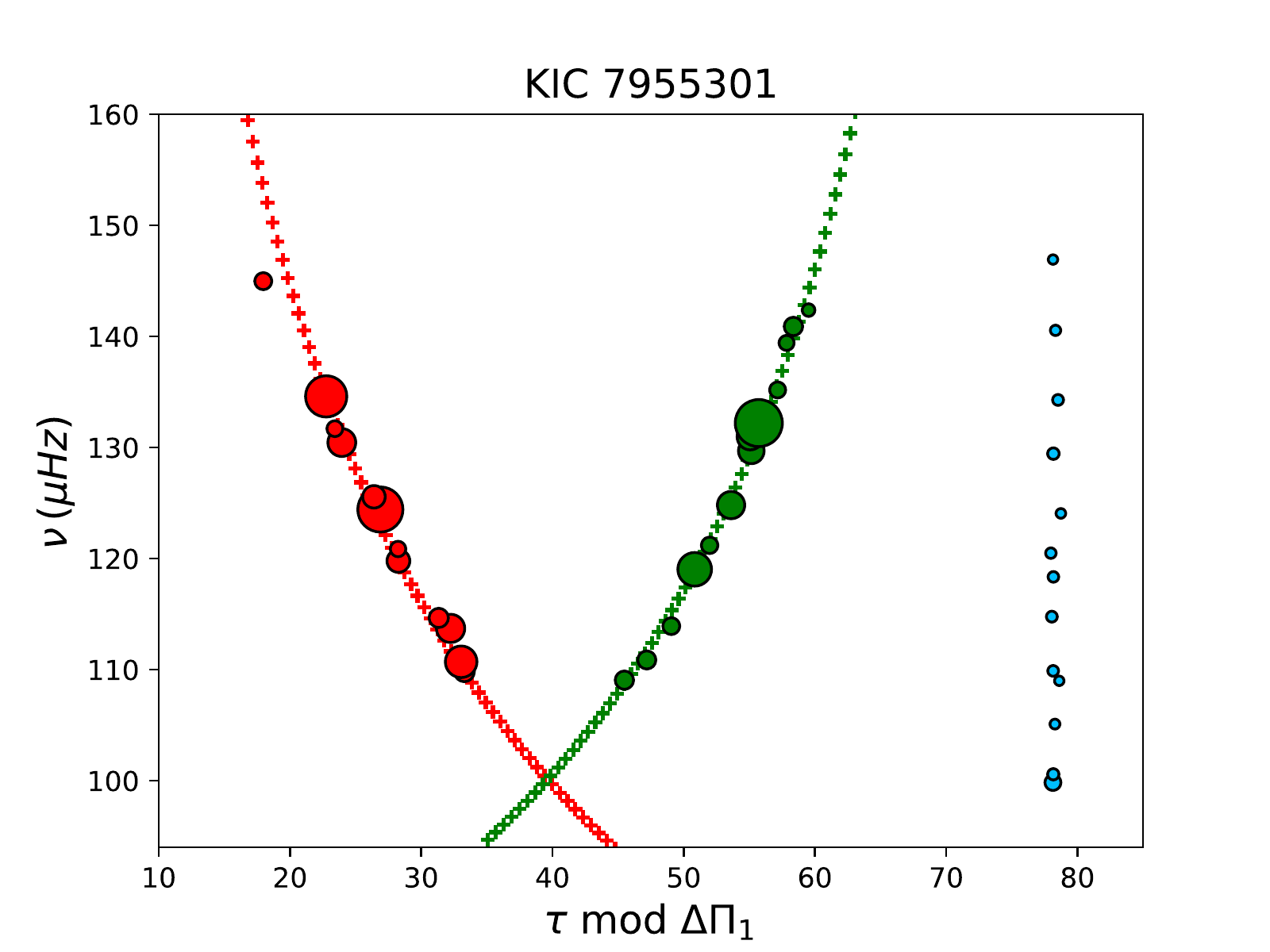} 
\caption{
Stretched period \'echelle diagram for dipole gravity-dominated mixed modes. Top panel is the work led by co-author Appourchaux with $\Delta\Pi_1=76.3$ s. Fitted ridges with $m=\{-1,1\}$ are represented as red and orange crosses, respectively; while $m=0$ is represented as a continuous white line.  The black stripes show the location of the $l=0,2$ mode power which has been removed for clarity. Bottom panel summarizes the finding of co-author Gehan, where peaks with a height-to-background ratio equal to or above 10 are represented. The symbol size varies as the measured power spectral density. 
Ridges with $m=\{-1,0,1\}$ are represented in green, light blue, and red, respectively. Red and green crosses represent the fit of the ridges with $m=\pm 1$. The location of the $m = 0$ ridge is identified considering that the $m = 0$ ridge is median with respect to $m = \pm 1$ ridges.}
\label{fig_echelle_stretched}
\end{figure}
\begin{figure}[t!]
\includegraphics[width=8.5cm]{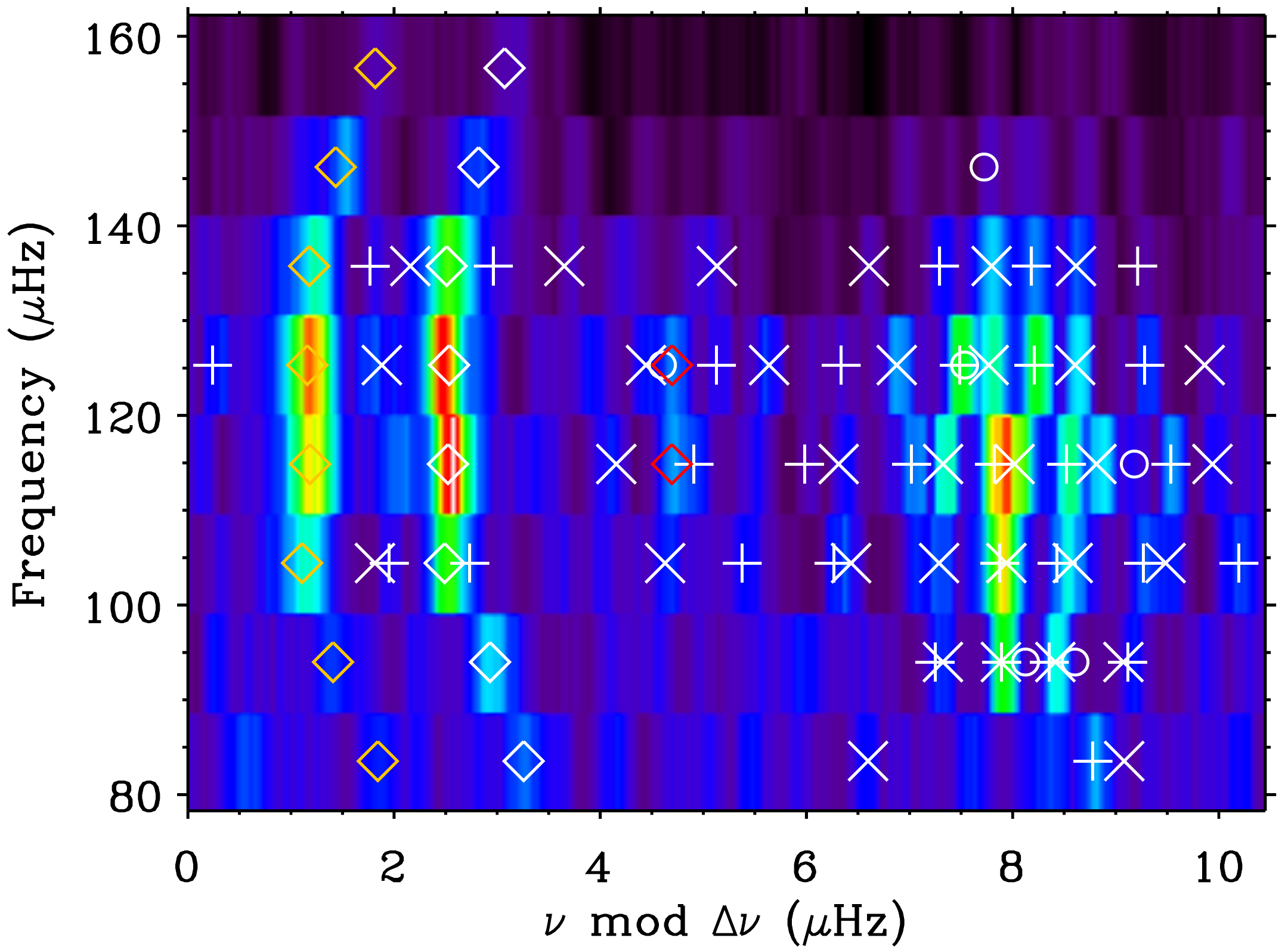}
\caption{Peak-bagging results including mixed-mode analysis by co-author Appourchaux displayed in the form of an \'echelle diagram of the power spectrum for $\Delta\nu=$10.44 $\mu$Hz. Fitted frequencies are shown: $l=0$ white diamonds, $l=2$ orange diamonds, $l=3$ red diamonds, $l=1, m=-1$ white crosses, $l=1, m=0$ white circles, $l=1, m=+1$ white pluses. }
\label{fig_echelle_fitted}
\end{figure}
\begin{figure*}[t!]
\center
\includegraphics[width=16cm]{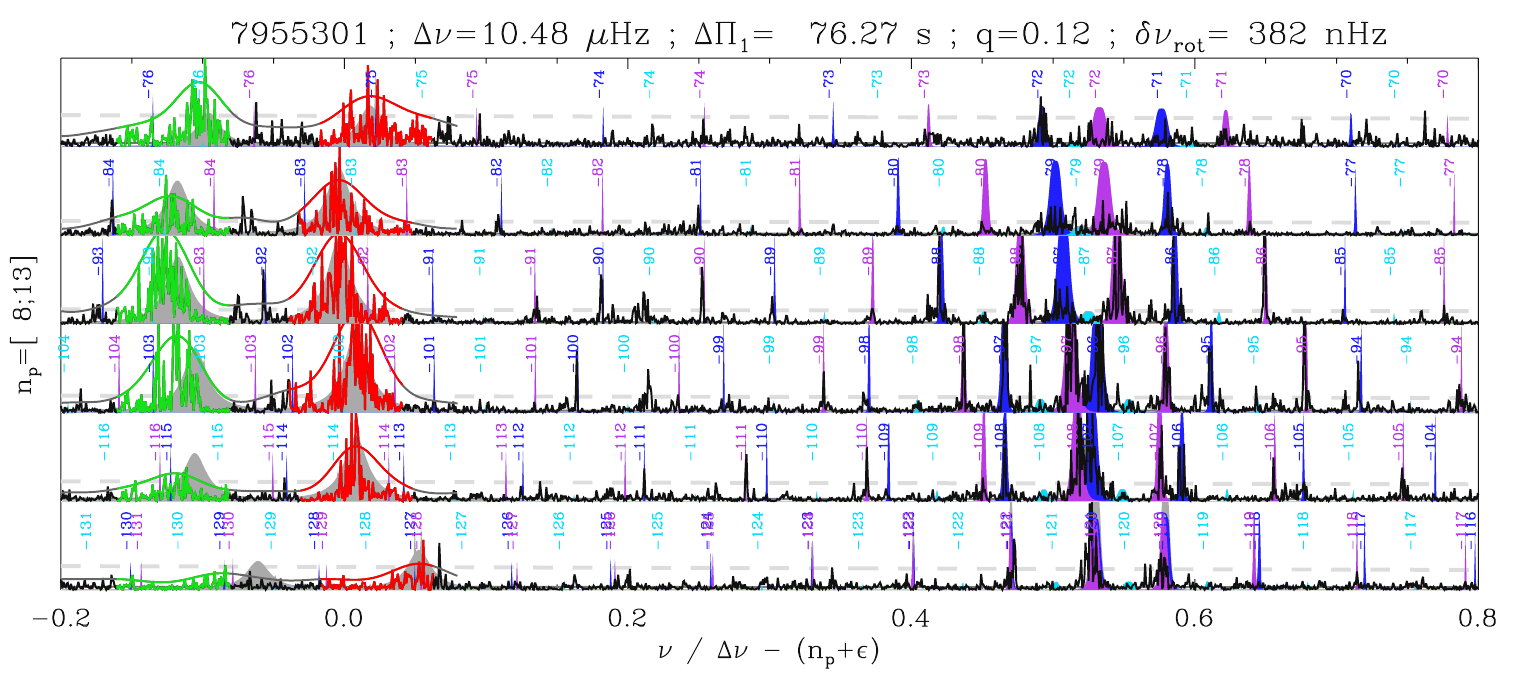}
\caption{Peak-bagging results including mixed-mode analysis by co-author Mosser in the form of a line-by-line \'echelle diagram, for radial pressure orders 8 to 13. Radial and quadrupole modes are highlighted in red and green, respectively. The expected locations of dipole mixed modes are labelled with their mixed radial orders, and indicated with a color depending on the azimuthal order: dark blue ($m = -1$), light blue ($m = 0$), or purple ($m = +1$).  $\ell= 3$ modes, not identified in this plot, are located near the abscissa 0.21. The gray dashed lines correspond to height-to-background ratios of 7 and 10.}
\label{fig_echelle_fitted_mos}
\end{figure*}

\begin{table*}
\caption{Global seismic parameters of the RG component.}
\label{tab:seis_param}
\centering
\begin{tabular}{l l l l l l l}
\hline
Parameter & & Gaulme &  Appourchaux & Mosser & Gehan & Vrard\\
\hline
$\nu\ind{max}$& [$\mu$Hz] & $124.9\pm0.4$ & & $122.7\pm24.1$ &  $122.5 \pm5.6$& \\  
$\Delta\nu$   & [$\mu$Hz] & $10.49\pm0.02$& $10.496 \pm 0.011$ &  $10.48\pm0.05$ & $10.45\pm0.27$& \\  
$\Delta\Pi_1$ & [s]       & & $76.297 \pm 0.003$ & $76.27\pm0.03$ & $76.3 \pm 0.1$& $76.09\pm0.72$\\
$q$             &           & & $0.098 \pm  0.002$ & $0.120\pm0.013$ & & $0.17\pm0.05$\\
$\delta_{01}$ &           & & $0.027 \pm 0.001$   &  $0.029\pm0.006$ & &\\
$\epsilon_p$  &           & & 1.233  & $1.194\pm0.052$ & &\\
$\epsilon_g$  &           & & $0.24 \pm 0.005$    & $0.246\pm0.032$ & &\\
$\nu\ind{core}$ & [nHz]   & & $382 \pm 2$        & $382\pm10$ & $380 \pm 6$ &\\
$\nu\ind{env}$ & [nHz]    & & $39 \pm 6$         &  & &\\
$i$ & [$^\circ$] & & 74.2$^{+7.7}_{-14.1}$ &&$75.8 ^{+14.2}_{-6.6}$ &\\
\hline
\end{tabular}
\tablefoot{Frequency at maximum amplitude $\nu\ind{max}$ ($\mu$Hz), mean large frequency spacing $\Delta\nu$ ($\mu$Hz), mean dipole mode period spacing $\Delta\Pi_1$ (s), coupling factor $q$, small frequency spacing $\delta\nu_{01}$, p- and g-mode phase offsets $\epsilon_p$ and $\epsilon_g$, respectively, rotational splittings of the core and envelope $\nu\ind{core}$ and $\nu\ind{env}$ (nHz), respectively, and the inclination $i$ (degree).}
\end{table*}

\subsection{Mixed modes}
\subsubsection{Formalism}
As indicated in the introduction, the RG component of KIC 7955301 shows a rich spectrum of dipolar mixed pressure and gravity modes. 
Here we briefly summarize some basic properties of the mixed modes (see \citealt{Mosser_2015}, and references therein).  The frequency of the mixed modes is an implicit expression given by
\begin{equation}
    \tan \theta_p=q \tan \theta_g,
\end{equation}
where $q$ is the coupling factor between the p and g modes, $\theta_p$ is the phase of the mixed modes with respect to the asymptotic p-mode frequencies separated by the large separation $\Delta\nu$, while $\theta_g$ the phase of the mixed modes with respect to the asymptotic g-mode periods separated by the period spacing $\Delta \Pi_1$. In other words,
\begin{equation}
    \theta_p=\pi \left(\frac{\nu}{\Delta\nu}-\frac{1}{2}-\epsilon_p\right)
    ,
\end{equation}
where $\nu$ is the mixed mode frequency and $\epsilon_p$ is a phase offset, and
\begin{equation}
    \theta_g=\pi \left(\frac{P}{\Delta\Pi_1}-\epsilon_g\right)
    ,
\end{equation}
where $P$ is the mixed mode period and $\epsilon_g$ is a gravity offset. 
Solving the implicit relation given by Eq.\,(3) provides the mixed mode frequencies.

The period separation $\Delta P$ between two consecutive mixed mode periods is given by
\begin{equation}
    \Delta P= \zeta(\nu) \,\ \Pi_1
    ,
\end{equation}
where $\zeta$ is related to the phases given above by
\begin{equation}
    \zeta(\nu)=\left[1+\frac{\nu^2\Delta \Pi_1}{q \Delta\nu}\frac{\cos^2\theta_g}{\cos^2\theta_p}\right]^{-1}
    .
\end{equation}
$\zeta$ is also useful for expressing the frequency splitting of these dipole modes $\delta\nu_{\rm rot}$, since we have
\begin{equation}
    \delta\nu\ind{rot}=\zeta(\nu) \delta\nu\ind{core}+(1-\zeta(\nu))\,\delta\nu\ind{env}
    ,
\end{equation}
where $\delta\nu\ind{core}$ is the rotation mainly sampled by the g modes, while $\delta\nu\ind{env}$ is the rotation mainly sampled by the p modes.  We stress that this expression  is not symmetrical in terms of the azimuthal order $m$ since it depends on the mixed mixed frequency
\begin{equation}
    \zeta(\nu+\delta\nu\ind{rot}) \ne \zeta(\nu-\delta\nu\ind{rot})
    ,
\end{equation}
as discussed in \citet{Mosser_2012}. 

Equations\,(3) and (8) provide a full description of the mode frequency and its associated rotational splitting.
To ensure a robust analysis of the mixed modes, four independent studies were led by co-authors Appouchaux, Gehan, Mosser, and Vrard. All followed the approach developed by \citet{Mosser_2015}, where the first step consists of stretching the oscillation spectrum to transform the mixed-mode pattern into a comb-like pattern based on the gravity period spacing $\Delta\Pi_1$. Despite a common approach, each fitter performed different analysis with independent routines. The following subsections provide details for all.

\subsubsection{Analysis adapted from \citet{Appourchaux_2020}}
The stretched power spectrum is interpolated onto a regular grid for producing the \'echelle diagram of Fig.~\ref{fig_echelle_stretched} (top panel). The fitted asymptotic frequencies of the multiplet of the dipole mixed modes are computed using the expressions of \cite{Mosser_2015} where $\zeta$ is replaced by the analytical formulation given above. We note again that the mode splitting is not symmetrical due to $\zeta$ depending on frequency. First guesses for $\Delta\Pi_1$, rotation and $q$ are provided by the results of \citet{Mosser_2014, Mosser_2015, Mosser_2017}, respectively.  These three parameters are then visually adjusted to obtain a figure similar to that of Fig.~\ref{fig_echelle_stretched}.  The asymptotic frequencies obtained by this procedure are very close to the actual peaks, within $0.1\ \mu$Hz, easing the fitting of the peaks by Maximum Likelihood Estimation (MLE).  Due to the dense forest of dipole mixed modes, the fitting window for these modes is $0.5\, \mu$Hz, while for $l=0,2$ and 3, it is 3 $\mu$Hz.  Fitted frequencies are displayed in Table \ref{tab:peak_bagging}.  The \'echelle diagram of the power spectrum with the fitted frequencies is shown on Fig.~\ref{fig_echelle_fitted}

In a second step, we determine all the parameters of the asymptotic expression of the dipole mode frequencies, i.e. $\Delta\Pi_1$, q, $\epsilon_g$, $\delta_{01}$, $\nu\ind{core}$ and $\nu\ind{env}$ by doing an unweighted least squares fit of the fitted frequencies. The optimization is performed by using the same algorithm as in \citet{Appourchaux_2020}. Data were trimmed to exclude three frequencies out of the 62 dipole mode frequencies (see frequencies in Table \ref{tab:peak_bagging}). The root mean square value of the optimized difference is $0.00 \pm 0.03$ $\mu$Hz. The error bars were then extracted by doing a Monte Carlo simulation using the error bars on the dipole mode frequencies extracted from MLE.  The optimization was then repeated 100 times.  The values obtained are reported in Table \ref{tab:seis_param}.

The inclination angle of the star was extracted using a single mixed mode (n=-94) for which the triplets was clearly fitted.  We used the ratio between heights of modes with different azimuthal orders \citep{Gizon_Solanki_2003}.  The ratio between the $m=0$ and the $m=\pm 1$ is 0.08$^{+0.257}_{-0.060}$ providing an angle of $i = 74.2^{+7.7}_{-14.1}$ degrees.

\subsubsection{Analysis following \citet{Gehan_2018}}
Before the mixed-mode analysis, the frequency at maximum amplitude $\nu\ind{max}$ was derived by using the FRA pipeline. It computes a smoothed  power spectrum, then fits a Gaussian envelope accounting for oscillations along with a background contribution to measure $\nu\ind{max}$. It also performs a likelihood ratio test to validate the measured $\nu\ind{max}$ and check whether oscillations are present. The FRA pipeline was used in \cite{Huber_2022} and is explained in detail in \citet{Gehan_2022}. The large spacing $\Delta\nu$ was measured by using the EACF.

The power spectrum was then stretched as in \cite{Mosser_2015} and \cite{Gehan_2018}. Pressure-dominated mixed modes were located using the RG universal oscillation pattern \citep{Mosser_2011} and excluded, in order to keep only gravity-dominated modes that are mostly sensitive to the core rotation rate.
The stretched period \'echelle diagram reveals that dipole gravity-dominated mixed modes line up along two ridges, associated with the azimuthal orders $m=\pm 1$ (Fig.~\ref{fig_echelle_stretched}, bottom). Fitting these ridges (Fig.~\ref{fig_echelle_stretched}) allows us to derive the mean core rotational splitting, which is found to be $\dnurotcore = 380 \pm 6$ nHz \citep[][for details]{Gehan_2018}. Once the azimuthal order of mixed modes is identified, we measure the inclination angle of the rotation axis from the ratio between heights of modes with different azimuthal orders \citep{Gizon_Solanki_2003}. Only the ridges with azimuthal orders $m=\pm 1$ are visible, but we know where to look for the missing $m=0$ ridge in the background, at the midpoint between the $m=+1$ and $m=-1$ ridges (Fig.~\ref{fig_echelle_stretched}, bottom). We obtain an inclination of $i = 75.8 ^{+14.2}_{-6.6}$ degrees. We refer to \cite{Gehan_2021} for the details of the inclination measurement procedure. We note that the maximal possible value according to the uncertainties is $i = 90^\circ$, so that the star is possibly seen equator-on. This is because the $m=0$ ridge is lost in the background, thus there is a possibility that the peaks with $m=0$ are completely absent and that the observed signal is only due to background noise.

\subsubsection{Analysis following \citet{Mosser_2018}}
Stretching the power spectrum first requires a precise fitting of the radial and quadrupole modes $l=0$ and 2, thanks to the RG universal oscillation pattern, in order to locate precisely the expected pure pressure dipole modes. Firstly, the modes with the largest height-to-background ratios (HBR) are used to determine the parameters of the fit. Secondly, we pick modes with lower HBRs, still larger than $5$, provided that they stick to the asymptotic pattern within a frequency range wide six times the spectral resolution. We applied the method of \citet{Mosser_2018} for analyzing the rotational splittings to overcome the difficulties arising from the overlap of rotational multiplets larger than period spacings. The global seismic parameters derived from the fit of the mixed-mode pattern, following the updated approach described by \citet{Pincon_2019}, are reported in Table \ref{tab:seis_param}. The frequency analysis is displayed in Fig. \ref{fig_echelle_fitted_mos}.

\subsubsection{Analysis adapted from \citet{Vrard_2016}}
The steps that follow the power-spectrum stretching differ from the analysis led by co-author Mosser. We started by computing the Fourier transform of the stretched spectrum to obtain a first $\Delta\Pi_1$ value following the work of \citet{Vrard_2016}. After that, the coupling parameter $q$ was determined following the method described in \cite{Mosser_2017}, which involves searching for the maximum of the Fourier Transform of the stretched oscillation spectrum. Finally, an iterative process was performed between $\Delta\Pi_1$ and $q$: the $\Delta\Pi_1$ measurement was redone with the measured $q$ value, then the inverse is performed until the $\Delta\Pi_1$ and $q$ values converge. The final measurement of $\Delta\Pi_1$ and $q$ is obtained when those values converge. The final fitted parameters lead to $\Delta\Pi_1 = 76.09 \pm 0.72$ for the gravity mode period spacing and $q = 0.17 \pm 0.05$ for the coupling parameter. 

\subsubsection{The big picture from the mixed mode analysis}
The four independent approaches used to analyse the mixed dipole modes agree (Table \ref{tab:seis_param}). Firstly, the measured mixed-mode period spacing $\Delta\Pi_1 \approx 76$ s associated with the large frequency p-mode spacing $\Delta\nu \approx 10.5\,\mu$Hz, definitely classifies the star as an RGB \citep{Mosser_2014}. Secondly, the fact that the $m=0$ modes are not detected indicates a large inclination angle $i$ of the RG rotation axis with respect to the line of sight. The two authors who carefully analysed the ratio of the amplitudes of the mixed $m=1$ to $m=0$ dipole modes report a very consistent inclination angle of $\approx75^\circ$, with a rather large uncertainty ranging from $60^\circ$ to $90^\circ$. Thirdly, according to \citet[][Eqs., 22 and 23]{Mosser_2012}, the frequency spacing measured on the mixed modes provides a proxy of the mean core rotation period with the relation $P\ind{core} \approx 1/2\delta\nu\ind{core} = 15.1\pm0.2$ days by assuming an uncertainty of 5 nHz on $\delta\nu\ind{core}$. The average splitting of the envelope determined by co-author Appourchaux is about 10 times smaller, leading to a mean rotation period of $148_{-20}^{+27}$ days. If we add the fact that no spot-related variability was measured from the \textit{Kepler} lightcurve and that the oscillation amplitude meet our expectations for that type of RG, we can safely conclude that the RG is a regular slow rotator that does not exhibit any sign of tidally increased spin.

\section{Dynamical model}
\label{sect_dyn_mod}

The combined photodynamical analysis of the \textit{Kepler} photometry, the eclipse timing variations of the inner binary, and the radial-velocity curve of the outer giant component offer a robust dynamical determination of the mass of the RG ($M\ind{A}$) and of the total mass of the inner binary ($M_\mathrm{bin}$). It can be seen as follows. The amplitude of the RV curve gives the spectroscopic mass function:
\begin{equation}
f(M_\mathrm{bin})=M_\mathrm{bin}\left(\frac{q_\mathrm{out}}{1+q_\mathrm{out}}\right)^2\sin^3i_\mathrm{out},
\end{equation}
where $q_\mathrm{out}=M_\mathrm{bin}/M_\mathrm{A}$ and $i_\mathrm{out}$ are the mass ratio and the inclination of the outer (or wide) binary formed by the RG and the center of mass of the inner eclipsing pair. In addition, \citet{Borkovits_2016} showed that the ETV of the eclipsing binary is strongly dominated by the gravitational three-body perturbations against the pure light-travel time effect. Detailed analytical investigations of third-body perturbed ETVs \citep{Borkovits_2011,Borkovits_2016} have shown that such curves tightly constrain almost all of the orbital elements of both the inner and outer orbits, together with the mutual inclination of the two orbital planes. Their amplitude is related to the outer mass ratio as
\begin{equation} 
\mathcal{A}_\mathrm{dyn}\propto\frac{1}{1+q_\mathrm{out}}.
\end{equation}
Finally, the variation of the eclipse depth and duration of the inner eclipsing binary offer a very precise determination of the varying inclination of both the inner orbit of the eclipsing pair and the outer orbit.

The combined photodynamical analysis was carried out with the software package {\sc Lightcurvefactory} \citep[see, e. g.][and further references therein]{Borkovits_2019}. This code contains a built-in numerical integrator to calculate the three-body perturbed coordinates and velocities of the three bodies, a multi-band light curve, ETV and RV curve emulators, and an MCMC-based parameter search routine for the inverse problem. The latter employs an implementation of the generic Metropolis-Hastings algorithm \citep[see e. g.][]{Ford_2005}. 

\begin{figure*}
\includegraphics[width=9.0cm]{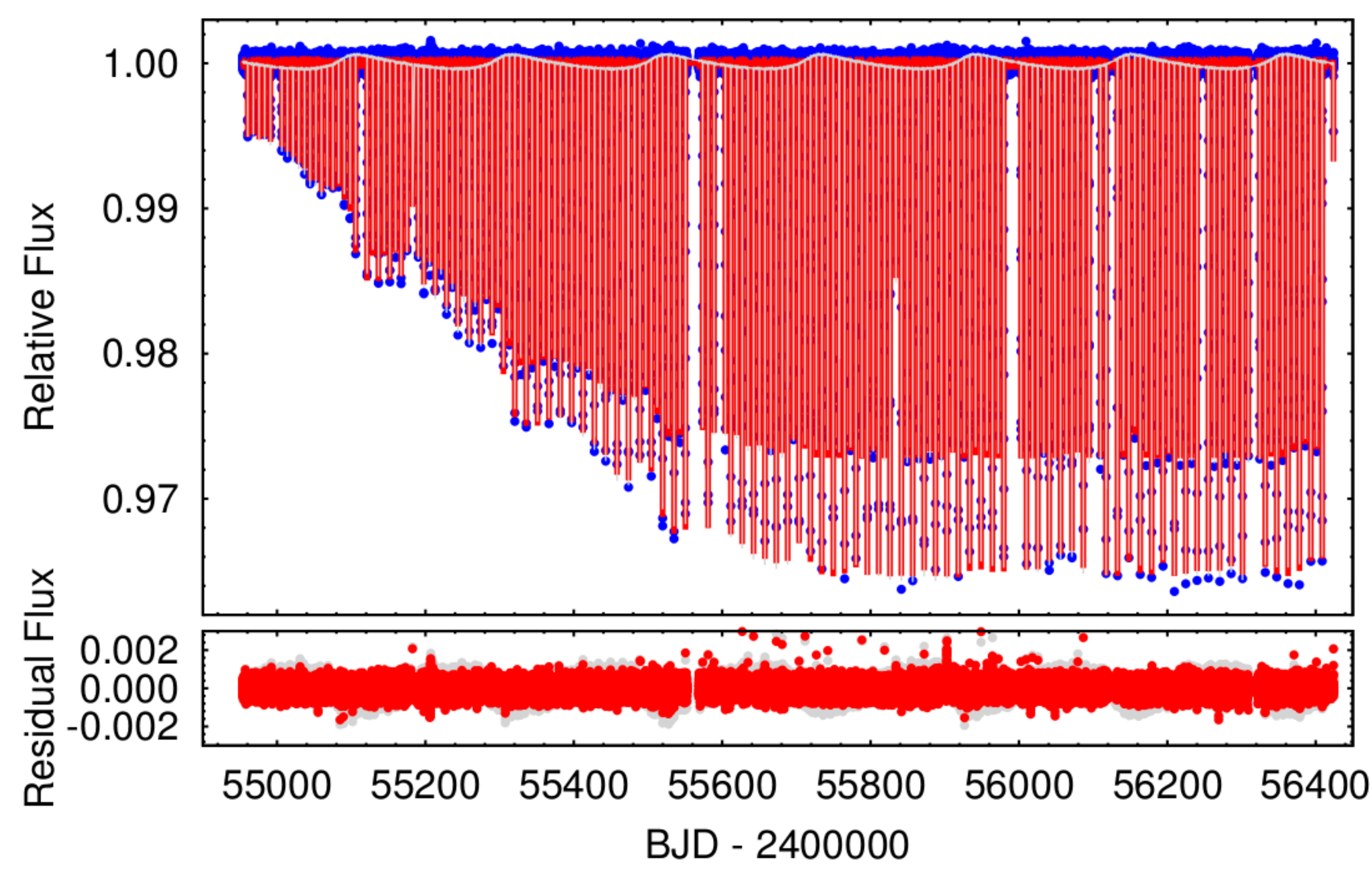}\includegraphics[width=9.0cm]{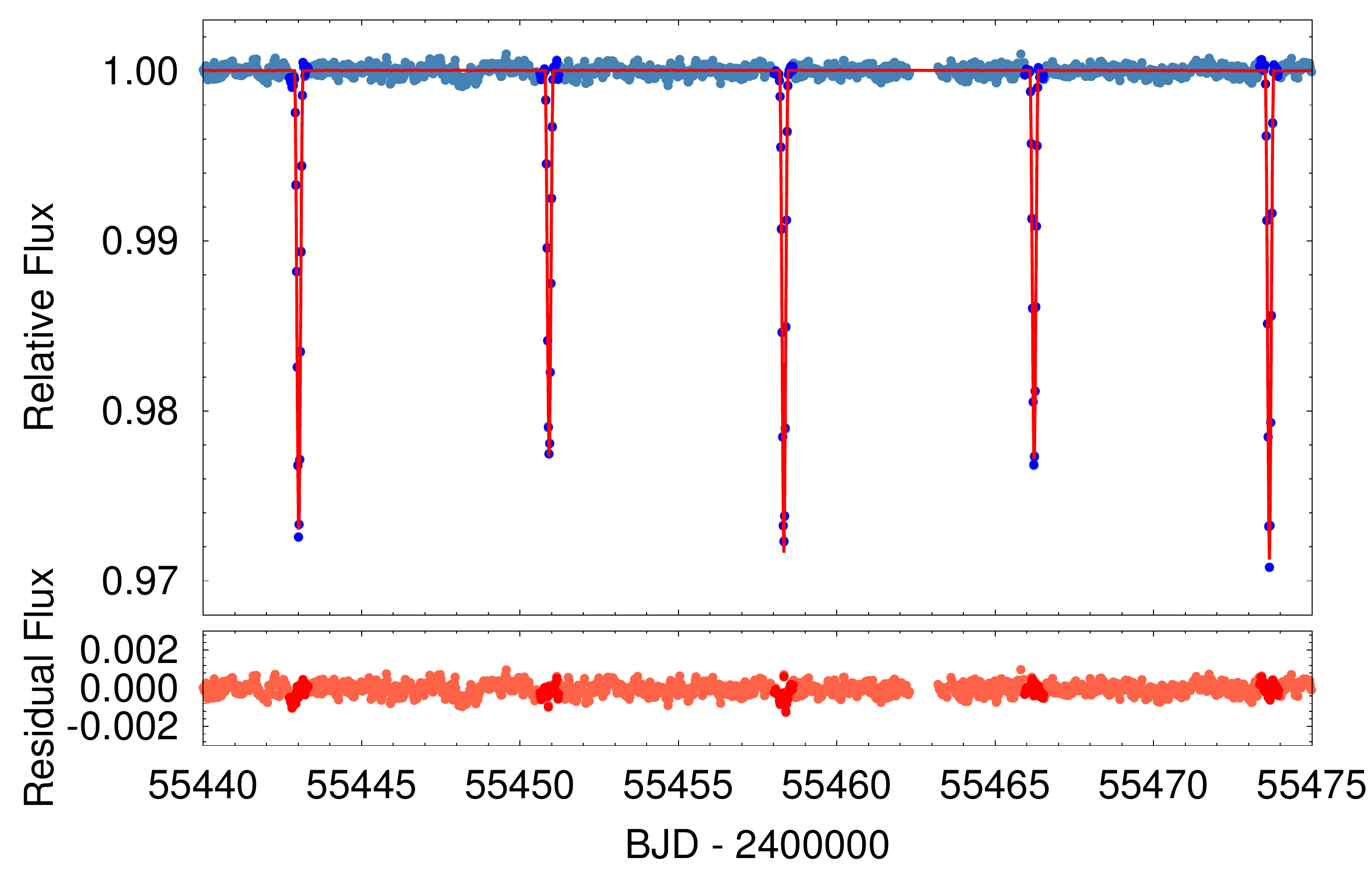}  
\caption{The long cadence \textit{Kepler} lightcurve of KIC\,7955301 together with the joint photodynamical model lightcurve solution. {\it Upper left panel:} The complete Q0-Q17 lightcurve (blue) with the bestfitted solution (red). This solution was obtained with the negligation of the Doppler-boosting effect. The grey curve show the same model after `switching on' the Doppler-boosting. While this effect, in theory, should have been observable with the accuracy of \textit{Kepler} photometry, the reasons of its negligence is discussed in the text. {\it Upper right panel:} A 1-month-long section of the \textit{Kepler} observations overplotted with the photodynamical solution. The pale blue circles represent each individual observations but, for the analysis only dark blue data (located within the $\pm0\fp02$ phase environment of each eclipses) were considered. {\it Lower panels} show the residual data.}
\label{fig_lcswithfit}
\end{figure*}

\begin{figure*}
\includegraphics[width=9.0cm]{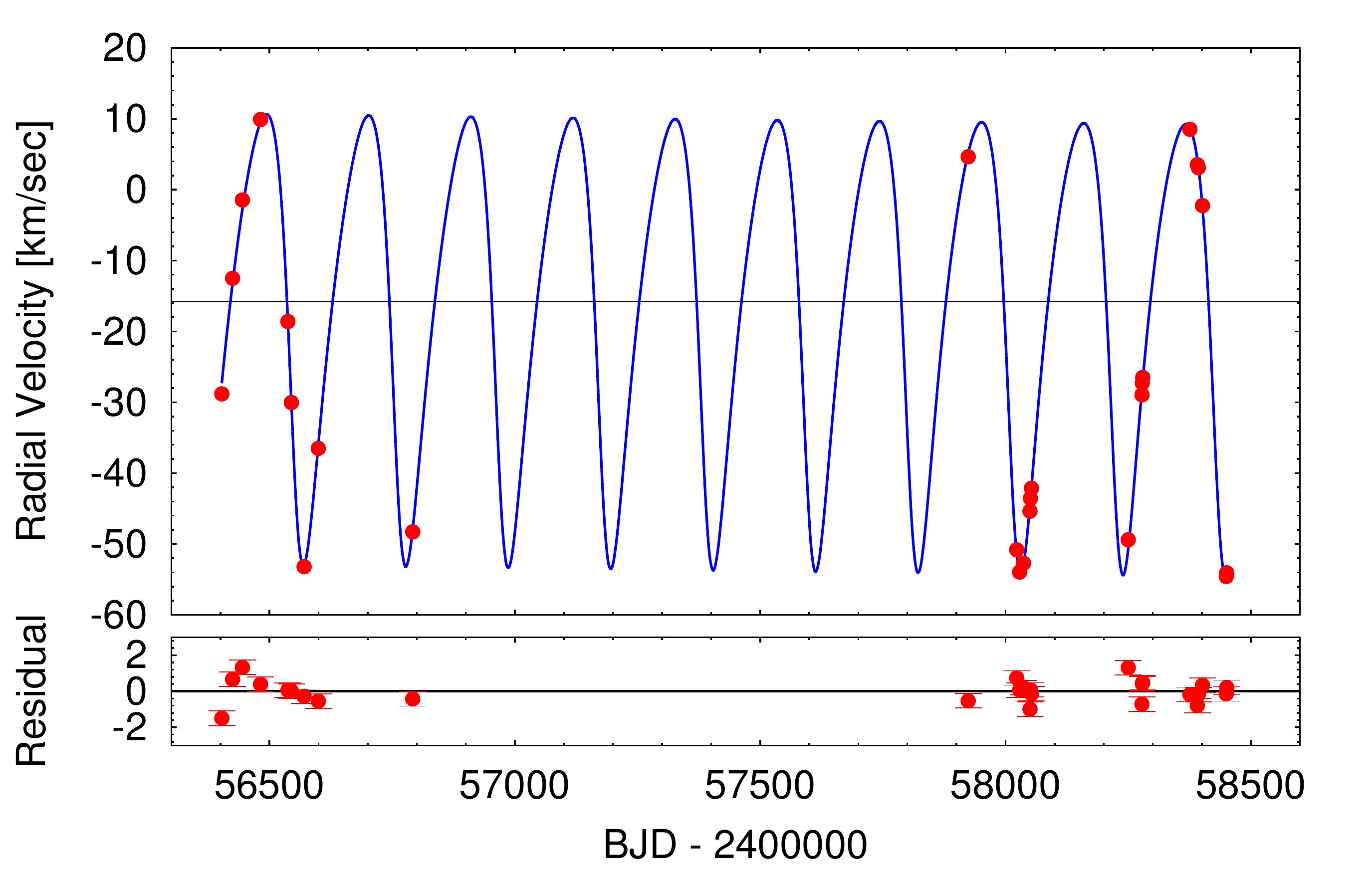}\includegraphics[width=9.0cm]{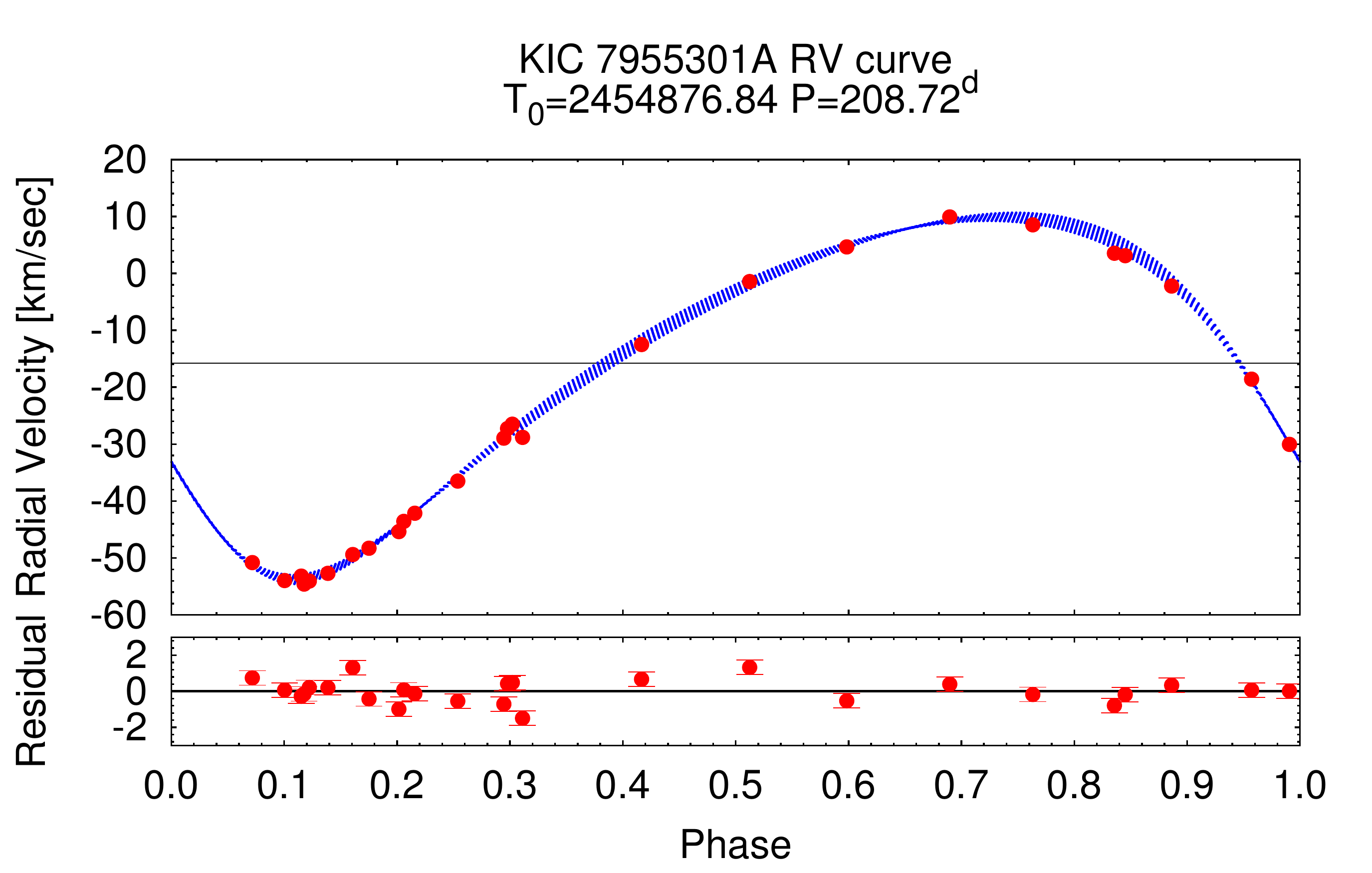}  
\caption{Left: radial velocity data (red) as a function of time expressed in modified Julian dates, with their best fit model (blue line). Right: radial-velocity data (red) folded over the RG orbital period as a function of the orbital phase. The dispersion of the best-fit curve in blue is caused by the varying shape of the RV curve as a function of time.}
\label{fig_RVwithfit}
\end{figure*}

\begin{figure*}
\includegraphics[width=9.0cm]{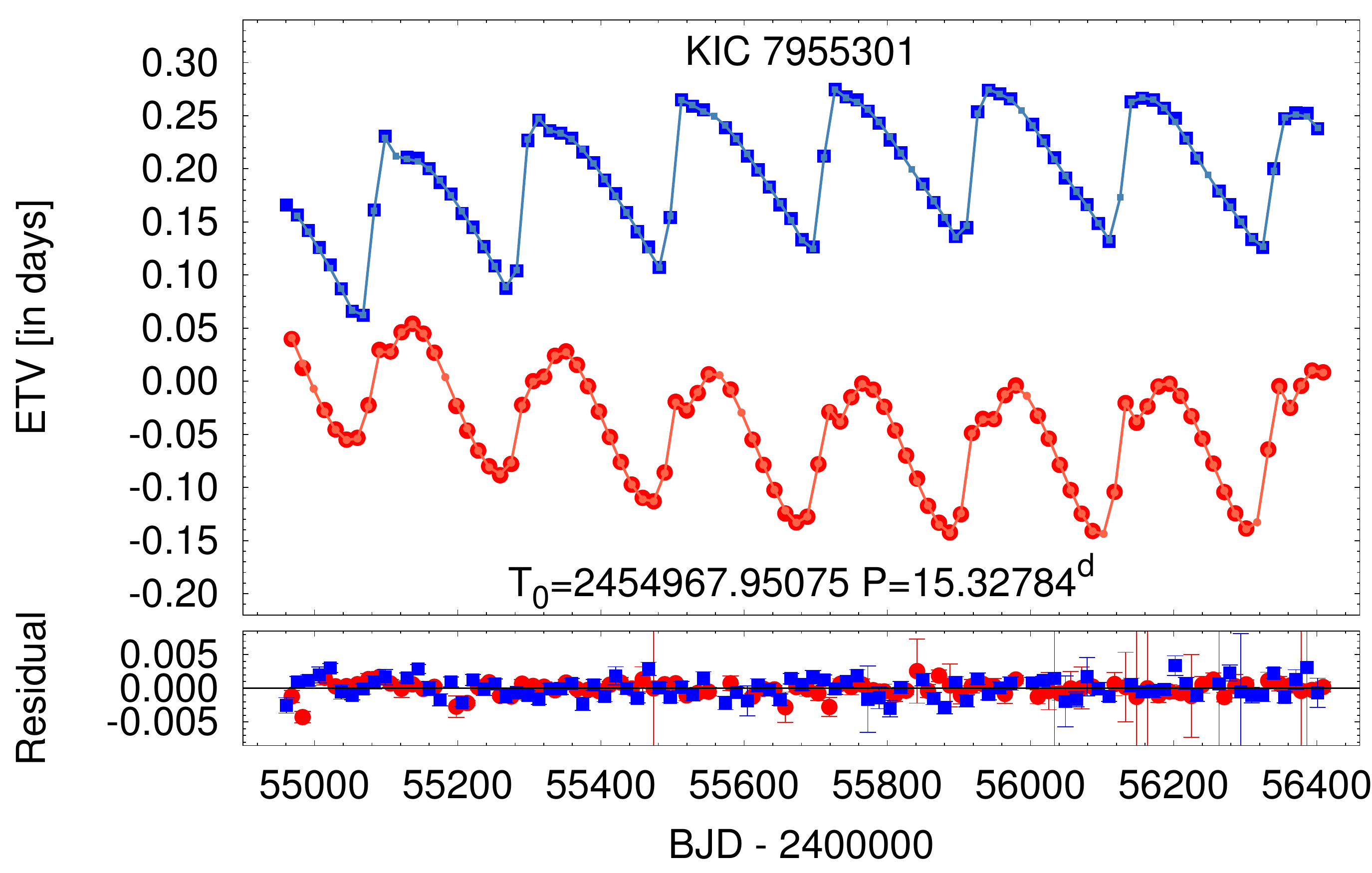}\includegraphics[width=9.0cm]{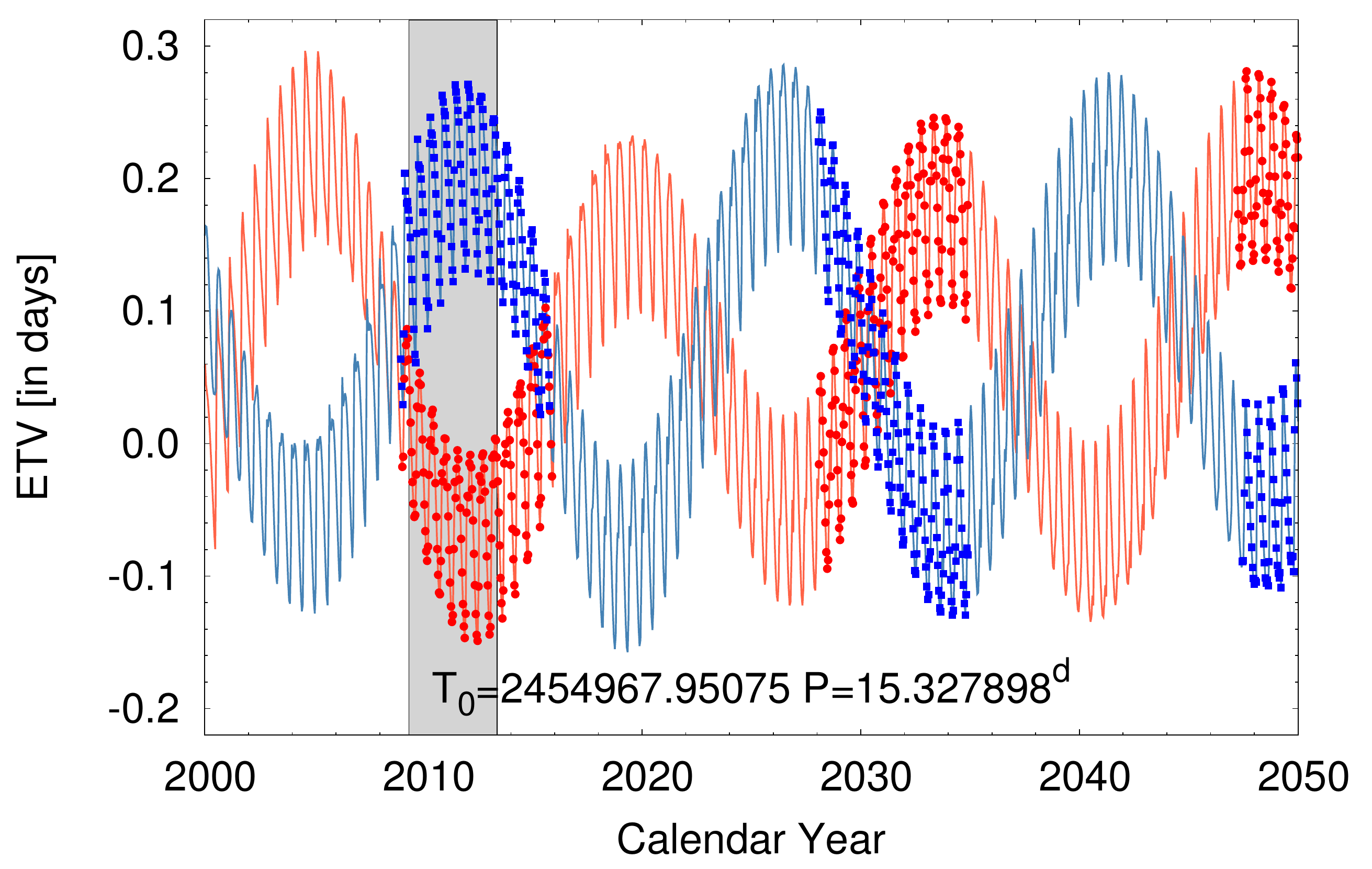}  
\caption{The eclipse timing variations of KIC\,7955301 on different timescales: Observations vs model and predictions. {\it Left:} The primary and secondary ETV curves derived from \textit{Kepler}-observations (red circles and blue boxes, respectively), together with the photodynamical model solution (smaller, pale red and blue symbols, connected with straight lines, respectively). {\it Right:} ETV-like curves derived from the times of the inferior and superior conjunctions of the secondary of the inner pair (pale red and blue lines), overplotted with `true', forecasted primary and secondary ETV points (larger red and blue symbols) during the eclipsing phases of the inner binary for the first half of the present century. The dynamically forced short-period apsidal motion of the inner binary is nicely visible. The grey-shaded area stands for the interval of the \textit{Kepler}-measurements. }
\label{fig_ETVs}
\end{figure*}
\begin{table*}
 \centering
 \caption{Orbital and astrophysical parameters of KIC\,7955301 from the joint photodynamical lightcurve, RV and ETV solution with and without the involvement of the stellar energy distribution and \texttt{PARSEC} isochrone fitting.}
 \label{tab: syntheticfit_KIC7955301}
\begin{tabular}{@{}lllllll}
\hline
 & \multicolumn{3}{c}{without SED+\texttt{PARSEC}} & \multicolumn{3}{c}{with SED+\texttt{PARSEC}} \\ 
\hline
\multicolumn{7}{c}{orbital elements} \\
\hline
   & \multicolumn{6}{c}{subsystem}  \\
   & \multicolumn{2}{c}{Ba--Bb} & A--B & \multicolumn{2}{c}{Ba--Bb} & A--B  \\
  $P$ [days] & \multicolumn{2}{c}{$15.31831_{-0.00025}^{+0.00024}$} & $209.760_{-0.020}^{+0.020}$ & \multicolumn{2}{c}{$15.31825_{-0.00023}^{+0.00023}$} & $209.761_{-0.018}^{+0.018}$   \\
  $a$ [R$_\odot$] & \multicolumn{2}{c}{$31.68_{-0.11}^{+0.19}$} & $217.0_{-0.9}^{+1.4}$ & \multicolumn{2}{c}{$31.68_{-0.08}^{+0.11}$} & $216.9_{-0.5}^{+1.1}$ \\
  $e$ & \multicolumn{2}{c}{$0.02763_{-0.00015}^{+0.00012}$} & $0.2733_{-0.0014}^{+0.0013}$ & \multicolumn{2}{c}{$0.02757_{-0.00007}^{+0.00007}$} & $0.2737_{-0.0014}^{+0.0013}$ \\
  $\omega$ [deg]& \multicolumn{2}{c}{$295.80_{-0.20}^{+0.22}$} & $117.02_{-0.40}^{+0.39}$ & \multicolumn{2}{c}{$295.90_{-0.19}^{+0.18}$} & $116.99_{-0.39}^{+0.39}$ \\ 
  $i$ [deg] & \multicolumn{2}{c}{$87.757_{-0.039}^{+0.037}$} & $84.23_{-0.25}^{+0.25}$ & \multicolumn{2}{c}{$87.768_{-0.038}^{+0.034}$} & $84.40_{-0.32}^{+0.23}$ \\
  $\tau$ [BJD - 2400000]& \multicolumn{2}{c}{$54961.5184_{-0.0083}^{+0.0090}$} & $54876.85_{-0.12}^{+0.12}$ & \multicolumn{2}{c}{$54961.5225_{-0.0081}^{+0.0076}$} & $54876.84_{-0.11}^{+0.12}$\\
  $\Omega$ [deg] & \multicolumn{2}{c}{$0.0$} & $-5.28_{-0.17}^{+0.16}$ & \multicolumn{2}{c}{$0.0$} & $-5.19_{-0.17}^{+0.16}$ \\
  $i_\mathrm{m}$ [deg] & \multicolumn{3}{c}{$6.34_{-0.24}^{+0.26}$} & \multicolumn{3}{c}{$6.18_{-0.23}^{+0.29}$} \\
  $\omega^\mathrm{dyn}$ [deg]& \multicolumn{2}{c}{$239.7_{-1.4}^{+1.1}$} & $240.6_{-1.4}^{+1.2}$ & \multicolumn{2}{c}{$239.2_{-1.4}^{+1.4}$} & $239.9_{-1.4}^{+1.5}$ \\
  $i^\mathrm{dyn}$ [deg] & \multicolumn{2}{c}{$5.36_{-0.21}^{+0.23}$} & $0.97_{-0.04}^{+0.04}$ & \multicolumn{2}{c}{$5.23_{-0.20}^{+0.25}$} & $0.95_{-0.04}^{+0.04}$ \\
  $\Omega^\mathrm{dyn}$ [deg] & \multicolumn{2}{c}{$236.4_{-1.1}^{+1.3}$} & $56.4_{-1.1}^{+1.3}$ & \multicolumn{2}{c}{$236.9_{-1.4}^{+1.4}$} & $56.9_{-1.4}^{+1.4}$ \\
  $i_\mathrm{inv}$ [deg] & \multicolumn{3}{c}{$84.77_{-0.22}^{+0.21}$} & \multicolumn{3}{c}{$84.91_{-0.27}^{+0.20}$} \\
  $\Omega_\mathrm{inv}$ [deg] & \multicolumn{3}{c}{$-4.47_{-0.15}^{+0.14}$} & \multicolumn{3}{c}{$-4.39_{-0.14}^{+0.13}$} \\
  \hline
  mass ratio $[q=M_\mathrm{sec}/M_\mathrm{pri}]$ & \multicolumn{2}{c}{$0.929_{-0.009}^{+0.010}$} & $1.399_{-0.012}^{+0.010}$ & \multicolumn{2}{c}{$0.933_{-0.003}^{+0.003}$} & $1.406_{-0.013}^{+0.010}$ \\
  $K_\mathrm{pri}$ [km\,s$^{-1}$] & \multicolumn{2}{c}{$50.44_{-0.32}^{+0.37}$} & $31.61_{-0.12}^{+0.13}$ & \multicolumn{2}{c}{$50.53_{-0.16}^{+0.13}$} & $31.65_{-0.08}^{+0.08}$ \\ 
  $K_\mathrm{sec}$ [km\,s$^{-1}$] & \multicolumn{2}{c}{$54.27_{-0.38}^{+0.46}$} & $22.58_{-0.16}^{+0.23}$ & \multicolumn{2}{c}{$54.13_{-0.16}^{+0.24}$} & $22.50_{-0.10}^{+0.23}$ \\ 
  $\gamma$ [km\,s$^{-1}$] & \multicolumn{3}{c}{$-15.779_{-0.048}^{+0.050}$} & \multicolumn{3}{c}{$-15.769_{-0.036}^{+0.036}$} \\
  \hline  
\multicolumn{7}{c}{stellar parameters} \\
\hline
   & Ba & Bb &  A  & Ba & Bb &  A  \\
  \hline
 \multicolumn{7}{c}{Relative quantities} \\
  \hline
 fractional radius [$R/a$] & $0.0285_{-0.0005}^{+0.0004}$ & $0.0256_{-0.0007}^{+0.0006}$  & $0.0241_{-0.0005}^{+0.0008}$ & $0.0285_{-0.0002}^{+0.0002}$ & $0.0259_{-0.0001}^{+0.0001}$  & $0.0263_{-0.0006}^{+0.0008}$ \\
 fractional flux [in $Kepler$-band] &   &     &  &   &     &  \\
 \hline
 \multicolumn{7}{c}{Physical Quantities} \\
  \hline 
 $M$ [$M_\odot$] & $0.941_{-0.011}^{+0.020}$ & $0.876_{-0.011}^{+0.012}$ & $1.298_{-0.019}^{+0.030}$ & $0.939_{-0.007}^{+0.011}$ & $0.877_{-0.008}^{+0.013}$ & $1.292_{-0.010}^{+0.026}$ \\
 $R$ [$R_\odot$] & $0.903_{-0.014}^{+0.012}$ & $0.809_{-0.022}^{+0.022}$ & $5.22_{-0.12}^{+0.22}$    & $0.901_{-0.007}^{+0.010}$ & $0.822_{-0.006}^{+0.006}$ & $5.70_{-0.13}^{+0.20}$ \\
 $T_\mathrm{eff}$ [K]& $5372_{-46}^{+84}$ & $5138_{-48}^{+73}$ & $4800$ & $5580_{-70}^{+44}$ & $5313_{-62}^{+38}$ & $4804_{-48}^{+26}$ \\
 $L_\mathrm{bol}$ [L$_\odot$] & $0.611_{-0.035}^{+0.043}$ & $0.408_{-0.026}^{+0.034}$ & $13.00_{-0.58}^{+1.12}$ & $0.704_{-0.038}^{+0.038}$ & $0.480_{-0.022}^{+0.022}$ & $15.55_{-0.75}^{+0.87}$ \\
 $M_\mathrm{bol}$ & $5.27_{-0.07}^{+0.06}$ & $5.71_{-0.09}^{+0.07}$ & $1.95_{-0.09}^{+0.05}$ & $5.15_{-0.06}^{+0.06}$ & $5.57_{-0.05}^{+0.05}$ & $1.79_{-0.06}^{+0.05}$ \\
 $M_V           $ & $5.45_{-0.09}^{+0.08}$ & $5.97_{-0.12}^{+0.08}$ & $2.36_{-0.09}^{+0.05}$ & $5.24_{-0.07}^{+0.07}$ & $5.73_{-0.06}^{+0.07}$ & $2.15_{-0.05}^{+0.06}$ \\
 $\log g$ [dex]   & $4.503_{-0.014}^{+0.014}$ & $4.565_{-0.019}^{+0.024}$ & $3.115_{-0.026}^{+0.017}$ & $4.500_{-0.005}^{+0.004}$ & $4.551_{-0.003}^{+0.003}$ & $3.034_{-0.023}^{+0.019}$ \\
 \hline
$\log$(age) [dex] & \multicolumn{3}{c}{$-$} & \multicolumn{3}{c}{$9.65_{-0.03}^{+0.02}$} \\
$[\mathrm{M/H}]$ [dex] & \multicolumn{3}{c}{$-$} & \multicolumn{3}{c}{$0.061_{-0.037}^{+0.028}$} \\
$E(B-V)$ [mag]    & \multicolumn{3}{c}{$-$} & \multicolumn{3}{c}{$0.09_{-0.02}^{+0.02}$} \\     
$(M_V)_\mathrm{tot}$ & \multicolumn{3}{c}{$2.26_{-0.09}^{+0.05}$} & \multicolumn{3}{c}{$2.05_{-0.05}^{+0.07}$} \\
distance [pc]     &\multicolumn{3}{c}{$-$} & \multicolumn{3}{c}{$1369_{-29}^{+44}$} \\
\hline
\end{tabular}
\tablefoot{Besides the usual observational system of reference related angular orbital elements ($\omega$, $i$, $\Omega$), their counterparts in the system's invariable plane related dynamical frame of reference are also given ($\omega^\mathrm{dyn}$, $i^\mathrm{dyn}$, $\Omega^\mathrm{dyn}$). Moreover, $i_\mathrm{m}$ denotes the mutual inclination of the two orbital planes, while $i_\mathrm{inv}$ and $\Omega_\mathrm{inv}$ give the position of the invariable plane with respect to the tangential plane of the sky (i.\,e., in the observational frame of reference). Note, the instantaneous, osculating orbital elements, are given for epoch $t_0=2454953.0000$ (BJD).}
\end{table*}

We simultaneously analysed the eclipse photometry from the \textit{Kepler} light curve, the primary and secondary ETVs deduced from the same light curve, and the RVs of the outer RG component. For the \textit{Kepler}-light curve, we only used the $\phi\pm0\fp02$ phase sections of each eclipses to reduce computational costs. Except for preliminary tests, the following parameters were adjusted with MCMC:
\begin{itemize}
\item[(i)] Nine of the twelve orbital parameters. Regarding the inner orbit of the eclipsing pair, we fit $e_1\cos\omega_1$, $e_1\sin\omega_1$, and $i_1$, where $e_1$ is the osculating eccentricity, $\omega_1$ the argument of periastron and $i_1$ the orbital inclination. About the wide outer orbit, besides $e_2\cos\omega_2$, $e_2\sin\omega_2$, and $i_2$, we fit the anomalistic period $P_2$, the periastron passage time $\tau_2$ and the longitude of the node $\Omega_2$. Regarding two of the three ``missing'' parameters describing the inner orbit, namely, the period $P_1$ and time of a primary eclipse or, with other words, the time of an inferior conjunction of the secondary component of the inner pair $\mathcal{T}^\mathrm{inf}_1$, they were constrained internally at each trial step via the ETV curves in the manner described in \citet{Borkovits_2019}. Finally, the third ``missing'' orbital parameter, the longitude of the node of the inner orbit $\Omega_1$ was set to zero. It can be done because, for the complete analysis, only the difference of the nodes, i. e. $\Delta\Omega=\Omega_2-\Omega_1$ has relevance. Therefore, in such a manner, by setting $\Omega_2$, one sets $\Delta\Omega$. Note that because the motion of the three stars are not purely Keplerian, none of the orbital parameters is constant. Therefore, all the trial values either adjusted, constrained, or fixed in the initializing phase of any trial steps refer only a given epoch $t_0$.
\item[(ii)] Three parameters connected to the masses of the three stars: the RG mass $M_\mathrm{A}$, the spectroscopic mass function $f(M_\mathrm{B})$, and the mass ratio of the eclipsing binary $q_1=M_\mathrm{Bb}/M_\mathrm{Ba}$.
\item[(iii)] Four parameters almost exclusively connected to the light curve. The first is the duration of the primary eclipse around the epoch $t_0$ ($\Delta t_\mathrm{pri}$), which is related to the sum of the fractional radii of the members of the inner eclipsing binary, that is, the radius divided by the semi major axis. The second is the ratio of the stellar radii of the inner pair $R_\mathrm{Bb}/R_\mathrm{Ba}$. The third is the RG radius $R_\mathrm{A}$. The fourth is the ratio of the effective temperatures of the eclipsing pair $T_\mathrm{Bb}/T_\mathrm{Ba}$. We note that in case of a single-band photometric observations, the EB lightcurve only carries information about the temperature ratio -- most strictly speaking, the passband-dependent surface brightness ratio -- of the two components rather than their true temperatures. Therefore, one of the temperatures should be taken from some external sources. In a single binary it can be obtained from spectroscopic analysis or from stellar energy distribution (SED) data \citep[see, e.~g.][]{Miller_2020}. In the present situation, most of the total flux comes from the third RG component which makes the classical approach to be irrelevant.
Therefore, we assume that the two components of the inner binary are still on the main sequence, by being less massive than the RG, so that at each trial step the software sets the effective temperature of the primary component of the eclipsing pair $T_\mathrm{Ba}$ accordingly to the mass -- radius and mass -- effective temperature relations of \citet{Tout_1996} for main sequence stars. About the RG component, we assume that its contribution to the light curve is only a constant extra flux\footnote{The validity and the limits of this assumption will be discussed later}. Therefore, we fixed its temperature to $T_\mathrm{A}=4\,800$\,K for all runs. In such a way, adjusting the RG radius $R_\mathrm{A}$ directly sets its bolometric luminosity, and indirectly its total flux in the \textit{Kepler} photometric band. Hence, the adjustment of $R_\mathrm{A}$ was simply used to set the amount of the extra flux to the eclipsing binary light curve or; in other terms, it served as a substitute of the usual third light parameter.
\end{itemize}
Regarding other smaller effects that are mainly light-curve related, we used a logarithmic limb-darkening law. The corresponding parameters were calculated internally at each trial step with the use of the publicly available passband-dependent tables of the {\sc Phoebe} software \citep{Prsa_2005}\footnote{These tables were downloaded from the Phoebe 1.0 Legacy page -- \url{http://phoebe-project.org/1.0/download}}. For the well detached nature of both the inner and the outer binaries and, therefore, for the practically spherical shape of the stars, the gravity darkening coefficients have no detectable effect on the light curve solution. Therefore, we did not consider the results of the recent, more sophisticated study of \citet{Claret_2011}, but  simply adopted the fixed value of $g=0.32$, which is based on the seminal model of \citet{Lucy_1967}. Finally, we neglected both the reradiation/illumination and the Doppler-boosting effects \citep{Loeb_2003,vanKerkwijk_2010}. While the first effect has really no any influence on the light curve of our system, this is not certainly true for the second one. In this regard, we note that, as one can see in the left panel of Fig.\,\ref{fig_lcswithfit}, Doppler-boosting would result in an $\sim1000$\,ppm amplitude variation in the out-of-eclipse flux level with the period of the outer orbit, which, in theory, should be detected with \textit{Kepler}. Despite this fact, it cannot be seen in the lightcurves, because this low amplitude variation, having period longer than two quarters was filtered out during the data processing, which fact justifies its negligence in the analysis.

The median values of the orbital and physical parameters of the triple system derived from the MCMC posteriors and their 1-$\sigma$ uncertainties are tabulated in Table~\ref{tab: syntheticfit_KIC7955301}. The observed versus model photometric, RV, and ETV curves are plotted in Figs.\,\ref{fig_lcswithfit}--\ref{fig_ETVs}.

\section{Stellar evolution model}
\label{sect_stellar_model}
We modeled the system in three independent manners by using parameters that are deduced from the observations. The first two model the RG component only, by considering the seismic information and atmospheric parameters. The first method (\texttt{PARAM}) makes use of the asteroseismic global parameters $\nu\ind{max}$, $\Delta\nu$, and $\Delta\Pi_1$ to optimize the model from a grid of stellar evolution models. The second method makes use of the individual frequencies of the radial modes and $\Delta\Pi_1$. The third method is a complete model of the triple system, which includes both the RG and the inner binary, but that does not make use of the asteroseismic constraints. 

\subsection{Red-giant model based on the global seismic parameters}
\label{sect_param}
We used the code \texttt{PARAM} \citep{Rodrigues_2017} to infer the radius, mass, and age of the RG component by using a combination of seismic and non-seismic constraints. On the one hand, the average large frequency separation is computed using the radial-mode frequencies of the models in the grid, not added as an a-posteriori correction to the scaling relation between $\Delta\nu$ and the square root of the stellar mean density.  On the other hand, $\nu\ind{max}$ in the model grid is computed using a simple scaling relation \citep{Kjeldsen_Bedding_1995}, by considering $\nu\ind{max,\odot}=3090\, \mu Hz$. The period spacing can also be included in the list of seismic constraints, in which case it is compared with the asymptotic value of $\Delta\Pi_1$ coming out of the models \citep[see][]{Rodrigues_2017}. The grid of stellar evolution models used in \texttt{PARAM} is the same as the reference grid adopted in \citet{Khan_2019} and \citet{Miglio_2021} (G2).

\begin{figure}
\includegraphics[width=.495\textwidth]{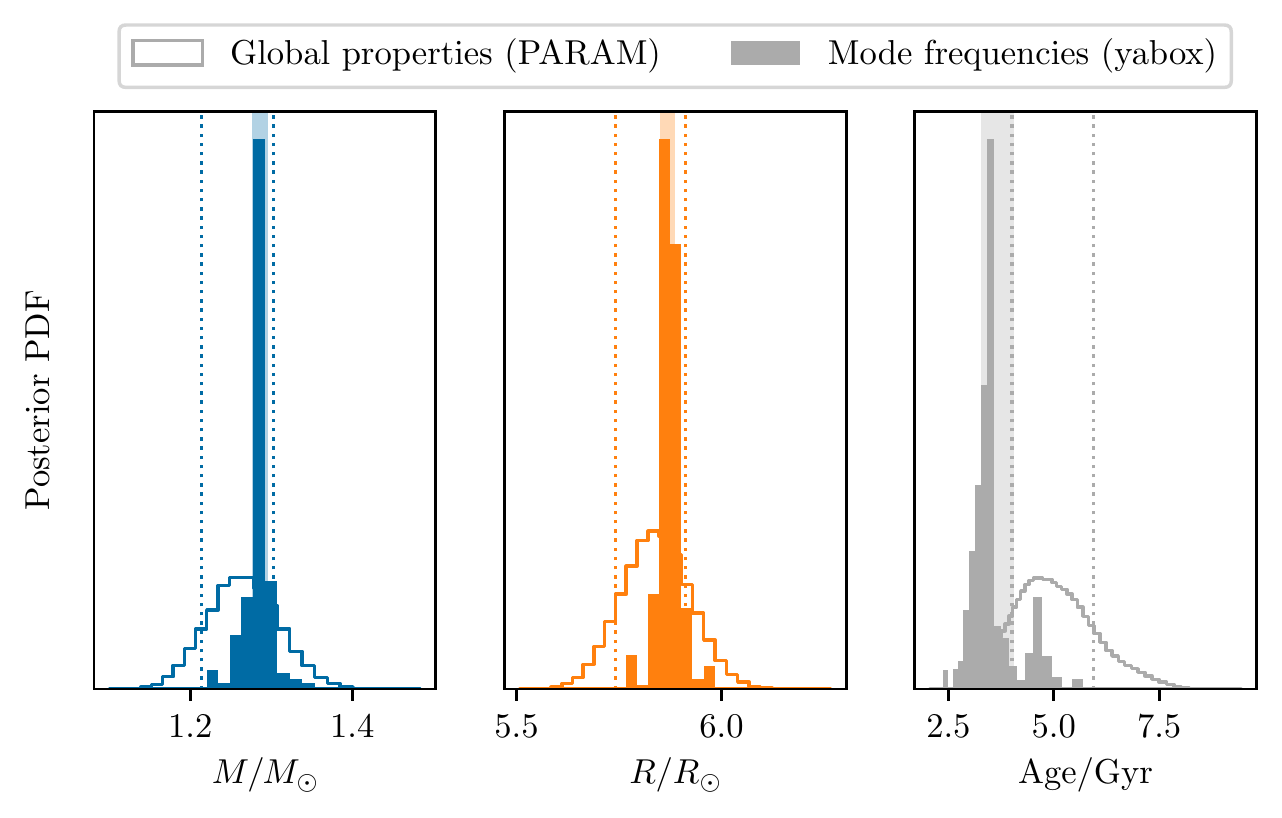}
\caption{Mass, radius and age posterior probability distribution functions derived from seismic constraints. Empty histograms show the distributions obtained using \texttt{PARAM}, and the dotted lines show their associated $\pm1\sigma$ credible regions. Solid histograms and shaded regions indicate the same quantities obtained using the individual mode frequencies.}
\label{fig_PARAM}
\end{figure}

We used the individual radial-mode frequencies in Table \ref{tab:peak_bagging} to compute an average $\Delta\nu$ that can be directly compared with that from the model grid, and obtain a value ($\Delta\nu=10.48$) compatible with that obtained using the EACF (see Sec. \ref{sect_astero}). With the observational constraints $\Delta\nu$, $\nu\ind{max}$, $T\ind{eff}$, and [M/H] we obtain $R\ind{A}=5.84 \pm 0.09\,R_\odot$, $M\ind{A}=1.27 \pm 0.05\,M_\odot$, and an age of $4.9\pm0.9$\,Gyr. In Fig. \ref{fig_PARAM} we show the posterior probability density function (PDF) of radius, mass and age, compared with the same resulting from the asteroseismic modeling based on individual frequencies that is exposed in the next section (Sect. \ref{sec:astero_indi_freq}). 

\subsection{Red-giant model based on the oscillation frequencies}
\label{sec:astero_indi_freq}
We generated models using the MESA code \citep{mesa_paper_1,mesa_paper_2,mesa_paper_3,mesa_paper_4,mesa_paper_5} with diffusion and settling of heavy elements, with no core overshoot, no mass loss, using an Eddington-grey atmosphere, and with chemical abundances evaluated relative to the solar values of \cite{grevesse_standard_1998}.

The nonradial mixed modes in red giants can be described as combinations of pure p- and g-modes --- or "$\pi$" and "$\gamma$-modes", in the sense of \cite{aizenman_avoided_1977}. We evaluate these pure $\pi$- and $\gamma$-mode frequencies using the modified wave operator construction described in \cite{ong_semianalytic_2020}, which splits the standard wave operator into two complementary p- and g-mode wave operators, each supporting only one type of wave propagation. We solve for these  mode frequencies numerically with the pulsation code GYRE \citep{townsend_gyre_2013}. The observed p-dominated mixed modes of degree $l \ge 2$ in red giants are very well approximated as pure $\pi$-modes, and as such we match the model $\pi$-modes against the observed modes of degree $l \ge 2$. Likewise, dipole modes far from the p-dominated mixed modes have period spacings close to those of the pure $\gamma$-modes; accordingly, rather than evaluating $\Delta\Pi_{1}$ using its asymptotic value, we compute it directly from the pure dipole $\gamma$-modes.

The mode frequencies returned from stellar models are known to be systematically offset from those which would be obtained from stars of identical interior structure, owing to systematic errors in the modeling of stellar surfaces --- this is referred to as the asteroseismic "surface term". As such, the model frequencies cannot be used directly to construct likelihood functions against the observed mode frequencies in the usual fashion. Instead, we use a nonparametric description --- ``$\epsilon$-matching'', via the algorithm of \cite{roxburgh_asteroseismic_2016} --- to characterize the discrepancy between the model and observed mode frequencies in a surface-independent fashion. For this purpose we use only the model $l = 0, 2, 3$ frequencies, since those modes which we observe are either radial p-modes, or can be approximated as being entirely p-dominated; here we use the mode frequencies reported by co-author Appourchaux. For the dipole modes, which are more strongly mixed, we used the period spacing only, also as reported by co-author Appourchaux. Since the g-mode cavity is well-localized into the stellar interior, no surface correction is needed.

To supplement these seismic constraints, we used
the spectroscopic constraints from scenario C ([M/H] and $T\ind{eff}$). These
were used to compute log-likelihood functions as
\begin{equation}
\chi^2 = \chi^2\ind{spec} + \left(\chi^2_\epsilon + \chi^2_{\Delta\Pi_1}\right)/f,   
\end{equation}
where $f$ is initially set to 1 when performing the optimization.
Optimization was performed using the differential-evolution algorithm as implemented in the \textit{yabox} package \citep{mier_yabox_2017}, using the initial mass $M$, initial helium and metal mass fractions $Y_i$ and $Z_i$, and the mixing-length efficiency parameter $\alpha_\text{MLT}$ as the independent variables for this optimization. After the optimization was complete, $f$ is set to the minimum value of $\chi^2\ind{seis}$ along the optimization trajectory, in order to ensure that the reported results are not entirely dominated by the seismic quantities, which are significantly more precisely constrained than the spectroscopic ones.

We may consider the optimization trajectory to yield a series of nonuniformly distributed samples of the likelihood function over the input parameter space. As such, if we were to report estimates of stellar properties by naively taking likelihood-weighted averages of the desired quantities over all the sampled points, or find posterior distributions by constructing likelihood-weighted histograms, we would be over-representing parts of the parameter space where the density of samples is high to begin with. Effectively, the sampling function of the optimization trajectory provides an implicit prior distribution, which we estimate using a Gaussian kernel density estimator and divide out to yield a nonuniformly sampled posterior distribution function assuming a uniform prior. Using this uniform-prior posterior distribution, we report the posterior median and $\pm1\sigma$ quantiles of the following of fundamental stellar properties:
\begin{eqnarray}
M\ind{A} &=& 1.281^{+0.015}_{-0.004}\,M_\odot\\
R\ind{A} &=& 5.863^{+0.024}_{-0.011}\,R_\odot\\
Y_i &=& 0.291^{+0.010}_{-0.013}\\
Z_i &=& 0.012^{+0.003}_{-0.001}\\
\alpha\ind{MLT} &=& 1.87^{+0.05}_{-0.09}\\
\rho&=& 0.0089^{+0.00003}_{-0.00003}\, \mbox{g cm}^{-3}\\
\mathrm{age}&=& 3.43^{+0.62}_{-0.16} \mbox{Gyr}
\end{eqnarray}
We compare these results with those obtained with the \texttt{PARAM} code, by plotting the posterior distributions of the mass, radius and age in Fig. \ref{fig_PARAM}. The mass and radius appear to be in agreement with the values obtained from
only $\Delta\nu$, $\nu\ind{max}$, and $\Delta\Pi_1$ with the \texttt{PARAM} routine, while the ages are somewhat different although still within the $1\sigma$ error. We discuss this difference in Sect. \ref{sect_disc_age}.

\subsection{Overall system model based on classical parameters}

\begin{figure*}[t!]
\includegraphics[width=0.5\textwidth]{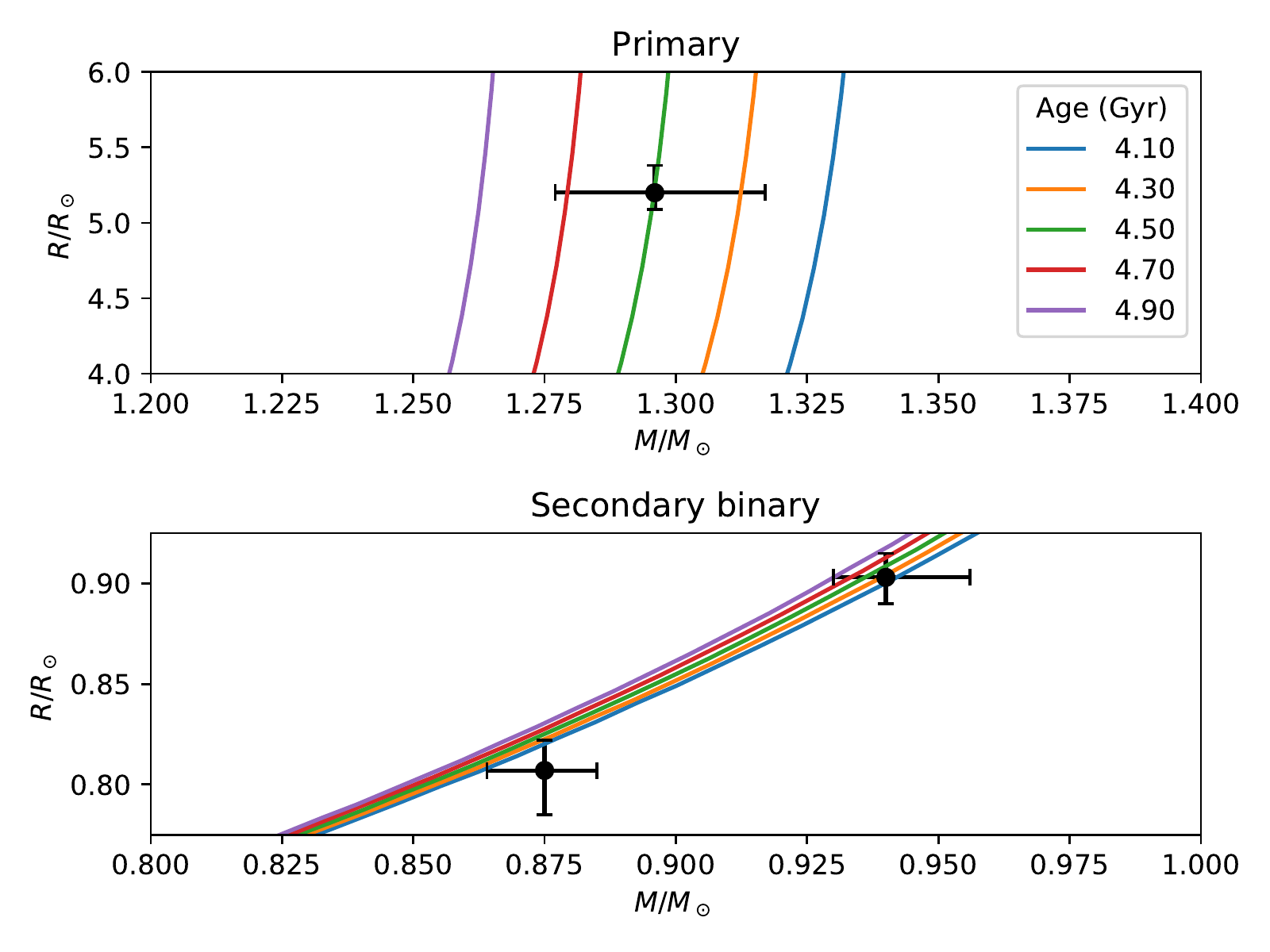}
\includegraphics[width=0.5\textwidth]{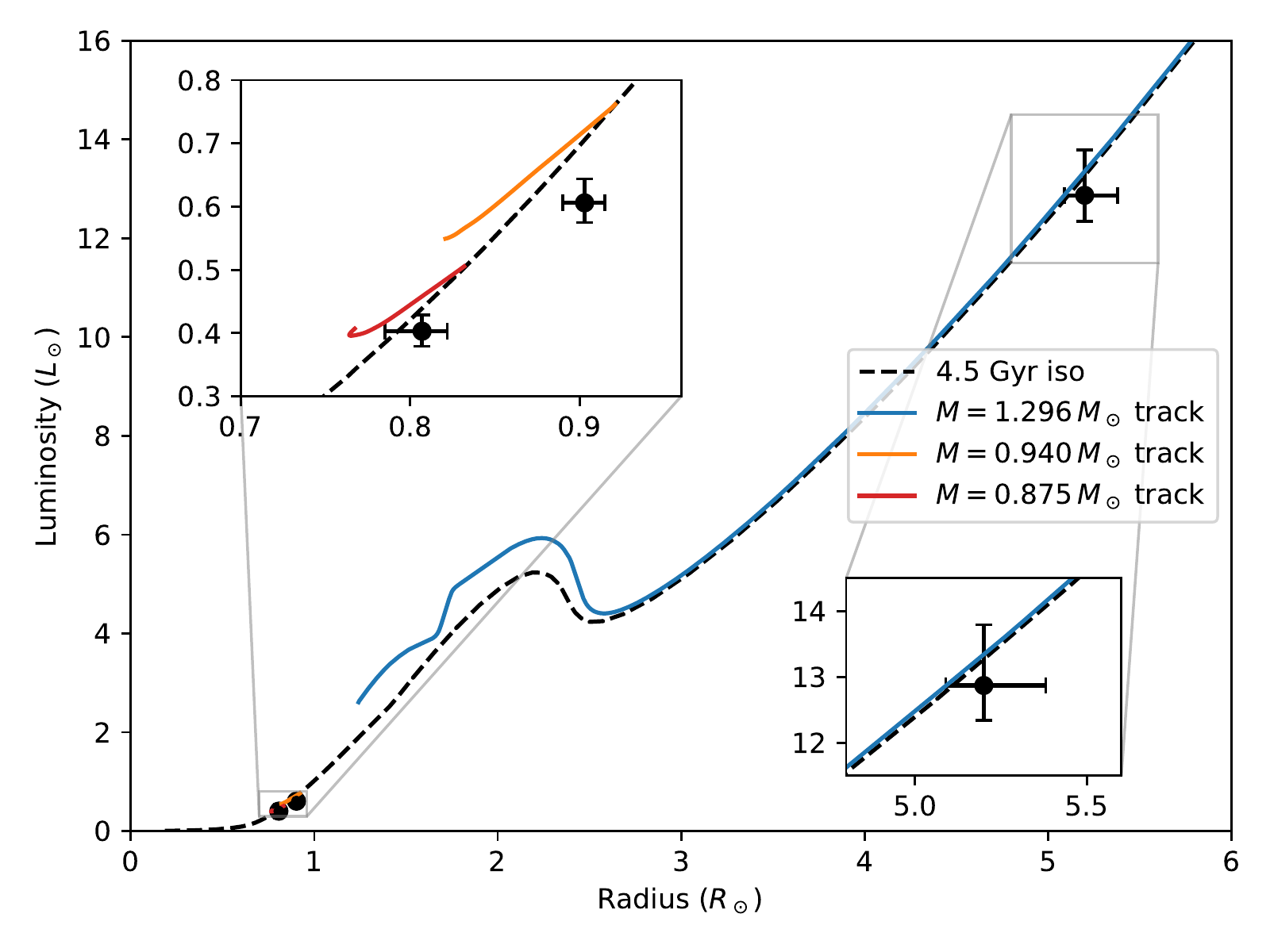}
\caption{Visual fit of observed classical parameters of the system with theoretical tracks and isochrones from the YaPSI database \citep{Spada_ea:2017}. Left: comparison in the mass radius plane with isochrones of different ages for the primary (top panel), and the two components of the inner binary (bottom). Right: comparison in the radius-luminosity plane; a 4.5-Gyr isochrone and evolutionary tracks of appropriate mass and metallicity are shown.}
\label{fig_yapsi}
\end{figure*}

We have compared the classical parameters of the three components of the system with the isochrones and tracks from the YaPSI database \citep{Spada_ea:2017}; the results are summarized in Figure \ref{fig_yapsi}.
In the two panels on the left of the Figure, the measured masses and radii are compared with several isochrones of appropriate metallicity, generated by interpolation from the YaPSI database.
The best fit is obtained for an age of $4.5\pm 0.2$ Gyr for the primary component (top left panel); the components of the inner binary are marginally compatible with this isochrone (bottom left).
In the right panel of Figure \ref{fig_yapsi}, we show the best-fitting isochrone, together with evolutionary tracks constructed with the same code and input physics used for the YaPSI models (see \citealt{Spada_ea:2017} for details). For the primary star, both the isochrone and the track are in good agreement with the observed radius and luminosity (note that the isochrone and track for component A essentially coincide on the RGB, as a result of the fast-paced evolution typical of this evolutionary phase).
The agreement is less good for the two components of the inner binary. 
This could be ascribed to non-standard effects not taken into account in our models.

\begin{figure*}[t!]
\includegraphics[width=9.0cm]{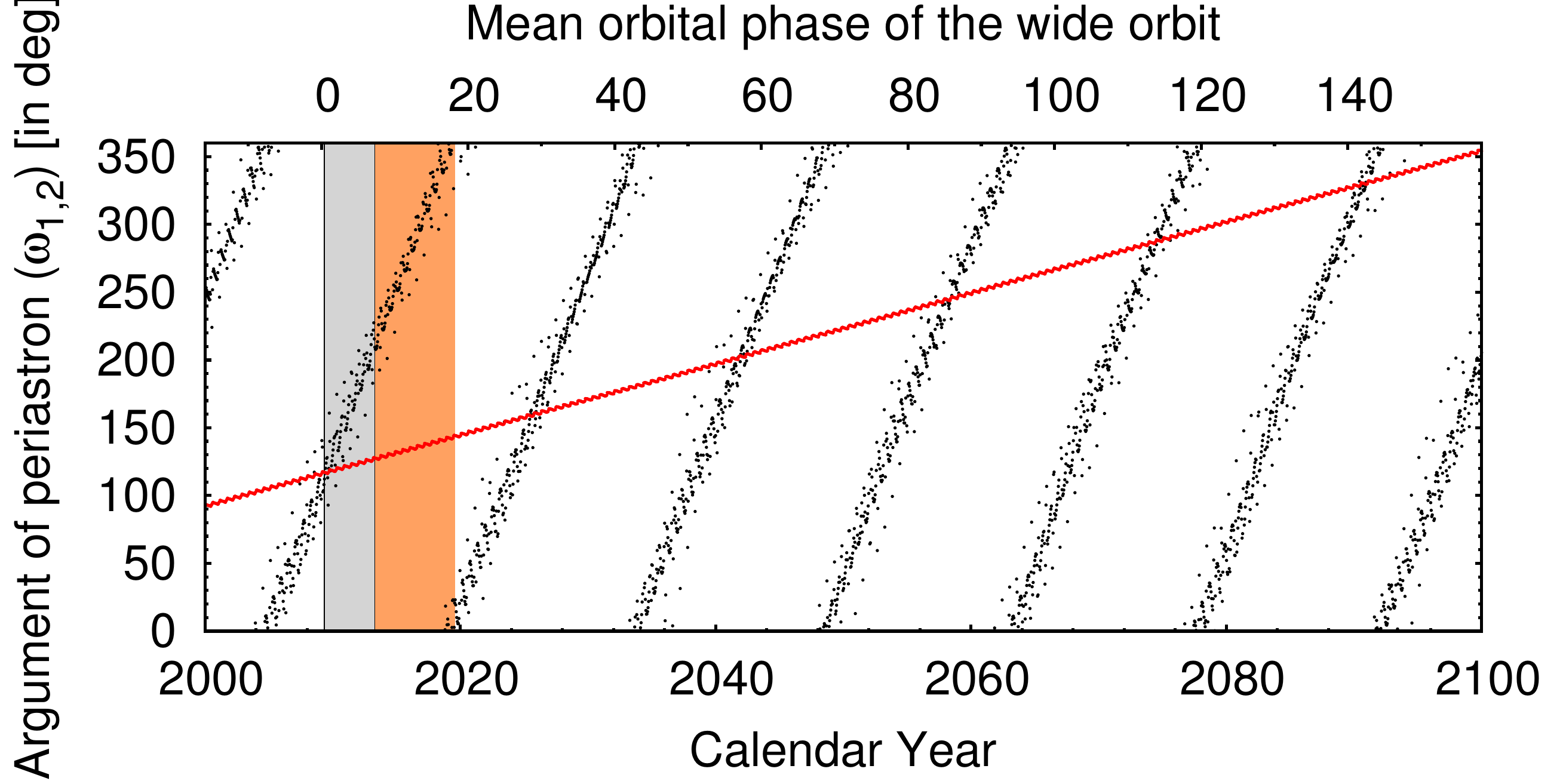}\includegraphics[width=9.0cm]{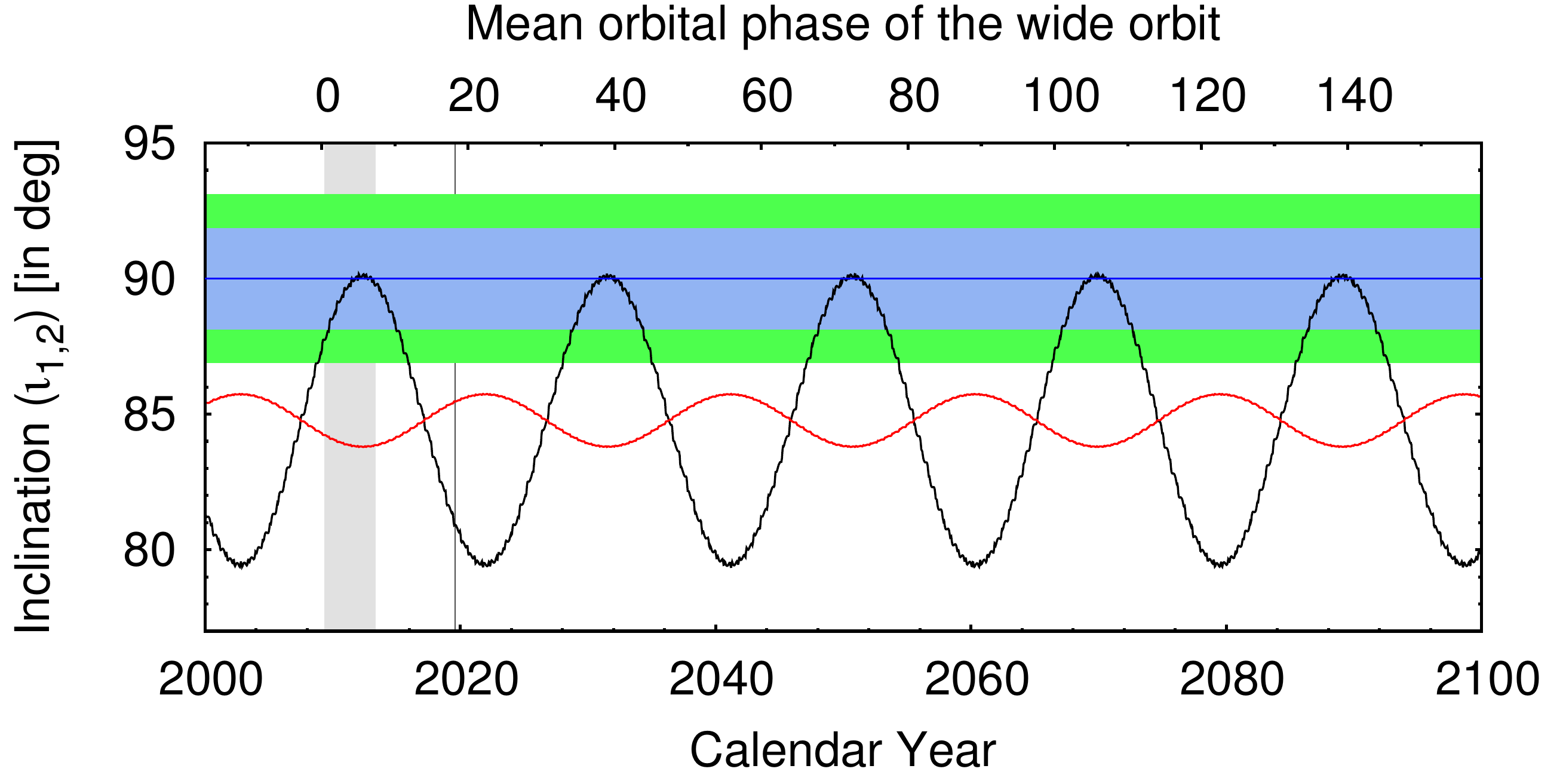}  
\caption{The evolutions of some of the orbital elements on a century-long timescale, obtained via numerical integrations. {\it Left:} The variations of the observable arguments of periastrons ($\omega_{1,2}$) of the inner and outer orbits (black and red, respectively). The dynamically forced apsidal motions of both orbits are well visible. The grey and orange shaded regions mark the intervals of the \textit{Kepler} and the ground-based spectroscopic follow up observations, respectively. {\it Right:} The inclinations of the inner and outer binaries (black and red curves, respectively). The green- and lightblue-shaded horizontal areas denote the inclination domains of the inner and outer orbits where eclipses can occur. The vertical grey area represents the domain of the \textit{Kepler}-observations, while the thin, black, vertical line denotes the interval of the 27-day-long \textit{TESS} observations (Sector 14). As one can see, during these measurements the inner inclination ($i_1$) was far below the green line, therefore, the inner pair did not present any eclipses. (Note, the dark blue line simply stands for $i_{1,2}=90\degr$ to guide the eye.)}
\label{fig_omincl100yr}
\end{figure*}

\begin{figure}
\includegraphics[width=9.3cm]{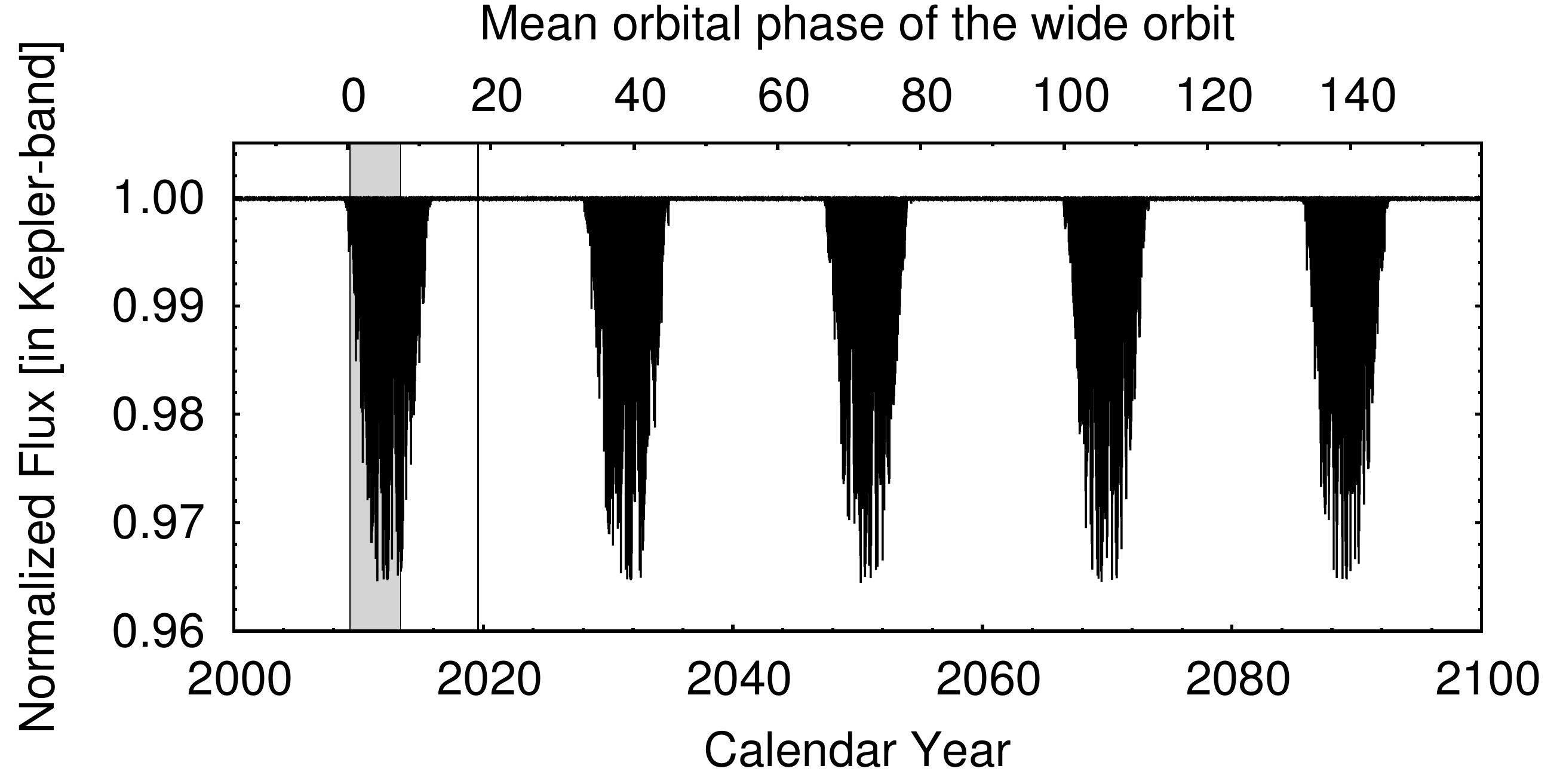}
\caption{Photodynamical model light curve of KIC\,7955301 for the present century. The $\sim19.2$\,yr-period precession cycle is clearly visible. Within a cycle the inner pair exhibits regular eclipses during a $\sim7.3$\,yr-long interval. The last eclipsing session has been finished at 2nd January 2016 with a short, $\sim80$\,ppm amplitude fading, while the next session is expected to begin at 18 January 2028. The grey vertical area represents the time of \textit{Kepler} observations, and \textit{TESS} Sector 14 observations are also denoted with a thin vertical line.}
\label{fig_lc100yr}
\end{figure}

\subsection{Model based on isochrones and SED}
\label{subsect:PARSECSEDetc}
We carried out a second spectro-photodynamical study with the software package {\sc Lightcurvefactory} that includes the lightcurve, ETVs and RVs, as earlier, but also the stellar energy distribution (SED), a stellar evolution code \citep[\texttt{PARSEC},][]{Bressan_2012}, and the \textit{Gaia}-based accurate trigonometric distance. The cumulative SED, which consists of pass-band magnitudes, drives the temperature of sources, or at least of the dominant one. That way, effective temperatures together with stellar masses and radii determine the locations of the stars on co-eval \texttt{PARSEC} evolutionary tracks. Besides the lightcurve, RVs, and ETVs, the publicly available multipassband magnitudes (see Table\,\ref{tbl:mags}) were simultaneously fitted against the theoretical passband magnitudes interpolated from theoretical \texttt{PARSEC} isochrone tables. We refer the reader to \citet{Borkovits_2020} for a detailed description of the method. 

With this new model, the free parameters partially differ from  Sect.\,\ref{sect_dyn_mod}. Firstly, three new free parameters are added: the logarithm of the age; the metallicity $[M/H]$ of the system; the interstellar extinction $E(B-V)$. Furthermore, the distance of the system is constrained a posteriori at each trial step,  by minimizing the value of $\chi^2_\mathrm{SED}$. Secondly, the stellar radii and temperatures that used to be free are no longer fitted. However, the code computes them internally by interpolating the appropriate \texttt{PARSEC} tables according to the trial values of the [mass, age, metallicity] triplets at each trial runs. The results are tabulated in the last columns of Table\,\ref{tab: syntheticfit_KIC7955301}.

The comparison of the results from the purely dynamical analysis, which is independent from stellar evolution models, and the analysis including dynamical analysis, SED, and isochrones, produce output parameters that lie within their 1-$\sigma$ uncertainties. This means that the inclusion of stellar evolution models into the complex analysis did not lead to any unavoided bias in the values of the physical properties of the system. It also means that our model-independent lightcurve solution has led to dimensionless quantities, like fractional radii and temperature ratios, that are fully consistent with stellar evolutionary models. 

The mass of the RG component $M_\mathrm{A}=1.30^{+0.03}_{-0.02}\,M_\odot$  is in excellent agreement with the asteroseismic analysis. The inner binary components appear to be on the main sequence with masses $M_\mathrm{Ba}=0.94_{-0.01}^{+0.02}\,\mathrm{M}_\odot$,  $M_\mathrm{Bb}=0.88\pm0.01\,\mathrm{M}_\odot$, and radii of $R_\mathrm{Ba}=0.90\pm0.01\,\mathrm{R}_\odot$, $R_\mathrm{Bb}=0.81\pm0.02\,\mathrm{R}_\odot$ in both solutions. 

In contrast, the radius of the RG component $R_\mathrm{A}$ and the temperatures $T_\mathrm{Ba}$, $T_\mathrm{Bb}$ of the inner stars differ by about $5-10\%$ between the two models. This is expected because these quantities are not functions of the lightcurve properties and of the dynamics of the system. Only the ratio of the effective temperatures of the inner binary $T_\mathrm{Bb}/T_\mathrm{Ba}$ is model-independent; this one is in good agreement (within $1\%$) between the two models. 
In the purely dynamical model (Sect. \ref{sect_dyn_mod}), the RG temperature was fixed to the available catalog value $T_\mathrm{A}=4800$\,K. Here, the combined dynamical, SED and isochrone model confirms this value by finding $T_\mathrm{A}=4804_{-48}^{+26}$\,K, which within the uncertainty of the value we measured from the disentangled optical spectra (Table \ref{tab:atm_param}). In the dynamical model, the temperature $T_\mathrm{Ba}$ of the primary star of the EB was internally set with the use of the zero age main sequence mass-radius and mass-luminosity relations of \citet{Tout_1996}, and the temperature ratio $T_\mathrm{Bb}/T_\mathrm{Ba}$ was an adjustable parameter. In the combined model they were constrained from the appropriate \texttt{PARSEC} isochrone. This latter model has resulted in higher temperatures by $\sim200$\,K in accordance with the fact that KIC\,7955301 is a quite old system, having an age of $4.5\pm0.2$\,Gyr according to the present model. Finally, the combined model predicts the RG radius to be of $R_\mathrm{A}=5.70_{-0.13}^{+0.20}\,\mathrm{R}_\odot$, which is in agreement with the asteroseismic value.

The stellar metallicity $[M/H]=0.06_{-0.04}^{+0.03}$ and interstellar extinction $E(B-V)=0.09\pm0.02$ obtained from the combined model are consistent with catalog values (see Table\,\ref{tbl:mags}). Finally, the SED fitting provides an estimate of the photometric distance of $d=1369_{-29}^{+44}$\,pc, which is in perfect agreement with trigonometric distance of $d=1375\pm35$\,pc derived from the Gaia EDR3 catalog by \citet{Bailer-Jones_2021}. However, we keep in mind that Gaia EDR3 data are not corrected for binarity and, might suffer from systematic errors for such a triple system.

About the dynamical evolution of the system, the small outer-to-inner period ratio $P_2/P_1\approx13.8$ puts the system among the category of compact hierarchical triple systems. That being said, the small mutual inclination of $i_\mathrm{m}=6\fdg2\pm0\fdg3$ and almost circular inner orbit ($e_1=0.0276\pm0.0001$), the orbital configuration of the system is stable in dynamical timescale. Consequently, the dynamical evolution of the system will be driven by the stellar evolution of the evolved RG component. Therefore, here we restrict our discussion for some short term effects, and their observational consequences.

\begin{figure}[t!]
\includegraphics[width=8.5cm]{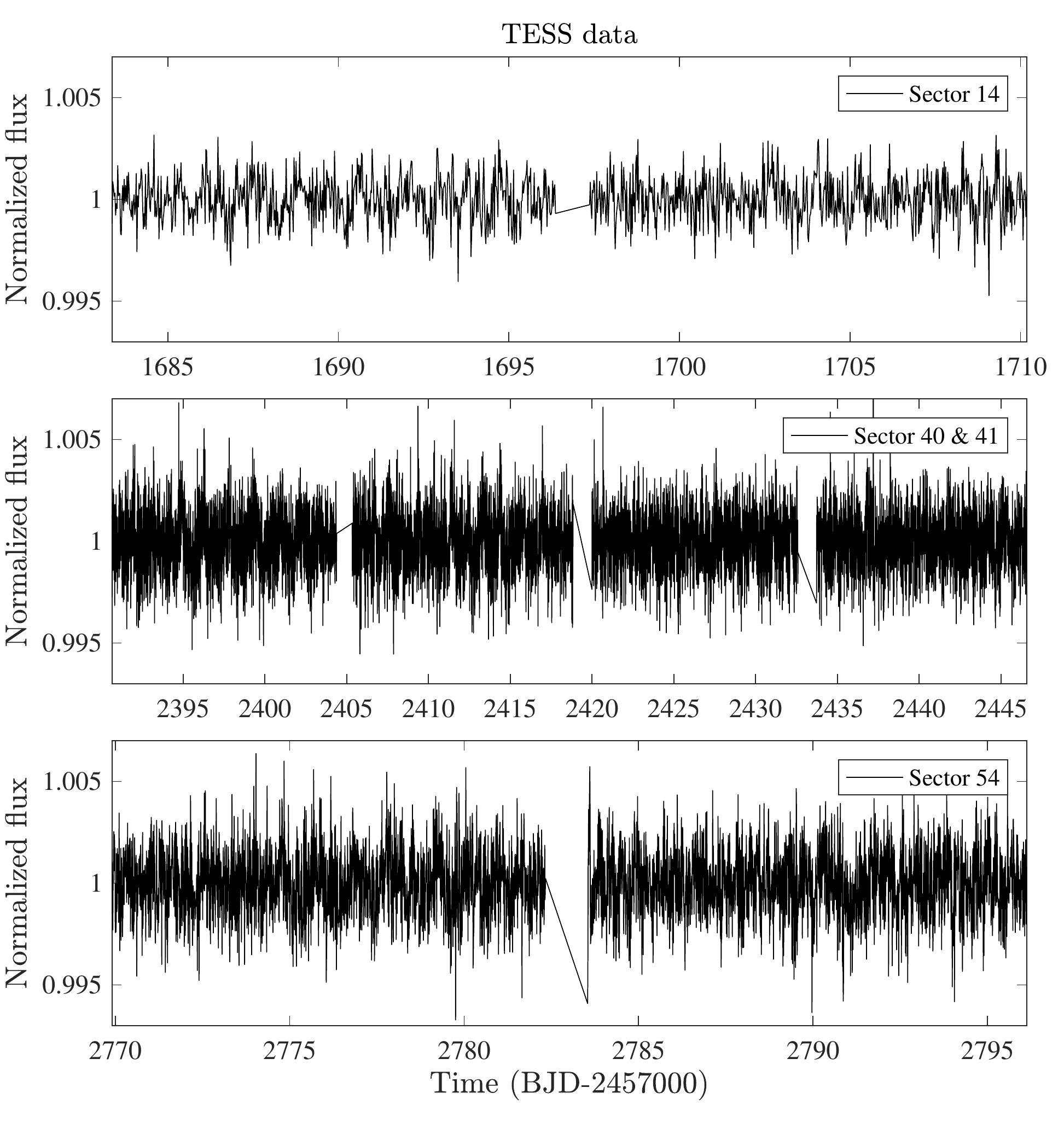}
\includegraphics[width=8.5cm]{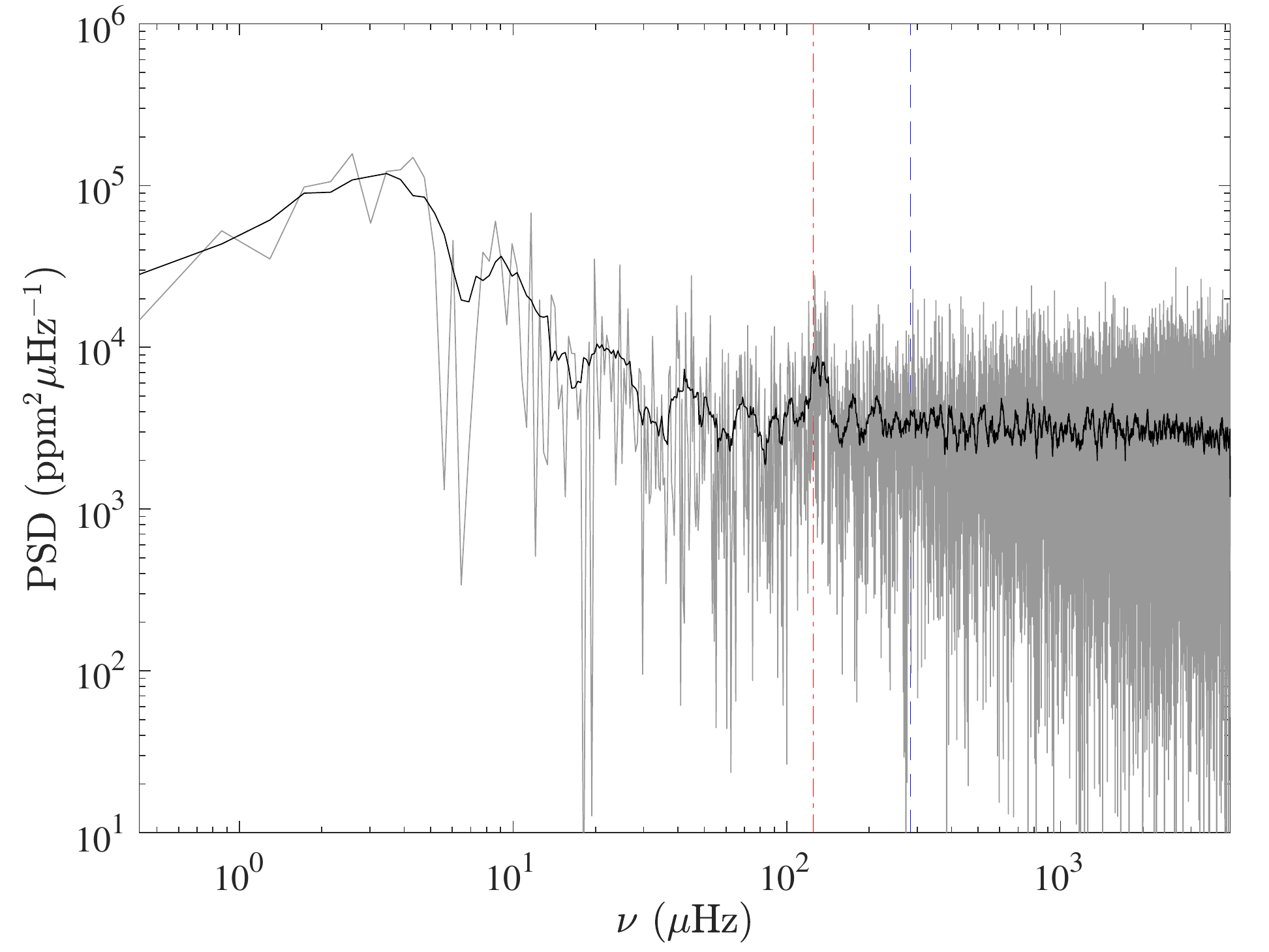} 
\caption{TESS observations of KIC 7955301 from 2019 to 2022. Top panel: time series processed with the software package FITSH \citep{Pal_2012}. Bottom panel: power spectral density of the times series for sector 14 only. The vertical red dot-dashed line indicates $\nu\ind{max}$ and the vertical blue dashed line indicates \kep's Nyquist frequency. 
}
\label{fig_TESS}
\end{figure}

In Fig.\,\ref{fig_omincl100yr} we plot the variations of the observable arguments of periastrons ($\omega_{1,2}$) and  inclinations ($i_{1,2}$) of the inner and outer orbits during the present century. The very fast, dynamically forced apsidal motion can nicely be seen on the ETV curves determined from the 4-yr-long \textit{Kepler}-observations (see Fig.\,\ref{fig_ETVs}). An observationally more dramatic effect caused by precession of the orbital plane of the inner binary is illustrated in Fig.\,\ref{fig_lc100yr}. As the orbital plane precesses with a half amplitude of $i^\mathrm{dyn}_1=5\fdg3\pm0\fdg2$ around the invariable plane of the triple (having an inclination of $i_\mathrm{inv}=84\fdg8\pm0\fdg3$) with a period of $P_\mathrm{node}\approx19.2$\,yr, the inner binary exhibits eclipses (with varying eclipse depths) during a $\approx7.3$\,yr-long interval, while in the remaining $\approx11.9$\,yr the system no longer exhibits eclipses. Fortunately, \textit{Kepler} observations caught the first half of the latest eclipsing session and it led to the discovery of this very exciting system. This eclipsing session has ended at the very beginning of 2016, and KIC\,7955301 will no exhibit any eclipses till the first days of 2028. In accordance with our findings, \textit{TESS} spacecraft did not observe any remarkable light variations of this target when the original \textit{Kepler}-field was revisited three times from 2019 to 2022 (Fig. \ref{fig_TESS}).

\section{Concluding discussion}
\label{sec:discussion}
\subsection{Dynamical versus seismic masses}

With this paper, we illustrate the first detailed analysis of a hierarchical triple system that includes an oscillating RG by combining asteroseismology, spectroscopy and dynamical measurements. This system is an important benchmark for calibrating the measurement of RG masses with asteroseismology. 
Thanks to the acquisition of 23 high-resolution spectra with the ARCES spectrometer at APO we were able to first classify the system as an SB1 and monitor the RV shift of the RG component along its orbit. Then, with the dynamical model solution, we could disentangle and co-add the optical spectra to derive the atmospheric parameters of the three components, even though the parameters of the inner binary have a low S/N. By fixing the RG surface gravity to that found by the complete {\sc Lightcurvefactory} model, we were able to determine the RG temperature and metallicity at $T\ind{A} = 4720\pm105$ K and [M/H]$= -0.01\pm0.12$ (Table \ref{tab:atm_param}).  

We computed the asteroseismic mass of the RG in three different manners. The first and most basic one, which consists of applying the asteroseismic scaling law (SL) tuned for RG stars by \citet{Mosser_2013} from the values of $\nu\ind{max}$, $\Delta\nu$, and $T\ind{A}$, leads to $M\ind{A,SL} = 1.27 \pm 0.05\ M_\odot$. The second makes use of the (\texttt{PARAM}) approach that fits the global asteroseismic parameters, including $\Delta\Pi_1$, and by including the metallicity on a grid of stellar-evolution models. The mass appears to be the exact same as with the asteroseismic scaling relation: $M\ind{A,\texttt{PARAM}}=1.27 \pm 0.05\,M_\odot$. Finally, the most sophisticated approach based on fitting individual oscillation frequencies leads to a mass of $M\ind{A} = 1.281^{+0.015}_{-0.004}\,M_\odot$.  

The RG mass that we derive from the purely dynamical modeling -- independent from both asteroseismology and stellar evolution models -- is $1.30^{+0.03}_{-0.02}\,M_\odot$. The uncertainty is thus  about 2\,\%, which is better than what can be obtained from asteroseismology or stellar evolution models based on atmospheric parameters, luminosity and parallaxes. This type of system,  a solar-like oscillator in an SB1 hierarchical triple system, should definitely be considered for helping to calibrate the asteroseismic masses. About ten of them are available in \citet{Gaulme_2013} and \citet{Gaulme_Guzik_2019}, even though none of them exhibit such clear ETVs, which means that a 2\,\% accuracy on mass may not be achievable with those. 

It is interesting to notice that in this system the asteroseismic mass matches the dynamical one, contrary to what was observed with most RGs in EBs \citep{Gaulme_2016a,Brogaard_2018,Benbakoura_2021}, where asteroseismology appeared to overestimate masses by about 15\,\% on average. This puzzling fact may just be a case that the sample of RG oscillators with accurate independent mass estimates is small and the dispersion of the observed overestimation is large, but it could also highlight issues with RGs in EBs where the companion is a MS dwarf star. For an RG in an EB, the mass of the RG is driven by the low S/N Doppler shift that is measured by tracking the absorption lines of the companion star. In most cases, the flux coming from the companion is less than 5\,\% of the total flux. We also note that a good agreement between dynamical and seismic masses was found with the double RG KIC 9246715 \citep{Rawls_2016}, where there were no issues of noisy RVs. 

In contrast, KIC 7955301 is a special target with respect to the bulk of RGs in EBs because it is the least evolved and  thus smallest RG ($\approx5.85\,R_\odot$) for which we have an independent mass measurement. The radii of all of the RGs in EBs listed in \citet{Benbakoura_2021}, for example, range from 8 to $14 R_\odot$. In addition a significant fraction of them are red clump stars. So far, no significant  overestimation of stellar masses and radii by asteroseismology was observed for MS and subgiant stars. We may have an intermediate case here, for which asteroseismic measurements are unbiased.  To clarify the question, our recommendation is to take new RV measurements of the RG/EBs listed in \citet{Benbakoura_2021} and \citet{Gaulme_Guzik_2019}  with larger S/N ratios, that is longer exposure times, and with spectrometers that are dedicated to high-precision RV measurements. 

\subsection{Age of the system}
\label{sect_disc_age}
One of the goals of the future ESA PLATO mission \citep{Rauer_2014} is to provide ages of solar-like MS stars hosting exoplanets with an accuracy of 10\,\%. Many papers dealing with stellar physics report ages arising from stellar evolution codes with a similar precision, especially when asteroseismic measurements are available. With this paper, we could have run a single stellar evolution code and provided an age with a relatively low uncertainty. We decided to let four different modelers estimate the age in an independent fashion. We obtain $4.9\pm0.9$\,Gyr with \texttt{PARAM}, $3.43^{+0.62}_{-0.16}$\,Gyr with MESA, $4.5\pm0.2$\,Gyr with YaPSI, and $4.5\pm0.2$\,Gyr with {\sc Lightcurvefactory}, that is 18.4, 22.7, 4.4, and 4.4\,\% respectively. Both the YaPSI and {\sc Lightcurvefactory} estimates have a better precision likely because the three components of the system are included into the optimization process. Ages determined for the RG only (component A), with the help of asteroseismic and classical parameters, have uncertainties of about 20\,\%, despite RGs are usually considered to be good cases for age estimate thanks to their fast evolution along the RG branch.

Our age estimate thus ranges from 3.3 to 5.8\,Gyr, which is quite broad given the unusually vast amount of information that we have. By looking more closely, the age is consistent between the three ``standard'' approaches that use the classical parameters (mass, metallicity, effective temperature), and the global asteroseismic parameters (PARAM only). On the contrary, the model based on reproducing the individual oscillation frequencies with MESA stands apart. Even though a deeper understanding of the differences between the age estimates goes beyond the scope of the paper, we can already draw some conclusions. A careful inspection of the posterior probability distribution (PDF) of the mass, radius and age (Fig. \ref{fig_PARAM}) shows a good agreement between the mass and radius obtained with PARAM and MESA, whereas the age PDF appears to be bimodal with MESA. In fact,  the secondary peak of the age distribution matches the PARAM estimate (at $\approx 4.5$ Gyr). A key aspect of the MESA-model based optimization is that the helium abundance $Y$ and metallicity $Z$ are two free independent parameters of the model. The resulting $Y$ value of 0.29 is quite large by considering the slightly sub-solar value found for the metallicity ($Z\approx 0.012$) and assuming a linear $Y=Y(Z)$ relation, and that the Sun falls on that relation \citep[see Fig A.4][]{Miglio_2021}. A difference of about 0.02 in $Y$ with respect to what we have in the grid used in PARAM, would mean a difference of $\sim15\,\%$ in age which, considering also the slight difference in mass, could explain the different age estimate.

\subsection{Formation of the system}
Physical properties and orbital configuration carry information about the formation process of the systems. We refer the reader to \citet{Borkovits_2022} and references therein, where a connection between dynamical and orbital parameters and formation scenarios is discussed. An outer eccentricity of 0.27 for a 209-day period outer orbit is far from being exceptional. For example, \citet{Borkovits_2016} reported 16 triple star candidates with smaller outer period than that of KIC 7955301 and, 6 of them have larger outer eccentricities. Since then, further highly eccentric systems amongst substantially more compact triply eclipsing triples were found in TESS data \citep{Borkovits_2022}. 

With an outer semi-major axis less than 10 AU, KIC 7955302 is considered to be a compact triple system. The formation of such type of systems involves either a fragmentation of the accretion disk, or a disk-mediated capture of the outer component. In both cases the outer star forms by being less massive than the inner binary. Then, thanks to its wider and eccentric orbit it sweeps out a larger orbit and thus accretes most of the infalling material from the accretion disk, which eventually leads $q\ind{out}$ toward larger values. Although it is expected that the total mass of the inner pair remains larger than the mass of the wide component by the end of the complete formation process, the third component may actually become the most massive star of the system \citep[][]{Tokovinin_2021}, which is the case of KIC 7955301.

\subsection{Evolution of the system}

In the short term, we have shown that the inner binary components eclipse each other over a period of about 7.3 years, and stop eclipsing for the next 11.9 years. The RG will never eclipse the inner binary as seen from the Earth. This is why we were able to study the eclipses of the inner binary from the \textit{Kepler} data and that we are not able to do so with TESS. The eclipses should be visible again during the ESA PLATO mission, whose nominal operations should happen between 2027 and 2031. The PLATO observations will certainly help fine-tune the model of the system. 

Over the long term and in general, stars are likely to merge in close multiple-star systems when the most massive component reaches the tip of the RG branch. 
This is especially true for low-mass stars that ignite helium only at the tip of the red giant branch, after reaching very large radii ($\sim200\ R_\odot$). 
RGs may swallow planets or stars, converting their orbital momentum into spin and friction loss. There is indirect observational evidence for it. For example, about 3\,\% of the solar-mass stars on the red-giant branch and 11\,\% of the helium-burning stars observed by \kep display unusually fast rotation and magnetic fields \citep[e.g.,][]{Tayar_2015,Ceillier_2017,Tayar_Pinsonneault_2018, Gaulme_2020}. This means that a fraction of the RGs in that mass range gains angular momentum between the RGB and the RC.  Until alternative explanations are provided, a likely possibility is that these stars have engulfed a stellar or substellar companion.  Recently, \citet{Price-Whelan_2020} reported a notable dearth of close companions around the red clump --
where companions may have been engulfed when the primary star ascended
the upper giant branch -- in the color-magnitude diagram produced by the APO Galactic Evolution Experiment \citep[APOGEE,][]{Majewski_2017}, which is part of the Sloan Digital Sky Survey IV \citep[e.g.,][]{Blanton_2017}.
The orbital configuration of KIC 7955301 represents a borderline case for this type of scenario. The current distance at periastron between the giant and the inner binary is about $\approx142\,R_\odot$ according to the data from Table \ref{tab: syntheticfit_KIC7955301}. 


\begin{figure}[t!]
\includegraphics[width=8.7cm]{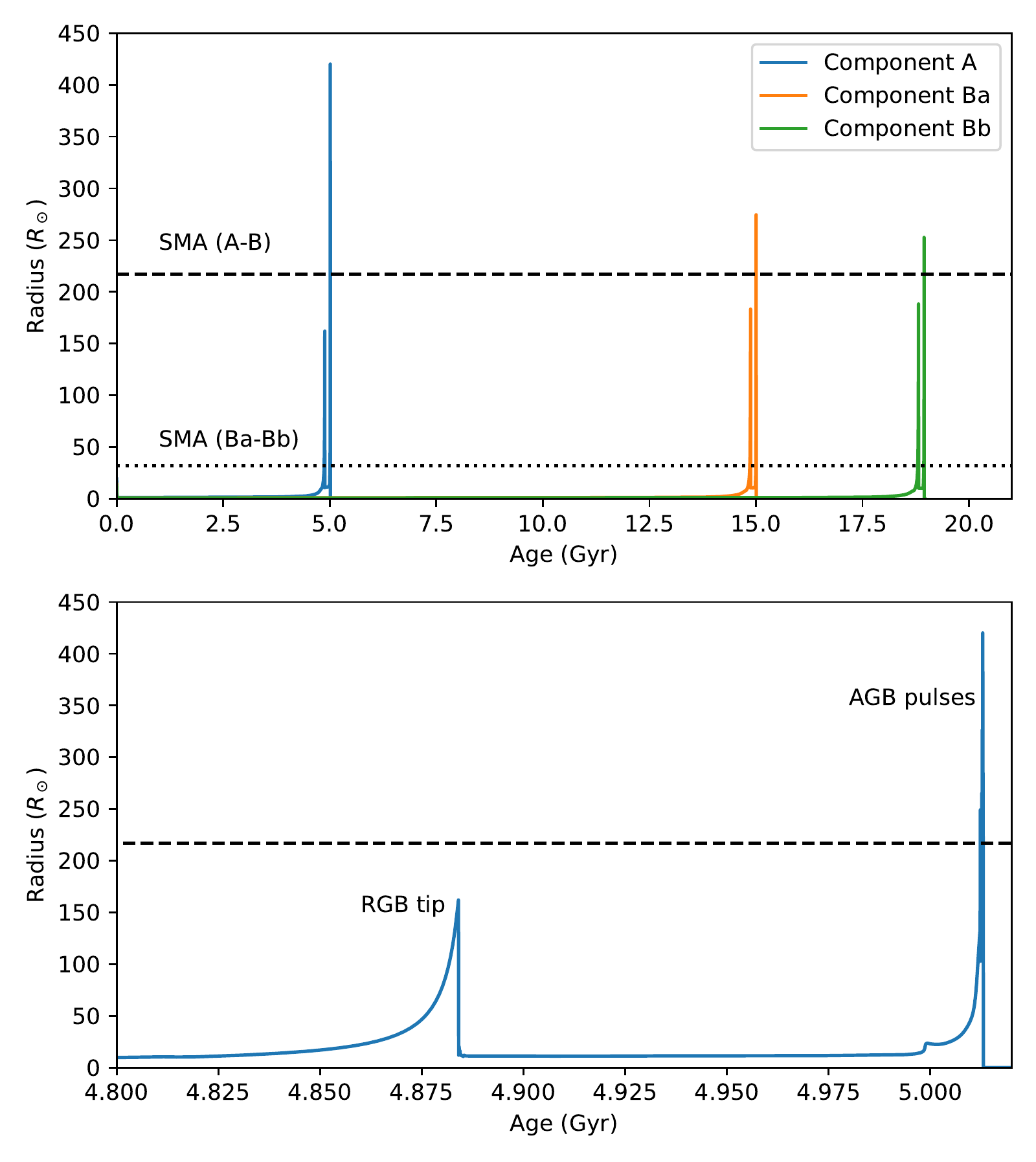}
\caption{Top panel: radius evolution of the three components according to the MIST tracks (with masses and metallicity set equal to their nominal values from Table \ref{tab: syntheticfit_KIC7955301}). The semi-major axis of the inner binary is indicated by the black dotted line, and the semi-major axis of the outer orbit with dashed lines. Bottom panel: zoom on the RG component around its RGB and AGB phases.}
\label{fig_mist}
\end{figure}

We can formulate some broad predictions for the long-term evolution of the system, assuming that the dynamical interaction among the components can be neglected until they come into contact with each other because of post-MS evolution. 
Figure \ref{fig_mist} shows the evolution of the radii of the three components according to evolutionary tracks generated with the MIST web interpolator \citep{Choi_ea:2016}, setting the masses to the central values given in Table \ref{tab: syntheticfit_KIC7955301} (1.29, 0.94, and 0.88 $M_\odot$ for components A, Ba, and Bb, respectively), and the metallicity to +0.06. Significant mass loss takes place during the RGB and AGB phases, and as a result all three stars reach a final mass of about $0.5\, M_\odot$ by the end of their respective AGB phases. 
Stellar rotation is not included in these models.
At an age of approximately 5 Gyr, component A (already a red giant at present time)
reaches the RGB tip, and shortly after (relative to the $10^9$ yr time scale) the AGB tip and thermal pulses phase. The peak radius reached at the tip of the RGB ($\approx 160\, R_\odot$) is mostly equal to the  distance at periastron, while the peak radius on some of the AGB pulses is significantly larger than the semi-major axis of the outer orbit, and thus merging can be expected to occur at one of those times. Should merging of component A with the inner binary be avoided because of some process not considered in this crude analysis, at around 15 Gyr component Ba will evolve onto the RGB and merge with Bb. In the light of this discussion, it is comforting that none of our age estimates is larger than 5 Gyr.


\begin{acknowledgements}
This paper includes data collected by the \textit{Kepler} mission. Funding for the \textit{Kepler} mission is provided by the NASA Science Mission directorate. Some of the data presented in this paper were obtained from the Mikulski Archive for Space Telescopes (MAST). STScI is operated by the Association of Universities for Research in Astronomy, Inc., under NASA contract NAS5-26555. This work is also based on observations obtained with the Apache Point Observatory 3.5-meter telescope, which is owned and operated by the Astrophysical Research Consortium. Part of our spectroscopic observations were done at the Observatoire de Haute-Provence. We acknowledge the technical team for their onsite support as well as the ``Programme National de Physique Stellaire'' (PNPS) of CNRS/INSU for their financial support. P.G. and F.S. were supported by the German space agency (Deutsches Zentrum für Luft- und Raumfahrt) under PLATO data grant 50OO1501. P.G. and J.J. acknowledge NASA grant NNX17AF74G for partial support.
The research leading to these results has (partially) received funding from the KU~Leuven Research Council (grant C16/18/005: PARADISE) and from the BELgian federal Science Policy Office (BELSPO) through PRODEX grant PLATO. M.B. acknowledges support from the ``Deutscher Akademischer Austauschdienst'' (DAAD) and the Université Paris Diderot. 
\end{acknowledgements}

\bibliographystyle{aa} 
 \bibliography{bibi_short.bib}





\appendix
\section{Radial velocity data}
Table~\ref{tab:radial_velocities} contains the radial velocities obtained at Apache Point and Haute-Provence observatories. 

\begin{table}
\caption{Radial velocity data obtained at Apache Point and Haute-Provence observatories. 
}
\label{tab:radial_velocities}
\centering
\begin{tabular}{ll}
\hline\hline
Date & RV \\
(KJD) & km s$^{-1}$  \\
\hline
1569.823526 & -28.82 \\
1591.832861 & -12.50 \\
1611.826703 &  -1.44 \\
1648.831533 &   9.92 \\
1704.641327 & -18.59 \\
1711.645679 & -30.04 \\
1737.619563 & -53.19 \\
1766.572107 & -36.50 \\
1958.878300 & -48.28 \\
3090.825370 &   4.65 \\
3189.617802 & -50.82 \\
3195.587236 & -53.95 \\
3203.595970 & -52.70 \\
3216.717021 & -45.36 \\
3217.631558 & -43.57 \\
3219.657836 & -42.14 \\
3416.917213 & -49.39 \\
3444.847760 & -28.94 \\
3445.534184$^\star$ & -27.26 \\
3446.421966$^\star$ & -26.49 \\
3542.669669 &   8.53 \\
3557.755987 &   3.53 \\
3559.754443 &   3.12 \\
3568.344058$^\star$ &  -2.24 \\
3616.635254 & -54.58 \\
3617.597411 & -54.06 \\
\hline
\end{tabular}
\tablefoot{Dates are mid-exposure times expressed in \kep Julian dates (KJD). \kep Julian dates KJD are related to barycentric Julian dates BJD by KJD = BJD $-$ 2,454,833 days. The first spectrum was taken on April 20th, 2013, and the last on November, 28th, 2018. Measurement errors are estimated to be 0.5 km s$^{-1}$. Date with an $\star$ sign are data taken at OHP.}
\end{table}

\section{Oscillation frequencies}
Table \ref{tab:peak_bagging} displays the oscillation frequencies of the RG component, as performed by co-authors Appourchaux and Mosser. 
\longtab[1]{
\begin{longtable}{lrlll|ll|lllll}
\caption{
Oscillation frequencies of KIC 7955301 by co-authors Appourchaux and Mosser (COR method).
The columns report $\nu\ind{as}$ as asymptotic fit ($\mu$Hz); $\nu\ind{obs}$ are the observed peak ($\mu$Hz) with uncertainties;  $nm$ is the proxy of the mixed order ($nm = np-ng$ ; 0 for p-modes); zeta  = zeta factor of mixed modes (Mosser+2015, 2018), and HBR is the height-to-background ratio).
}
\label{tab:peak_bagging}\\
\hline
 \multicolumn{5}{c}{} & \multicolumn{2}{|c|}{Appourchaux}&\multicolumn{5}{c}{Mosser} \\
\hline
$l$ & $m$ & $n$  & $nm\ind{TA}$ & $nm\ind{BM}$ &   $\nu\ind{obs,TA}$ & err  &  $\nu\ind{obs,BM}$  & err & $\nu\ind{as}$ & $\zeta$  & HBR \\
    &     &      &  &  &  $\mu$Hz & $\mu$Hz  &  $\mu$Hz & $\mu$Hz & $\mu$Hz &  & \\
\hline
\endfirsthead
\caption{Continued.} \\
\hline
 \multicolumn{5}{c}{} & \multicolumn{2}{|c|}{Appourchaux}&\multicolumn{5}{c}{Mosser} \\
\hline
$l$ & $m$ & $n$  & $nm\ind{TA}$ & $nm\ind{BM}$ &   $\nu\ind{obs,TA}$ & err  &  $\nu\ind{obs,BM}$  & err & $\nu\ind{as}$ & $\zeta$  & HBR \\
\hline 
\endhead
\hline
\endfoot
\hline
\endlastfoot
\hline

       0 &        0 &        7 &        8 &       87 &    0.057 &      ... &      ... &      ... &      ... &      ... &      ... \\ 
       0 &        0 &        8 &        9 &        0 &   96.918 &    0.027 &   96.905 &    0.063 &   96.593 &    0.000 &     17.4 \\ 
       0 &        0 &        9 &       10 &        0 &  106.921 &    0.009 &  106.917 &    0.009 &  106.920 &    0.000 &     47.4 \\ 
       0 &        0 &       10 &       11 &        0 &  117.396 &    0.013 &  117.402 &    0.018 &  117.323 &    0.000 &    110.2 \\ 
       0 &        0 &       11 &       12 &        0 &  127.851 &    0.167 &  127.753 &    0.013 &  127.802 &    0.000 &    127.9 \\ 
       0 &        0 &       12 &       13 &        0 &  138.268 &    0.018 &  138.230 &    0.027 &  138.357 &    0.000 &     50.2 \\ 
       0 &        0 &       13 &       14 &        0 &  149.021 &    0.004 &  148.943 &    0.013 &  148.987 &    0.000 &     21.1 \\ 
       0 &        0 &       14 &       15 &      ... &  159.718 &    0.059 &      ... &      ... &      ... &      ... &      ... \\ 
       1 &       -1 &      ... &     -136 &      ... &   90.142 &    0.002 &      ... &      ... &      ... &      ... &      ... \\ 
       1 &        1 &      ... &     -132 &      ... &   92.323 &    0.022 &      ... &      ... &      ... &      ... &      ... \\ 
       1 &       -1 &      ... &     -132 &      ... &   92.626 &    0.052 &      ... &      ... &      ... &      ... &      ... \\ 
       1 &        1 &        8 &      ... &     -126 &      ... &      ... &   98.377 &    0.014 &   98.371 &    0.989 &      5.4 \\ 
       1 &        1 &        8 &     -120 &      ... &  101.235 &    0.004 &      ... &      ... &      ... &      ... &      ... \\ 
       1 &       -1 &        8 &     -120 &     -121 &  101.310 &    0.004 &  101.305 &    0.017 &  101.277 &    0.825 &     19.5 \\ 
       1 &       -1 &        8 &     -119 &     -120 &  101.874 &    0.006 &  101.880 &    0.026 &  101.831 &    0.614 &     64.9 \\ 
       1 &        1 &        8 &     -119 &      ... &  101.881 &    0.006 &      ... &      ... &      ... &      ... &      ... \\ 
       1 &        0 &        8 &     -119 &      ... &  102.112 &    0.001 &      ... &      ... &      ... &      ... &      ... \\ 
       1 &        1 &        8 &     -118 &      ... &  102.342 &    0.015 &      ... &      ... &      ... &      ... &      ... \\ 
       1 &       -1 &        8 &     -118 &     -119 &  102.401 &    0.006 &  102.391 &    0.011 &  102.390 &    0.825 &     14.5 \\ 
       1 &       -1 &        8 &     -117 &      ... &  103.061 &    0.002 &      ... &      ... &      ... &      ... &      ... \\ 
       1 &        1 &        8 &     -117 &      ... &  103.103 &    0.005 &      ... &      ... &      ... &      ... &      ... \\ 
       1 &        1 &        8 &      ... &     -118 &      ... &      ... &  103.840 &    0.014 &  103.853 &    0.973 &      7.3 \\ 
       1 &       -1 &      ... &     -113 &      ... &  106.245 &    0.000 &      ... &      ... &      ... &      ... &      ... \\ 
       1 &        1 &      ... &     -113 &      ... &  106.384 &    0.009 &      ... &      ... &      ... &      ... &      ... \\ 
       1 &        1 &      ... &     -112 &      ... &  107.164 &    0.012 &      ... &      ... &      ... &      ... &      ... \\ 
       1 &       -1 &        9 &     -110 &     -111 &  109.059 &    0.001 &  109.051 &    0.012 &  109.066 &    0.985 &     14.1 \\ 
       1 &        1 &        9 &     -109 &     -111 &  109.807 &    0.002 &  109.806 &    0.010 &  109.816 &    0.977 &     20.5 \\ 
       1 &        1 &        9 &     -108 &     -110 &  110.698 &    0.001 &  110.696 &    0.010 &  110.705 &    0.949 &     24.2 \\ 
       1 &       -1 &        9 &     -108 &     -109 &  110.863 &    0.002 &  110.837 &    0.021 &  110.868 &    0.938 &      5.6 \\ 
       1 &        1 &        9 &      ... &     -109 &      ... &      ... &  111.554 &    0.012 &  111.550 &    0.823 &     10.6 \\ 
       1 &       -1 &        9 &     -107 &     -108 &  111.717 &    0.001 &  111.719 &    0.016 &  111.693 &    0.770 &     34.8 \\ 
       1 &        1 &        9 &     -107 &     -108 &  112.305 &    0.013 &  112.254 &    0.035 &  112.188 &    0.567 &     59.1 \\ 
       1 &       -1 &        9 &     -106 &     -107 &  112.369 &    0.016 &  112.325 &    0.016 &  112.299 &    0.584 &     59.3 \\ 
       1 &        0 &        9 &     -118 &     -107 &  102.592 &    0.001 &  112.514 &    0.021 &  112.547 &    0.699 &      9.2 \\ 
       1 &        1 &        9 &     -106 &     -107 &  112.862 &    0.004 &  112.876 &    0.022 &  112.836 &    0.822 &     34.9 \\ 
       1 &       -1 &        9 &     -105 &     -106 &  113.013 &    0.011 &  113.041 &    0.016 &  113.014 &    0.870 &     26.1 \\ 
       1 &        1 &        9 &     -105 &     -106 &  113.699 &    0.010 &  113.703 &    0.012 &  113.718 &    0.949 &     20.3 \\ 
       1 &       -1 &        9 &     -104 &     -105 &  113.913 &    0.007 &  113.931 &    0.011 &  113.933 &    0.959 &      9.1 \\ 
       1 &        1 &        9 &     -104 &     -105 &  114.624 &    0.002 &  114.655 &    0.013 &  114.673 &    0.976 &     15.8 \\ 
       1 &       -1 &        9 &      ... &     -103 &      ... &      ... &  115.938 &    0.014 &  115.914 &    0.986 &     45.1 \\ 
       1 &        1 &        9 &      ... &     -103 &      ... &      ... &  116.647 &    0.017 &  116.668 &    0.988 &      6.1 \\ 
       1 &       -1 &       10 &      ... &     -102 &      ... &      ... &  116.953 &    0.011 &  116.941 &    0.988 &     14.6 \\ 
       1 &       -1 &       10 &      ... &     -101 &      ... &      ... &  117.977 &    0.015 &  117.988 &    0.988 &      5.2 \\ 
       1 &        1 &       10 &      ... &     -101 &      ... &      ... &  118.717 &    0.018 &  118.741 &    0.986 &      5.4 \\ 
       1 &       -1 &       10 &      -99 &     -100 &  119.028 &    0.002 &  119.032 &    0.013 &  119.051 &    0.984 &     31.0 \\ 
       1 &        1 &       10 &      -99 &     -100 &  119.782 &    0.002 &  119.779 &    0.015 &  119.801 &    0.978 &      9.1 \\ 
       1 &       -1 &       10 &      ... &      -99 &      ... &      ... &  120.157 &    0.019 &  120.127 &    0.974 &      9.3 \\ 
       1 &        1 &       10 &      -98 &      -99 &  120.854 &    0.003 &  120.858 &    0.010 &  120.864 &    0.954 &     21.1 \\ 
       1 &       -1 &       10 &      -97 &      -98 &  121.186 &    0.008 &  121.180 &    0.015 &  121.199 &    0.935 &      7.1 \\ 
       1 &        0 &       10 &      ... &      -98 &      ... &      ... &  121.558 &    0.014 &  121.549 &    0.899 &      6.3 \\ 
       1 &        1 &       10 &      -97 &      -98 &  121.893 &    0.004 &  121.897 &    0.011 &  121.881 &    0.831 &     94.6 \\ 
       1 &       -1 &       10 &      -96 &      -97 &  122.198 &    0.004 &  122.196 &    0.015 &  122.170 &    0.717 &    110.0 \\ 
       1 &        0 &       10 &      ... &      -97 &      ... &      ... &  122.385 &    0.020 &  122.417 &    0.586 &     25.2 \\ 
       1 &        1 &       10 &      -96 &      -97 &  122.695 &    0.007 &  122.708 &    0.044 &  122.622 &    0.523 &    275.3 \\ 
       1 &       -1 &       10 &      -95 &      ... &  122.891 &    0.009 &      ... &      ... &      ... &      ... &      ... \\ 
       1 &        0 &       10 &      ... &      -96 &      ... &      ... &  123.030 &    0.027 &  123.078 &    0.707 &     15.7 \\ 
       1 &        1 &       10 &      -95 &      -96 &  123.394 &    0.000 &  123.384 &    0.012 &  123.369 &    0.823 &     45.7 \\ 
       1 &       -1 &       10 &      -94 &      -95 &  123.685 &    0.022 &  123.723 &    0.009 &  123.721 &    0.897 &     35.3 \\ 
       1 &        0 &       10 &      -94 &      -95 &  124.054 &    0.006 &  124.054 &    0.013 &  124.069 &    0.933 &     14.4 \\ 
       1 &        1 &       10 &      -94 &      -95 &  124.409 &    0.001 &  124.416 &    0.010 &  124.430 &    0.953 &     72.8 \\ 
       1 &       -1 &       10 &      -93 &      -94 &  124.812 &    0.001 &  124.809 &    0.015 &  124.834 &    0.966 &     28.1 \\ 
       1 &        0 &       10 &      ... &      -94 &      ... &      ... &  125.187 &    0.017 &  125.204 &    0.973 &      5.1 \\ 
       1 &        1 &       10 &      -93 &      -94 &  125.555 &    0.007 &  125.557 &    0.014 &  125.577 &    0.978 &     15.7 \\ 
       1 &       -1 &       10 &      ... &      -93 &      ... &      ... &  125.982 &    0.018 &  126.011 &    0.981 &     10.8 \\ 
       1 &       -1 &       10 &      -91 &      -92 &  127.197 &    0.002 &  127.194 &    0.016 &  127.221 &    0.986 &     30.5 \\ 
       1 &        0 &       10 &      ... &      -92 &      ... &      ... &  127.627 &    0.017 &  127.598 &    0.986 &     40.7 \\ 
       1 &        1 &       11 &      ... &      -91 &      ... &      ... &  129.217 &    0.010 &  129.209 &    0.983 &     16.2 \\ 
       1 &       -1 &       11 &      -89 &      -90 &  129.760 &    0.051 &  129.689 &    0.015 &  129.714 &    0.980 &     28.7 \\ 
       1 &        0 &       11 &      -89 &      ... &  129.919 &    0.000 &      ... &      ... &      ... &      ... &      ... \\ 
       1 &        1 &       11 &      -89 &      -90 &  130.442 &    0.002 &  130.437 &    0.014 &  130.460 &    0.972 &     33.4 \\ 
       1 &       -1 &       11 &      -88 &      -89 &  130.954 &    0.005 &  130.957 &    0.016 &  130.982 &    0.961 &     16.4 \\ 
       1 &        1 &       11 &      -88 &      -89 &  131.653 &    0.026 &  131.697 &    0.012 &  131.704 &    0.924 &      9.5 \\ 
       1 &       -1 &       11 &      -87 &      -88 &  132.191 &    0.001 &  132.193 &    0.013 &  132.212 &    0.854 &    101.6 \\ 
       1 &        0 &       11 &      -87 &      -88 &  132.850 &    0.024 &  132.531 &    0.015 &  132.518 &    0.760 &      7.4 \\ 
       1 &        1 &       11 &      -87 &      -88 &  132.800 &    0.009 &  132.807 &    0.014 &  132.785 &    0.628 &    114.0 \\ 
       1 &       -1 &       11 &      -86 &      -87 &  133.083 &    0.017 &  133.051 &    0.025 &  133.095 &    0.486 &     27.4 \\ 
       1 &        0 &       11 &      ... &      -87 &      ... &      ... &  133.137 &    0.075 &  133.286 &    0.509 &     22.9 \\ 
       1 &        1 &       11 &      -86 &      -87 &  133.524 &    0.030 &  133.523 &    0.017 &  133.494 &    0.613 &     71.9 \\ 
       1 &       -1 &       11 &      -85 &      -86 &  133.923 &    0.001 &  133.924 &    0.010 &  133.934 &    0.811 &    101.8 \\ 
       1 &        1 &       11 &      -85 &      -86 &  134.598 &    0.003 &  134.593 &    0.011 &  134.608 &    0.923 &     66.9 \\ 
       1 &       -1 &       11 &      -84 &      -85 &  135.179 &    0.004 &  135.176 &    0.014 &  135.191 &    0.955 &      9.3 \\ 
       1 &        1 &       11 &      ... &      -85 &      ... &      ... &  135.924 &    0.010 &  135.928 &    0.972 &     11.9 \\ 
       1 &        1 &      ... &      -83 &      ... &  137.527 &    0.003 &      ... &      ... &      ... &      ... &      ... \\ 
       1 &       -1 &      ... &      -82 &      ... &  137.917 &    0.409 &      ... &      ... &      ... &      ... &      ... \\ 
       1 &        1 &       12 &      -82 &      -83 &  138.722 &    0.003 &  138.718 &    0.014 &  138.733 &    0.983 &      8.5 \\ 
       1 &       -1 &       12 &      -81 &      -82 &  139.410 &    0.004 &  139.411 &    0.015 &  139.432 &    0.982 &     12.3 \\ 
       1 &       -1 &       12 &      -80 &      -81 &  140.891 &    0.002 &  140.890 &    0.012 &  140.905 &    0.970 &     18.0 \\ 
       1 &        0 &       12 &      ... &      -81 &      ... &      ... &  141.292 &    0.016 &  141.274 &    0.963 &      6.1 \\ 
       1 &       -1 &       12 &      -79 &      -80 &  142.366 &    0.007 &  142.362 &    0.014 &  142.356 &    0.906 &      5.7 \\ 
       1 &        1 &       12 &      -79 &      -80 &  143.050 &    0.024 &  142.984 &    0.018 &  143.008 &    0.770 &      8.7 \\ 
       1 &       -1 &       12 &      -78 &      -79 &  143.555 &    0.099 &  143.527 &    0.014 &  143.515 &    0.512 &     18.1 \\ 
       1 &        0 &       12 &      ... &      -79 &      ... &      ... &  143.527 &    0.087 &  143.698 &    0.447 &     18.1 \\ 
       1 &        1 &       12 &      -78 &      -79 &  143.940 &    0.003 &  143.882 &    0.013 &  143.874 &    0.469 &     21.5 \\ 
       1 &       -1 &       12 &      -77 &      -78 &  144.373 &    0.001 &  144.362 &    0.011 &  144.351 &    0.719 &     26.7 \\ 
       1 &        0 &       12 &      ... &      -78 &      ... &      ... &  144.661 &    0.014 &  144.643 &    0.820 &     14.1 \\ 
       1 &        1 &       12 &      -77 &      -78 &  144.972 &    0.002 &  144.968 &    0.012 &  144.963 &    0.882 &      8.9 \\ 
       1 &       -1 &       12 &      ... &      -76 &      ... &      ... &  147.329 &    0.013 &  147.339 &    0.976 &      7.8 \\ 
       1 &        1 &       12 &      ... &      -76 &      ... &      ... &  148.101 &    0.013 &  148.086 &    0.980 &     11.5 \\ 
       1 &        0 &       13 &      ... &      -75 &      ... &      ... &  149.337 &    0.019 &  149.362 &    0.981 &      5.0 \\ 
       1 &        1 &       13 &      ... &      -75 &      ... &      ... &  149.754 &    0.014 &  149.736 &    0.980 &      8.8 \\ 
       1 &        0 &       13 &      ... &      -74 &      ... &      ... &  151.029 &    0.015 &  151.044 &    0.972 &      6.1 \\ 
       1 &        1 &       13 &      ... &      -74 &      ... &      ... &  151.407 &    0.013 &  151.415 &    0.968 &      7.5 \\ 
       1 &       -1 &       13 &      ... &      -72 &      ... &      ... &  153.902 &    0.017 &  153.879 &    0.647 &     13.8 \\ 
       1 &        0 &       13 &      -71 &      -72 &  153.925 &    0.010 &  154.185 &    0.046 &  154.102 &    0.518 &      5.8 \\ 
       1 &        1 &       13 &      ... &      -72 &      ... &      ... &  154.398 &    0.060 &  154.283 &    0.431 &      7.9 \\ 
       1 &       -1 &       13 &      ... &      -71 &      ... &      ... &  154.823 &    0.040 &  154.750 &    0.539 &      7.8 \\ 
       1 &        0 &       13 &      ... &      -71 &      ... &      ... &  155.020 &    0.026 &  154.979 &    0.666 &      5.6 \\ 
       1 &        1 &       13 &      ... &      -71 &      ... &      ... &  155.272 &    0.018 &  155.254 &    0.779 &      5.1 \\ 
       2 &        0 &        6 &        7 &      ... &   85.387 &    0.120 &      ... &      ... &      ... &      ... &      ... \\ 
       2 &        0 &        7 &        8 &        0 &   95.395 &    0.036 &   95.449 &    0.135 &   94.778 &    0.000 &      8.4 \\ 
       2 &        0 &        8 &        9 &        0 &  105.537 &    0.026 &  105.579 &    0.095 &  105.106 &    0.000 &     14.0 \\ 
       2 &        0 &        9 &       10 &        0 &  116.058 &    0.004 &  116.072 &    0.112 &  115.512 &    0.000 &     95.1 \\ 
       2 &        0 &       10 &       11 &        0 &  126.474 &    0.018 &  126.462 &    0.095 &  125.988 &    0.000 &     68.9 \\ 
       2 &        0 &       11 &       12 &        0 &  136.935 &    0.028 &  136.986 &    0.089 &  136.544 &    0.000 &     34.0 \\ 
       2 &        0 &       12 &       13 &        0 &  147.637 &    0.035 &  147.676 &    0.102 &  147.170 &    0.000 &     22.7 \\ 
       2 &        0 &       13 &       14 &      ... &  158.462 &    0.099 &      ... &      ... &      ... &      ... &      ... \\ 
       3 &        0 &        9 &        9 &      ... &  119.568 &    0.098 &      ... &      ... &      ... &      ... &      ... \\ 
       3 &        0 &       10 &       10 &      ... &  130.013 &    0.026 &      ... &      ... &      ... &      ... &      ... \\ 
\hline
\end{longtable}
}

\section{Eclipse properties}
Table \ref{tab:eclipse_ppties} presents the eclipses properties obtained with the \textsc{Lightcurvefactory}.

\longtab[1]{
\begin{longtable}{ccccccccc}
\caption{
Eclipse properties.
}
\label{tab:eclipse_ppties}\\
\hline\hline
n          & $t_1$ & $O_1-nP - t_0$ & $D_1$ & $W_1$ & $t_2$ & $O_2-nP - t_0$ & $D_2$ & $W_2$\\ 
           &(KJD) & day          & \%    & hour &(KJD) & day          & \%    & hour \\
\hline
\endfirsthead
\caption{Continued.} \\
\hline\hline
n          & $t_1$ & $O_1-nP - t_0$ & $D_1$ & $W_1$ & $t_2$ & $O_2-nP - t_0$ & $D_2$ & $W_2$\\ 
           &(KJD) & day          & \%    & hour &(KJD) & day          & \%    & hour \\
\hline
\endhead
\hline
\endfoot
\hline
\endlastfoot
\hline\hline
\hline
8 & 127.4560 & 4.4034 & -0.55 & 2.10 & 134.9905 & 11.9379 & -0.50 & 2.24 \\ 
9 & 142.8022 & 4.3911 & -0.52 & 2.15 & 150.3368 & 11.9121 & -0.51 & 2.20 \\ 
10 & 158.1280 & 4.3778 & -0.53 & 2.32 & ... & ... & ... & ... \\ 
11 & 174.1688 & 4.3612 & -0.62 & 2.48 & 181.7034 & 11.8727 & -0.66 & 2.18 \\ 
12 & 188.7792 & 4.3464 & -0.65 & 2.38 & 196.3137 & 11.8560 & -0.72 & 2.39 \\ 
13 & 204.1046 & 4.3251 & -0.77 & 2.31 & 211.6391 & 11.8481 & -0.88 & 2.53 \\ 
14 & 219.4298 & 4.3058 & -0.84 & 2.40 & 226.9644 & 11.8495 & -0.96 & 2.57 \\ 
15 & 235.7970 & 4.3025 & -0.87 & 2.40 & 243.3315 & 11.8809 & -0.91 & 2.48 \\ 
16 & 250.0798 & 4.4026 & -0.89 & 2.43 & 257.6144 & 11.9343 & -1.01 & 2.57 \\ 
17 & 265.4046 & 4.4707 & -1.15 & 2.67 & 272.9391 & 11.9343 & -1.39 & 2.78 \\ 
18 & ... & ... & ... & ... & 288.2638 & 11.9523 & -1.56 & 2.69 \\ 
19 & 296.0539 & 4.4538 & -1.38 & 2.67 & 303.5884 & 11.9615 & -1.62 & 2.71 \\ 
20 & 311.3988 & 4.4533 & -1.38 & 2.79 & 318.9334 & 11.9532 & -1.58 & 2.80 \\ 
21 & 327.9493 & 4.4437 & -1.40 & 2.79 & 335.4839 & 11.9355 & -1.53 & 2.81 \\ 
22 & 342.0480 & 4.4350 & -1.35 & 2.71 & ... & ... & ... & ... \\ 
23 & 357.3727 & 4.4217 & -1.42 & 2.63 & 364.9072 & 11.8882 & -1.68 & 2.79 \\ 
24 & 372.6975 & 4.4051 & -1.39 & 2.55 & 380.2320 & 11.8657 & -1.71 & 2.70 \\ 
25 & 388.7172 & 4.3921 & -1.52 & 2.81 & 396.2517 & 11.8462 & -1.81 & 2.79 \\ 
26 & 405.2275 & 4.3752 & -1.60 & 2.75 & 412.7621 & 11.8327 & -1.93 & 2.87 \\ 
27 & 418.6730 & 4.3593 & -1.64 & 2.70 & 426.2075 & 11.8260 & -2.04 & 2.79 \\ 
28 & 433.9985 & 4.3395 & -1.75 & 2.65 & 441.5330 & 11.8378 & -1.98 & 2.78 \\ 
29 & 449.3241 & 4.3562 & -1.68 & 2.73 & 456.8586 & 11.8931 & -1.94 & 2.95 \\ 
30 & 464.6703 & 4.4809 & -1.73 & 2.74 & 472.2048 & 11.9169 & -2.15 & 2.94 \\ 
31 & 480.6909 & 4.5004 & -1.97 & 2.85 & 488.2254 & 11.9218 & -2.52 & 2.98 \\ 
32 & 495.3219 & 4.4902 & -2.16 & 2.88 & 502.8565 & 11.9427 & -2.58 & 2.93 \\ 
33 & 510.6478 & 4.4881 & -2.15 & 2.79 & 518.1823 & 11.9468 & -2.53 & 2.92 \\ 
34 & 525.9735 & 4.4836 & -2.15 & 2.80 & 533.5081 & 11.9353 & -2.61 & 2.85 \\ 
35 & 541.2992 & 4.4736 & -2.10 & 2.81 & 548.8337 & 11.9158 & -2.50 & 2.93 \\ 
36 & 556.6248 & 4.4625 & -2.13 & 2.79 & 564.1593 & 11.8930 & -2.54 & 2.85 \\ 
37 & 571.9502 & 4.4480 & -2.11 & 2.81 & 579.4847 & 11.8694 & -2.63 & 2.90 \\ 
38 & 587.2755 & 4.4327 & -2.21 & 2.79 & 594.8100 & 11.8469 & -2.72 & 2.88 \\ 
39 & 602.6006 & 4.4184 & -2.26 & 2.82 & 610.1351 & 11.8267 & -2.79 & 2.91 \\ 
40 & 617.9255 & 4.4025 & -2.34 & 2.82 & 625.4601 & 11.8149 & -2.88 & 2.98 \\ 
41 & 633.2504 & 4.3860 & -2.35 & 2.82 & 640.7849 & 11.8126 & -2.94 & 2.85 \\ 
42 & 648.5955 & 4.3689 & -2.40 & 2.78 & 656.1300 & 11.8403 & -2.81 & 2.87 \\ 
43 & 663.9201 & 4.4192 & -2.28 & 2.77 & 671.4546 & 11.9076 & -2.86 & 2.85 \\ 
44 & 679.2446 & 4.5282 & -2.46 & 2.82 & 686.7792 & 11.9015 & -3.18 & 2.86 \\ 
45 & 694.5692 & 4.5233 & -2.60 & 2.87 & 702.1037 & 11.9192 & -3.28 & 2.84 \\ 
46 & 709.8938 & 4.5202 & -2.61 & 2.83 & 717.4283 & 11.9373 & -3.21 & 2.83 \\ 
47 & ... & ... & ... & ... & ... & ... & ... & ... \\ 
48 & 740.5432 & 4.5077 & -2.61 & 2.83 & 748.0777 & 11.9246 & -3.15 & 2.85 \\ 
49 & 755.8682 & 4.4959 & -2.57 & 2.83 & ... & ... & ... & ... \\ 
50 & 771.1933 & 4.4826 & -2.66 & 2.77 & 778.7278 & 11.8789 & -3.26 & 2.86 \\ 
51 & 786.5186 & 4.4682 & -2.60 & 2.81 & 794.0531 & 11.8559 & -3.27 & 2.83 \\ 
52 & 801.8440 & 4.4541 & -2.64 & 2.82 & 809.3786 & 11.8329 & -3.30 & 2.86 \\ 
53 & 817.1901 & 4.4394 & -2.68 & 2.80 & 824.7246 & 11.8117 & -3.26 & 2.82 \\ 
54 & 832.5158 & 4.4247 & -2.68 & 2.88 & 840.0503 & 11.8028 & -3.35 & 2.82 \\ 
55 & 847.8416 & 4.4062 & -2.67 & 2.85 & 855.3761 & 11.8107 & -3.34 & 2.85 \\ 
56 & 863.1674 & 4.3991 & -2.65 & 2.81 & 870.7020 & 11.8615 & -3.31 & 2.84 \\ 
57 & 878.4933 & 4.4867 & -2.63 & 2.79 & 886.0278 & 11.9122 & -3.37 & 2.82 \\ 
58 & 893.8191 & 4.5508 & -2.71 & 2.83 & 901.3536 & 11.9018 & -3.52 & 2.82 \\ 
59 & 909.1448 & 4.5441 & -2.71 & 2.83 & 916.6793 & 11.9259 & -3.40 & 2.80 \\ 
60 & 924.4703 & 4.5415 & -2.71 & 2.83 & 932.0049 & 11.9388 & -3.46 & 2.78 \\ 
61 & 939.7958 & 4.5342 & -2.72 & 2.85 & 947.3303 & 11.9356 & -3.36 & 2.83 \\ 
62 & 955.1211 & 4.5223 & -2.72 & 2.78 & 962.6557 & 11.9193 & -3.40 & 2.78 \\ 
63 & 970.4463 & 4.5100 & -2.63 & 2.66 & 977.9808 & 11.8984 & -3.42 & 2.79 \\ 
64 & 985.7713 & 4.4957 & -2.73 & 2.77 & 993.3058 & 11.8751 & -3.40 & 2.80 \\ 
65 & ... & ... & ... & ... & 1008.6307 & 11.8528 & -3.60 & 2.87 \\ 
66 & 1016.4413 & 4.4668 & -2.79 & 2.83 & 1023.9758 & 11.8293 & -3.54 & 2.78 \\ 
67 & 1031.7659 & 4.4523 & -2.74 & 2.79 & 1039.3004 & 11.8137 & -3.44 & 2.78 \\ 
68 & 1047.0905 & 4.4373 & -2.75 & 2.85 & 1054.6250 & 11.8063 & -3.44 & 2.79 \\ 
69 & 1062.4150 & 4.4218 & -2.71 & 2.86 & 1069.9495 & 11.8256 & -3.51 & 2.78 \\ 
70 & 1077.7396 & 4.4325 & -2.60 & 2.78 & 1085.2741 & 11.9016 & -3.38 & 2.81 \\ 
71 & 1093.0642 & 4.5402 & -2.74 & 2.80 & 1100.5987 & 11.9150 & -3.39 & 2.83 \\ 
72 & 1108.3889 & 4.5620 & -2.70 & 2.88 & 1115.9235 & 11.9160 & -3.42 & 2.85 \\ 
73 & 1123.7138 & 4.5594 & -2.70 & 2.78 & 1131.2484 & 11.9406 & -3.36 & 2.83 \\ 
74 & 1139.0389 & 4.5551 & -2.74 & 2.85 & 1146.5734 & 11.9484 & -3.48 & 2.78 \\ 
75 & ... & ... & ... & ... & ... & ... & ... & ... \\ 
76 & 1169.7100 & 4.5329 & -2.74 & 2.82 & 1177.2446 & 11.9232 & -3.41 & 2.79 \\ 
77 & 1185.0356 & 4.5174 & -2.71 & 2.83 & 1192.5701 & 11.9026 & -3.45 & 2.77 \\ 
78 & 1200.3613 & 4.5009 & -2.71 & 2.81 & 1207.8958 & 11.8782 & -3.48 & 2.82 \\ 
79 & 1215.6870 & 4.4898 & -2.61 & 2.64 & 1223.2216 & 11.8551 & -3.42 & 2.83 \\ 
80 & 1231.0129 & 4.4730 & -2.66 & 2.83 & 1238.5474 & 11.8341 & -3.45 & 2.78 \\ 
81 & 1246.3387 & 4.4609 & -2.68 & 2.85 & 1253.8732 & 11.8182 & -3.38 & 2.76 \\ 
82 & 1261.6645 & 4.4451 & -2.77 & 2.91 & ... & ... & ... & ... \\ 
83 & 1276.9903 & 4.4303 & -2.79 & 2.83 & 1284.5248 & 11.8558 & -3.42 & 2.72 \\ 
84 & ... & ... & ... & ... & 1299.8504 & 11.9417 & -3.43 & 2.72 \\ 
85 & 1307.6414 & 4.5621 & -2.64 & 2.90 & 1315.1759 & 11.9248 & -3.34 & 2.83 \\ 
86 & 1322.9667 & 4.5669 & -2.66 & 2.96 & 1330.5013 & 11.9406 & -3.40 & 2.75 \\ 
87 & 1338.2920 & 4.5667 & -2.74 & 2.87 & 1345.8265 & 11.9605 & -3.44 & 2.75 \\ 
88 & 1353.6170 & 4.5586 & -2.74 & 2.83 & 1361.1515 & 11.9625 & -3.41 & 2.75 \\ 
89 & 1368.9623 & 4.5468 & -2.74 & 2.85 & 1376.4969 & 11.9525 & -3.62 & 2.77 \\ 
90 & 1384.2871 & 4.5313 & -2.83 & 2.85 & 1391.8216 & 11.9353 & -3.55 & 2.70 \\ 
91 & 1399.6117 & 4.5150 & -2.82 & 2.79 & 1407.1463 & 11.9131 & -3.55 & 2.75 \\ 
92 & ... & ... & ... & ... & 1422.4708 & 11.8905 & -3.43 & 2.74 \\ 
93 & 1430.2608 & 4.4844 & -2.72 & 2.84 & 1437.7954 & 11.8663 & -3.43 & 2.76 \\ 
94 & 1445.5854 & 4.4706 & -2.73 & 2.86 & 1453.1199 & 11.8456 & -3.43 & 2.78 \\ 
95 & 1460.9100 & 4.4564 & -2.71 & 2.91 & 1468.4445 & 11.8321 & -3.41 & 2.73 \\ 
96 & 1476.5208 & 4.4392 & -2.71 & 2.89 & 1484.0553 & 11.9078 & -3.43 & 2.70 \\ 
97 & ... & ... & ... & ... & ... & ... & ... & ... \\ 
98 & 1506.9050 & 4.5093 & -2.77 & 2.85 & 1514.4395 & 11.9682 & -3.47 & 2.78 \\ 
99 & 1522.2302 & 4.5590 & -2.73 & 2.92 & 1529.7647 & 11.9494 & -3.47 & 2.75 \\ 
100 & 1537.5556 & 4.5621 & -2.64 & 2.94 & 1545.0901 & 11.9717 & -3.49 & 2.77 \\ 
101 & 1552.8811 & 4.5624 & -2.69 & 2.90 & 1560.4156 & 11.9858 & -3.36 & 2.72 \\ 
102 & 1568.2067 & 4.5518 & -2.66 & 2.89 & 1575.7413 & 11.9851 & -3.37 & 2.74 
\end{longtable}
}

\end{document}